\documentclass{ap-jnmp}

\usepackage{epsf}
\usepackage{epsfig}
\usepackage{floatflt}
\usepackage{float}
\usepackage{pdfpages}

\copyrightauthor{}

\title{Novel \textit{solvable} many-body problems}

\author{\footnotesize Oksana Bihun}

\address{Department of Mathematics, University of Colorado, Colorado Springs,\\
1420 Austin Bluffs Pkwy,\\
 Colorado Springs, CO  80918, USA\\ 
\email{obihun@uccs.edu}
}

\author{Francesco Calogero}

\address{
Physics Department, University of Rome \textquotedblleft La Sapienza'' \\
p. Aldo Moro, I-00185 Roma, Italy\\
Istituto Nazionale di Fisica Nucleare, Sezione di Roma, Italy\\
\email{francesco.calogero@roma1.infn.it, francesco.calogero@uniroma1.it}}

\begin{document}

\maketitle
\thispagestyle{empty}

\vphantom{\vbox{%
\begin{history}
\received{(Day Month Year)}
\revised{(Day Month Year)}
\accepted{(Day Month Year)}
\end{history}
}}

\begin{abstract}
Novel classes of dynamical systems are introduced, including many-body
problems characterized by nonlinear equations of motion of Newtonian type
(``acceleration equals forces'') which determine the motion of points in the 
\textit{complex} plane. These models are \textit{solvable}, namely their
configuration at any time can be obtained from the initial data by \textit{%
algebraic} operations, amounting to the determination of the \textit{zeros}
of a known time-dependent polynomial in the independent variable $z$. Some
of these models are \textit{multiply periodic}, \textit{isochronous} or 
\textit{asymptotically isochronous}; others display \textit{scattering}
phenomena.
\end{abstract}

\keywords{New solvable many-body problems; zeros and coefficients
of monic polynomials; generations of monic polynomials.}
\ccode{2000 Mathematics Subject Classification: 12D99, 70F10, 70K42. }

\section{Introduction}

\textbf{Notation 1.1}. Unless otherwise indicated, hereafter $N$
is an \textit{arbitrary positive integer}, $N\geq 2$, indices such as $n,$ $%
m,$ $\ell ,$ $...$ run over the \textit{integers} from $1$ to $N$, and
superimposed arrows denote $N$-vectors: for instance the vector $\vec{c}$
has the $N$ components $c_{m}$. We use instead a superimposed tilde to
denote an \textit{unordered} set of $N$ numbers: for instance the notation $%
\tilde{z}$ denotes the \textit{unordered} set of $N$ numbers $z_{n}$, say, the $N$ zeros
of a polynomial of degree $N$ in $z$.
Upper-case \textbf{boldface} letters denote $N\times N$ matrices: for
instance the matrix $\mathbf{M}$ features the $N^{2}$ elements $M_{nm}$. The
numbers we use are generally assumed to be \textit{complex}; except for
those restricted to be \textit{positive integers} (see above), which
generally play the role of indices; and except for the \textquotedblleft
time\textquotedblright~~variable, see below. The \textit{imaginary unit} is hereafter
denoted as $\mathbf{i}$, implying of course $\mathbf{i}^{2}=-1$. For
quantities depending on the \textit{real} independent variable $t$
(\textquotedblleft time\textquotedblright),  superimposed dots indicate
differentiation with respect to it: so, for instance, $\dot{z}_{n}\left(
t\right) \equiv dz_{n}\left( t\right) /dt$, $\ddot{z}_{n}\equiv
d^{2}z_{n}/dt^{2}$; but often the $t$-dependence is not explicitly
indicated, whenever this is unlikely to cause any misunderstanding (as, for
instance, in the second formula we just wrote and below in (\ref{Iden1})).
The Kronecker symbol $\delta _{nm}$ has the usual meaning: $\delta _{nm}=1$
if $n=m$, $\delta _{nm}=0$ if $n\neq m$; and we denote below as $\mathbf{I}$
the \textit{unit} $N\times N$ matrix the elements of which are $\delta _{nm}$%
. We adopt throughout the usual convention according to which a void sum
vanishes and a void product equals unity: $\sum_{j=J}^{K}f_{j}=0,$ $%
\prod\nolimits_{j=J}^{K}f_{j}=1$ if $J>K$. Moreover we introduce the
following convenient notations: 
\begin{subequations}
\label{sigma}
\begin{equation}
\sigma _{m}\left( \vec{z}\right) =\sum_{1\leq s_{1}<s_{2}<...<s_{m}\leq
N}\left( z_{s_{1}}z_{s_{2}}\cdots z_{s_{m}}\right) ~,  \label{sigma0}
\end{equation}%
\begin{equation}
\sigma _{n,m}\left( \vec{z}\right) =\delta _{1m}+\sum_{{\footnotesize 1\leq
s_{1}<s_{2}<\ldots <s_{m-1}\leq N~;~s_{j}\neq n,~j=1,...m-1}}\left(
z_{s_{1}}z_{s_{2}}\cdots z_{s_{m-1}}\right) ~,  \label{sigma1}
\end{equation}%
\begin{equation}
\sigma _{n_{1}n_{2},m}\left( \vec{z}\right) =\delta _{2m}+\sum_{\substack{ 
{\footnotesize 1\leq s_{1}<s_{2}<\ldots <s_{m-2}\leq N~;} \\ {\footnotesize %
s_{j}\neq n}_{1},~{\footnotesize s_{j}\neq n_{2},~j=1,...m-2}}}\left(
z_{s_{1}}z_{s_{2}}\cdots z_{s_{m-2}}\right) ~,  \label{sigma2}
\end{equation}%
\begin{equation}
\sigma _{n_{1}n_{2}n_{3},m}\left( \vec{z}\right) =\delta _{3m}+\sum
_{\substack{ {\footnotesize 1\leq s_{1}<s_{2}<\ldots <s_{m-3}\leq N~;} \\ 
{\footnotesize s_{j}\neq n}_{1}{\footnotesize ,~s_{j}\neq n_{2},~s_{j}\neq
n_{3},~j=1,...m-3}}}\left( z_{s_{1}}z_{s_{2}}\cdots z_{s_{m-3}}\right) ~,
\label{sigma3}
\end{equation}%
where of course the symbol $\sum_{1\leq s_{1}<s_{2}<...<s_{m}\leq N}$
denotes the sum from $1$ to $N$ over the $m$ integer indices $%
s_{1},s_{2},\ldots ,s_{m}$ with the restriction that $s_{1}<s_{2}<\ldots
<s_{m}$, while the symbol $\sum_{{\footnotesize 1\leq s_{1}<s_{2}<\ldots
<s_{m-1}\leq N~;~s_{j}\neq n,~j=1,...m-1}}$ denotes the sum from $1$ to $N$
over the $m-1$ indices $s_{1},s_{2},\ldots ,s_{m-1}$ with the restriction $%
s_{1}<s_{2}<\ldots <s_{m-1}$ and moreover the requirement that \textit{all
these indices be different from }$n;$ and likewise for the symbols $\sum_{%
{\footnotesize 1\leq s_{1}<s_{2}<\ldots <s_{m-2}\leq N~;~s_{j}\neq n}_{1},~%
{\footnotesize s_{j}\neq n_{2},~j=1,...m-2}}$ and $\sum_{{\footnotesize %
1\leq s_{1}<s_{2}<\ldots <s_{m-3}\leq N~;~s_{j}\neq n}_{1}{\footnotesize %
,~s_{j}\neq n_{2},~s_{j}\neq n_{3},~j=1,...m-3}}$. Note that---according to
the convention (see above) that a sum over an empty set of indices equals
zero---these definitions imply $\sigma _{n,1}(\vec{z})=1,$ $\sigma
_{n_{1}n_{2},1}\left( \vec{z}\right) =0$ and $\sigma _{n_{1}n_{2},2}\left( 
\vec{z}\right) =1,$ and $\sigma _{n_{1}n_{2}n_{3},m}\left( \vec{z}\right) =0$
for $m\leq 2$ while $\sigma _{n_{1}n_{2}n_{3},3}\left( \vec{z}\right) =1$.
Finally, the prime appended to a sum (see for instance below (\ref{Iden1}) and also note the simplification it would imply for~(\ref{sigma1}))
indicates that the sum runs---over the indicated indices, in the identified
range---with the additional restrictions that these indices be \textit{all
different among themselves} and moreover \textit{all different from the
``outside'' index} (which is for instance $n$ in (\ref{Iden1})); note that
this sum becomes void hence vanishes identically if $N$ is small enough, so
for instance the last sum in the left-hand side of (\ref{zndot4}) vanishes
for $N\leq 3$, and more generally the ``primed'' sum from $1$ to $N$ over $k$
indices $\ell _{1},$ $\ell _{2},$..., $\ell _{k}$ vanishes identically if $%
N\leq k$. $\blacksquare $

\textbf{Remark 1.1}. Note that the notation $\sigma _{m}\left( \tilde{z}%
\right) $ (instead of $\sigma _{m}\left( \vec{z}\right) $) is equally meaningful, 
since this quantity, see (\ref{sigma0}), only depends on \textit{%
symmetrical sums} of the $N$ components $z_{m}$ of the $N$-vector $\vec{z},$
hence it is independent of the ordering of the $N$ elements $z_{n}$ of the 
\textit{unordered} set $\tilde{z}$. The notations $\sigma _{n,m}\left( 
\tilde{z}\right) $, $\sigma _{n_{1}n_{2},m}\left( \tilde{z}\right) $, $%
\sigma _{n_{1}n_{2}n_{3},m}\left( \tilde{z}\right) ,$ see (\ref{sigma}), are
instead ill-defined and cannot therefore be used; except in the context of
expressions which remain valid for \textit{any} ordering of the $N$ numbers $%
z_{n}$, i. e., for any assignments of the $N$ different integer labels $n$
(in the range $1\leq n\leq N)$ to the $N$ elements of the \textit{unordered}
set $\tilde{z}$; provided of course that assignment is maintained throughout
that expression (in which case the relevant expression amounts in fact to $%
N! $ different formulas; assuming, as we generally do, that the $N$ numbers $%
z_{n}$ are \textit{all different among themselves}). This remark is of
course equally valid for any function $f\left( \tilde{z}\right) $. $%
\blacksquare $

The main protagonists of this paper are formulas relating the time-evolution of
the $N$ zeros $z_{n}\left( t\right) $ of a time-dependent monic polynomial
of degree $N$ in the independent variable $z,$%
\end{subequations}
\begin{subequations}
\label{Pol}
\begin{equation}
p_{N}\left( z;~\vec{c}\left( t\right) ,~\tilde{z}\left( t\right) \right)
=\prod\limits_{n=1}^{N}\left[ z-z_{n}\left( t\right) \right] ~,
\label{Polzn}
\end{equation}%
to the time-evolution of its $N$ coefficients $c_{m}\left( t\right) ,$%
\begin{equation}
p_{N}\left( z;~\vec{c}\left( t\right) ,~\tilde{z}\left( t\right) \right)
=z^{N}+\sum_{m=1}^{N}\left[ c_{m}\left( t\right) ~z^{N-m}\right] ~.
\label{Polcm}
\end{equation}%
The first two of these formulas read as follows \cite{C2015a}: 
\end{subequations}
\begin{subequations}
\label{Iden1}
\begin{equation}
\dot{z}_{n}=-\left[ \prod\limits_{\ell =1,~\ell \neq n}^{N}\left(
z_{n}-z_{\ell }\right) ^{-1}\right] ~\sum_{m=1}^{N}\left[ \dot{c}_{m}~\left(
z_{n}\right) ^{N-m}\right] ~,  \label{zndot1}
\end{equation}%
\begin{equation}
\ddot{z}_{n}-\sum_{\ell =1}^{N}{}^{\prime }\left( \frac{2~\dot{z}_{n}~\dot{z}%
_{\ell }}{z_{n}-z_{\ell }}\right) =-\left[ \prod\limits_{\ell =1,~\ell \neq
n}^{N}\left( z_{n}-z_{\ell }\right) ^{-1}\right] ~\sum_{m=1}^{N}\left[ \ddot{%
c}_{m}~\left( z_{n}\right) ^{N-m}\right] ~.  \label{zndot2}
\end{equation}%
In the present paper we report two additional formulas of this kind:%
\begin{eqnarray}
&&\dddot{z}_{n}-3~\sum_{\ell =1}^{N}{}^{\prime }\left( \frac{\ddot{z}_{n}~%
\dot{z}_{\ell }+\ddot{z}_{\ell }~\dot{z}_{n}}{z_{n}-z_{\ell }}\right)
+3~\sum_{\ell _{1},\ell _{2}=1}^{N}{}^{\prime }\left[ \frac{\dot{z}_{n}~\dot{%
z}_{\ell _{1}}~\dot{z}_{\ell _{2}}}{\left( z_{n}-z_{\ell _{1}}\right)
~\left( z_{n}-z_{\ell _{2}}\right) }\right]   \notag \\
&=&-\left[ \prod\limits_{\ell =1,~\ell \neq n}^{N}\left( z_{n}-z_{\ell
}\right) ^{-1}\right] ~\sum_{m=1}^{N}\left[ \dddot{c}_{m}~\left(
z_{n}\right) ^{N-m}\right] ~,  \label{zndot3}
\end{eqnarray}%
\begin{eqnarray}
&&\ddddot{z}_{n}-\sum_{\ell =1}^{N}{}^{\prime }\left( \frac{4~\dddot{z}_{n}~%
\dot{z}_{\ell }+4~\dddot{z}_{\ell }~\dot{z}_{n}+6~\ddot{z}_{n}~\ddot{z}%
_{\ell }}{z_{n}-z_{\ell }}\right)   \notag \\
&&+6~\sum_{\ell _{1},\ell _{2}=1}^{N}{}^{\prime }\left[ \frac{\ddot{z}_{n}~%
\dot{z}_{\ell _{1}}~\dot{z}_{\ell _{2}}+2~\ddot{z}_{\ell _{1}}~\dot{z}_{\ell
_{2}}~\dot{z}_{n}}{\left( z_{n}-z_{\ell _{1}}\right) ~\left( z_{n}-z_{\ell
_{2}}\right) }\right]   \notag \\
&&-4~\sum_{\ell _{1},\ell _{2},~\ell _{3}=1}^{N}{}^{\prime }\left[ \frac{%
\dot{z}_{n}~\dot{z}_{\ell _{1}}~\dot{z}_{\ell _{2}}~\dot{z}_{\ell _{3}}}{%
\left( z_{n}-z_{\ell _{1}}\right) ~\left( z_{n}-z_{\ell _{2}}\right) ~\left(
z_{n}-z_{\ell _{3}}\right) }\right]   \notag \\
&=&-\left[ \prod\limits_{\ell =1,~\ell \neq n}^{N}\left( z_{n}-z_{\ell
}\right) ^{-1}\right] ~\sum_{m=1}^{N}\left[ \ddddot{c}_{m}~\left(
z_{n}\right) ^{N-m}\right] ~.  \label{zndot4}
\end{eqnarray}%
A terse outline of the proof of these identities is reported in Appendix~A.

The first two of the formulas (\ref{Iden1}) have recently allowed the
identification of (endless sequences of) new \textit{solvable} many-body
problems characterized by nonlinear equations of motion of Newtonian type
(``acceleration equals forces'') determining the motion of $N$ points in the 
\textit{complex} $z$-plane \cite{C2015a, BC2015a, C2015b, BC2015b}. In the present paper we show how the last two of the formulas (%
\ref{Iden1}) allow the identification of additional endless sequences of new 
\textit{solvable} dynamical systems determining the motion of points in the 
\textit{complex} $z$-plane---also including many-body problems characterized
by nonlinear equations of motion of Newtonian type (``acceleration equals
forces'').

Note that the notation (\ref{Pol}), which we employ for polynomials, is somewhat
redundant, since they are equally well defined by the (time-dependent) $N$%
-vector $\vec{c}\left( t\right) $ the $N$ components of which are the $N$ 
\textit{coefficients} $c_{m}\left( t\right) $ of the polynomial (see (\ref%
{Polcm})), as by the (time-dependent) unordered set $\tilde{z}\left(
t\right) $ the $N$ elements of which are the $N$ \textit{zeros} $z_{n}\left(
t\right) $ of the polynomial (see (\ref{Polzn})). Indeed the $N$ \textit{%
coefficients} $c_{m}\left( t\right) $ can be \textit{explicitly} expressed
in terms of the $N$ \textit{zeros }$z_{n}\left( t\right) $ as follows: 
\end{subequations}
\begin{equation}
c_{m}=\left( -1\right) ^{m}~\sigma _{m}\left( \vec{z}\right) \equiv \left(
-1\right) ^{m}~\sigma _{m}\left( \tilde{z}\right)   \label{cm}
\end{equation}%
(see \textbf{Notation 1.1} and \textbf{Remark 1.1}). While the $N$ \textit{%
zeros }$z_{n}\left( t\right) $ are likewise uniquely determined (up to
permutations) by the $N$ \textit{coefficients} $c_{m}\left( t\right) $, but
of course \textit{explicit} expressions to this effect are generally
available only for $N\leq 4$.

There holds moreover the following identity: 
\begin{subequations}
\label{Id12}
\begin{equation}
\left( z_{n}\right) ^{N}+\sum_{m=1}^{N}\left[ c_{m}~\left( z_{n}\right)
^{N-m}\right] =0~,  \label{Identity1}
\end{equation}%
which is an obvious consequence of (\ref{Pol}), and via (\ref{cm}) it implies%
\begin{equation}
\left( z_{n}\right) ^{N}+\sum_{m=1}^{N}\left[ \left( -1\right) ^{m}~\sigma
_{m}\left( \tilde{z}\right) ~\left( z_{n}\right) ^{N-m}\right] =0~.
\label{Identity2}
\end{equation}%
Note that, while the formula (\ref{Identity1}) is an identity valid for the $%
N$ \textit{coefficients} $c_{m}$ and the $N$ \textit{zeros }$z_{n}$ of any
polynomial, see (\ref{Pol}), the identity (\ref{Identity2}) is clearly valid
for \textit{any} \textit{arbitrary} assignment of the $N$ elements $z_{n}$
of the unordered set $\tilde{z}$.

Likewise, there holds the following formula that is also clearly valid for 
\textit{any} assignment of the $N$ elements $z_{n}$ of the unordered set $%
\tilde{z}$ (see \textbf{Notation 1.1} and \textbf{Remark 1.1}): 
\end{subequations}
\begin{subequations}
\label{Id3}
\begin{equation}
-\left[ \prod\limits_{\ell =1,~\ell \neq n}^{N}\left( z_{n}-z_{\ell }\right)
^{-1}\right] ~\sum_{j=1}^{N}\left[ \left( -1\right) ^{j}~\left( z_{n}\right)
^{N-j}~\sigma _{m,j}\left( \tilde{z}\right) \right] =\delta _{nm}~;
\end{equation}%
and note that this formula can also be rewritten in the following $\left(
N\times N\right) $-matrix version:%
\begin{equation}
\left[ \mathbf{R}\left( \tilde{z}\right) \right] _{nm}\equiv R_{nm}\left( 
\tilde{z}\right) =-\left[ \prod\limits_{\ell =1,~\ell \neq n}^{N}\left(
z_{n}-z_{\ell }\right) ^{-1}\right] ~\left( z_{n}\right) ^{N-m}~,  \label{R}
\end{equation}%
\begin{equation}
\left[ \mathbf{R}^{-1}\left( \tilde{z}\right) \right] _{nm}\equiv \left[
R^{-1}\left( \tilde{z}\right) \right] _{nm}=\left( -1\right) ^{n}~\sigma
_{n,m}\left( \tilde{z}\right) ~,  \label{Rinv}
\end{equation}%
implying of course (see \textbf{Notation 1.1} and \textbf{Remark 1.1}) 
\begin{equation}
\mathbf{R}\left( \tilde{z}\right) ~\mathbf{R}^{-1}\left( \tilde{z}\right) =%
\mathbf{R}^{-1}\left( \tilde{z}\right) ~\mathbf{R}\left( \tilde{z}\right) =%
\mathbf{I~.}
\end{equation}

Finally let us report $3$ additional identities which are obvious
consequences of the definitions (\ref{cm}) and (\ref{sigma}) (see \textbf{%
Notation 1.1} and \textbf{Remark 1.1}): 
\end{subequations}
\begin{subequations}
\label{cmdot}
\begin{equation}
\dot{c}_{m}=\left( -1\right) ^{m}~\dot{\sigma}_{m}\left( \vec{z}\right)
\equiv \left( -1\right) ^{m}~\sum_{n=1}^{N}\left[ \sigma _{n,m}\left( \tilde{%
z}\right) ~\dot{z}_{n}\right] ~,  \label{cmdot1}
\end{equation}%
\begin{eqnarray}
&&\ddot{c}_{m}=\left( -1\right) ^{m}~\ddot{\sigma}_{m}\left( \vec{z}\right) 
\notag \\
&&\equiv \left( -1\right) ^{m}~\left\{ \sum_{n=1}^{N}\left[ \sigma
_{n,m}\left( \tilde{z}\right) ~\ddot{z}_{n}\right] +\sum_{n_{1},n_{2}=1, n_1 \neq n_2}^{N}%
\left[ \sigma _{n_{1}n_{2},m}\left( \tilde{z}\right) ~\dot{z}_{n_{1}}~\dot{z}%
_{n_{2}}\right] \right\} ~,  \label{cmdot2}
\end{eqnarray}%
\begin{eqnarray}
&&\dddot{c}_{m}=\left( -1\right) ^{m}~\dddot{\sigma}_{m}\left( \vec{z}%
\right) \equiv \left( -1\right) ^{m}~\left\{ \sum_{n=1}^{N}\left[ \sigma
_{n,m}\left( \tilde{z}\right) ~\dddot{z}_{n}\right] \right.  \notag \\
&&\left. +3~\sum_{n_{1},n_{2}=1, n_1 \neq n_2}^{N}\left[ \sigma _{n_{1}n_{2},m}\left( 
\tilde{z}\right) ~\ddot{z}_{n_{1}}~\dot{z}_{n_{2}}\right] \right.  \notag \\
&&\left. +\sum_{n_{1},n_{2},n_{3}=1, n_1\neq n_2\neq n_3}^{N}\left[ \sigma
_{n_{1}n_{2}n_{3},m}\left( \tilde{z}\right) ~\dot{z}_{n_{1}}~\dot{z}_{n_{2}}~%
\dot{z}_{n_{3}}\right] \right\} ~,  \label{cmdot3}
\end{eqnarray}
where the indices $n_1, n_2, n_3$ in the last sum all different among themselves.

In Section 3 it is indicated how these polynomial properties are
instrumental to identify endless classes of \textit{solvable} dynamical
systems including many-body problems of Newtonian type, one of which is
immediately reported in the following Section 2, while some of its solutions are 
displayed in Appendix B. The paper is then concluded by a section entitled
\textquotedblleft Outlook\textquotedblright , where further investigations
are tersely outlined.

\bigskip

\section{Display and discussion of a novel \textit{solvable} many-body
problem}

In this section we provide and discuss an instance of the novel \textit{%
solvable} many-body problems of Newtonian type identified in this paper. Its
equations of motion, characterizing the time-evolution of the $2N$ complex
dependent variables $z_{n}\equiv z_{n}\left( t\right) $ and $w_{n}\equiv
w_{n}\left( t\right) ,$ read as follows: 
\end{subequations}
\begin{subequations}
\label{ManyBodyNewt}
\begin{equation}
\ddot{z}_{n}=w_{n}~,  \label{zttw}
\end{equation}%
\begin{eqnarray}
&&\ddot{w}_{n}=\sum_{\ell =1}^{N}{}^{\prime }\left( \frac{4~\dot{w}_{n}~\dot{%
z}_{\ell }+4~\dot{w}_{\ell }~\dot{z}_{n}+6~w_{n}~w_{\ell }}{z_{n}-z_{\ell }}%
\right)  \notag \\
&&-6~\sum_{\ell _{1},\ell _{2}=1}^{N}{}^{\prime }\left[ \frac{w_{n}~\dot{z}%
_{\ell _{1}}~\dot{z}_{\ell _{2}}+2~w_{\ell _{1}}~\dot{z}_{n}~\dot{z}_{\ell
_{2}}}{\left( z_{n}-z_{\ell _{1}}\right) ~\left( z_{n}-z_{\ell _{2}}\right) }%
\right]  \notag \\
&&+4~\sum_{\ell _{1},\ell _{2},~\ell _{3}=1}^{N}{}^{\prime }\left[ \frac{%
\dot{z}_{n}~\dot{z}_{\ell _{1}}~\dot{z}_{\ell _{2}}~\dot{z}_{\ell _{3}}}{%
\left( z_{n}-z_{\ell _{1}}\right) ~\left( z_{n}-z_{\ell _{2}}\right) ~\left(
z_{n}-z_{\ell _{3}}\right) }\right] -\left[ \prod\limits_{\ell =1,~\ell \neq
n}^{N}\left( z_{n}-z_{\ell }\right) ^{-1}\right] \cdot  \notag \\
&&\cdot \sum_{m=1}^{N}\left[ \left( \alpha _{m}~\dddot{c}_{m}+\beta _{m}~%
\ddot{c}_{m}+\gamma _{m}~\dot{c}_{m}+\delta _{m}~c_{m}\right) ~\left(
z_{n}\right) ^{N-m}\right] ~,  \label{wdotdot}
\end{eqnarray}%
with $c_{m}$, $\dot{c}_{m}$ expressed in terms of $z_{n}$ and $\dot{z}_{n}$
by (\ref{cmdot1}) and (\ref{cm}) and $\ddot{c}_{m}$, $\dddot{c}_{m}$
expressed in terms of the dependent variables $z_{n}$, $w_{n}$ and their
time derivatives $\dot{z}_{n}$, $\dot{w}_{n}$ as follows (see \textbf{%
Notation 1.1} and \textbf{Remark 1.1}), 
\begin{equation}
\ddot{c}_{m}=\left( -1\right) ^{m}~\left\{ \sum_{n=1}^{N}\left[ \sigma
_{n,m}\left( \tilde{z}\right) ~w_{n}\right] +\sum_{n_{1},n_{2}=1}^{N}\left[
\sigma _{n_{1}n_{2},m}\left( \tilde{z}\right) ~\dot{z}_{n_{1}}~\dot{z}%
_{n_{2}}\right] \right\} ~,  \label{c2zw}
\end{equation}%
\begin{eqnarray}
&&\dddot{c}_{m}=\left( -1\right) ^{m}~\left\{ \sum_{n=1}^{N}\left[ \sigma
_{n,m}\left( \tilde{z}\right) ~\dot{w}_{n}\right] +3~\sum_{n_{1},n_{2}=1}^{N}%
\left[ \sigma _{n_{1}n_{2},m}\left( \tilde{z}\right) ~w_{n_{1}}~\dot{z}%
_{n_{2}}\right] \right.  \notag \\
&&\left. +\sum_{n_{1},n_{2},n_{3}=1}^{N}\left[ \sigma
_{n_{1}n_{2}n_{3},m}\left( \tilde{z}\right) ~\dot{z}_{n_{1}}~\dot{z}_{n_{2}}~%
\dot{z}_{n_{3}}\right] \right\} ~.  \label{c3zw}
\end{eqnarray}%
In (\ref{wdotdot}) the parameters $\alpha _{m},~\beta _{m},~\gamma
_{m},~\delta _{m}$ are $4N$ arbitrary \textit{complex} numbers, which may be
conveniently related to the $8N$ \textit{real} parameters $a_{m}^{\left(
1\right) }$, $a_{m}^{\left( 2\right) }$, $a_{m}^{\left( 3\right) }$, $%
a_{m}^{\left( 4\right) }$, $\omega _{m}^{\left( 1\right) }$, $\omega
_{m}^{\left( 2\right) }$, $\omega _{m}^{\left( 3\right) }$, $\omega
_{m}^{\left( 4\right) }$ (for their role see below eq. (\ref{cmt})) by the
following formulas 
\end{subequations}
\begin{subequations}
\label{abcd}
\begin{equation}
\alpha _{m}=-a_m^{(1)}- a_m^{(2)} - a_m^{(3)} - a_m^{(4)} + \mathbf{i}\Big[ \omega_m^{(1)} + \omega_m^{(2)} +  \omega_m^{(3)} + \omega_m^{(4)}\Big]~,
\end{equation}
\begin{eqnarray}
&&\beta _{m}=-a_m^{(1)} a_m^{(2)} - a_m^{(1)} a_m^{(3)} - a_m^{(2)} a_m^{(3)} - a_m^{(1)} a_m^{(4)} - a_m^{(2)} a_m^{(4)} - 
 a_m^{(3)} a_m^{(4)} \notag\\
 &&+ \omega_m^{(1)} \omega_m^{(2)} + \omega_m^{(1)} \omega_m^{(3)} + \omega_m^{(2)} 
\omega_m^{(3)} + \omega_m^{(1)} \omega_m^{(4)} + \omega_m^{(2)} \omega_m^{(4)} + \
\omega_m^{(3)} \omega_m^{(4)} \notag\\
&&+ 
\mathbf{i}\Big[a_m^{(2)} \omega_m^{(1)} + a_m^{(3)} \omega_m^{(1)} + a_m^{(4)} \omega_m^{(1)} 
    + a_m^{(1)} \omega_m^{(2)} + a_m^{(3)} \omega_m^{(2)} + a_m^{(4)} \omega_m^{(2)} \notag\\
    &&+ a_m^{(1)} \omega_m^{(3)} + a_m^{(2)} \omega_m^{(3)} + a_m^{(4)} \omega_m^{(3)}
     + a_m^{(1)} \omega_m^{(4)} + a_m^{(2)} \omega_m^{(4)} + a_m^{(3)} \omega_m^{(4)}\Big]~,
\end{eqnarray}%
\begin{eqnarray}
&&\gamma _{m}=-a_m^{(1)} a_m^{(2)} a_m^{(3)} - a_m^{(1)} a_m^{(2)} a_m^{(4)} - a_m^{(1)} a_m^{(3)} a_m^{(4)} - a_m^{(2)} a_m^{(3)} a_m^{(4)} \notag\\
&&+ a_m^{(1)} \omega_m^{(2)} \omega_m^{(3)} + a_m^{(1)} \omega_m^{(2)} \omega_m^{(4)} + 
 a_m^{(1)} \omega_m^{(3)} \omega_m^{(4)} + a_m^{(2)} \omega_m^{(1)} \omega_m^{(3)} \notag\\
 &&+ 
 a_m^{(2)} \omega_m^{(1)} \omega_m^{(4)} + a_m^{(2)} \omega_m^{(3)} \omega_m^{(4)} + 
 a_m^{(3)} \omega_m^{(1)} \omega_m^{(2)} + a_m^{(3)} \omega_m^{(1)} \omega_m^{(4)}\notag\\
 && + 
 a_m^{(3)} \omega_m^{(2)} \omega_m^{(4)} + a_m^{(4)} \omega_m^{(1)} \omega_m^{(2)} + 
 a_m^{(4)} \omega_m^{(1)} \omega_m^{(3)} + a_m^{(4)} \omega_m^{(2)} \omega_m^{(3)} \notag\\
 &&+ 
 \mathbf{i}\Big[a_m^{(2)} a_m^{(3)} \omega_m^{(1)} + a_m^{(2)} a_m^{(4)} \omega_m^{(1)} + a_m^{(3)} a_m^{(4)} \omega_m^{(1)}+ 
    a_m^{(1)} a_m^{(3)} \omega_m^{(2)} \notag\\
 &&+ a_m^{(1)} a_m^{(4)} \omega_m^{(2)} + a_m^{(3)} a_m^{(4)} \omega_m^{(2)} + 
    a_m^{(1)} a_m^{(2)} \omega_m^{(3)} + a_m^{(1)} a_m^{(4)} \omega_m^{(3)} \notag\\
 &&+ a_m^{(2)} a_m^{(4)} \omega_m^{(3)}+ 
    a_m^{(1)} a_m^{(2)} \omega_m^{(4)} + a_m^{(1)} a_m^{(3)} \omega_m^{(4)} + 
    a_m^{(2)} a_m^{(3)} \omega_m^{(4)}\notag\\
    && - \omega_m^{(1)} \omega_m^{(2)} \omega_m^{(3)} - 
\omega_m^{(1)} \omega_m^{(2)} \omega_m^{(4)} - \omega_m^{(1)} \omega_m^{(3)} \omega_m^{(4)} - 
\omega_m^{(2)} \omega_m^{(3)} \omega_m^{(4)}\Big]~,
\end{eqnarray}%
\begin{eqnarray}
&&\delta _{m}=-a_m^{(1)} a_m^{(2)} a_m^{(3)} a_m^{(4)} + a_m^{(3)} a_m^{(4)} \omega_m^{(1)} \omega_m^{(2)} + 
 a_m^{(2)} a_m^{(4)} \omega_m^{(1)} \omega_m^{(3)} + a_m^{(2)} a_m^{(3)} \omega_m^{(1)} \omega_m^{(4)} \notag\\
 &&+ 
 a_m^{(1)} a_m^{(4)} \omega_m^{(2)} \omega_m^{(3)} + a_m^{(1)} a_m^{(3)} \omega_m^{(2)} \omega_m^{(4)} + 
 a_m^{(1)} a_m^{(2)} \omega_m^{(3)} \omega_m^{(4)} - \omega_m^{(1)} \omega_m^{(2)} \omega_m^{(3)} 
\omega_m^{(4)}\notag\\
&&+\mathbf{i} \Big[ a_m^{(2)} a_m^{(3)} a_m^{(4)} \omega_m^{(1)} + a_m^{(1)} a_m^{(3)} a_m^{(4)} \omega_m^{(2)} + 
   a_m^{(1)} a_m^{(2)} a_m^{(4)} \omega_m^{(3)} \notag\\
 &&+ a_m^{(1)} a_m^{(2)} a_m^{(3)} \omega_m^{(4)}- 
   a_m^{(4)} \omega_m^{(1)} \omega_m^{(2)} \omega_m^{(3)} - 
   a_m^{(3)} \omega_m^{(1)} \omega_m^{(2)} \omega_m^{(4)} \notag\\
 &&- 
   a_m^{(2)} \omega_m^{(1)} \omega_m^{(3)} \omega_m^{(4)} - 
   a_m^{(1)} \omega_m^{(2)} \omega_m^{(3)} \omega_m^{(4)}\Big]~.
\end{eqnarray}%
As explained in the following Section 3, it is also possible to invert these
equations, i. e. to write formulas expressing the $8N$ \textit{real}
parameters $a_{m}^{\left( 1\right) }$, $a_{m}^{\left( 2\right) }$, $%
a_{m}^{\left( 3\right) }$, $a_{m}^{\left( 4\right) }$, $\omega _{m}^{\left(
1\right) }$, $\omega _{m}^{\left( 2\right) }$, $\omega _{m}^{\left( 3\right)
}$, $\omega _{m}^{\left( 4\right) }$ in terms of the $4N$ \textit{complex} parameters $\alpha
_{m},~\beta _{m},~\gamma _{m},~\delta _{m}$, but in view of the complicated
nature of these expressions---essentially based on the solution of algebraic
equations of fourth degree---this does not seem useful (see below \textbf{%
Remark 3.2}).

As shown in the following Section 3, the \textit{general solution} of this $%
\left( 2N\right) $-body problem is provided by the following prescription:
the values of the coordinates $w_{n}\left( t\right) $ are of course provided
by (\ref{zttw}), while the values of the $N$ coordinates $z_{n}\left(
t\right) $ are the $N$ zeros of the monic polynomial (\ref{Polcm}) where the 
$N$ coefficients $c_{m}\left( t\right) $---being the solutions of the 
\textit{solvable} dynamical system (\ref{Systcm})---are provided by the
following formulas: 
\end{subequations}
\begin{equation}
c_{m}\left( t\right) =\sum_{k=1}^{4}\left\{ b_{m}^{\left( k\right) }~\exp 
\left[ \left( -a_{m}^{\left( k\right) }+\mathbf{i}~\omega _{m}^{\left(
k\right) }\right) ~t\right] \right\} ~.  \label{cmt}
\end{equation}%
Here the coefficients $b_{m}^{\left( k\right) }$ are $4N$ \textit{a priori}
arbitrary \textit{complex} parameters. And the solution of the \textit{%
initial value problem} for this $\left( 2N\right) $-body problem, (\ref%
{ManyBodyNewt}), is obtained by determining the $4N$ coefficients $%
b_{m}^{\left( k\right) }$ as solutions, for every value of the parameter $m$%
, of the system of $4$ \textit{linear} algebraic equations%
\begin{equation}
\sum_{k=1}^{4}\left[ b_{m}^{\left( k\right) }~\left( -a_{m}^{\left( k\right)
}+\mathbf{i}~\omega _{m}^{\left( k\right) }\right) ^{s}\right] =\left. \frac{%
d^{s}c_{m}\left( t\right) }{dt^{s}}\right\vert _{t=0}~,~~~s=0,1,2,3~,
\end{equation}%
with, in the right-hand side, $c_{m}\left( 0\right) $, $\dot{c}_{m}\left(
0\right) $ expressed in terms of the initial data $z_{n}\left( 0\right) $, $%
\dot{z}_{n}\left( 0\right) $ by (\ref{cm}) and (\ref{cmdot1}) (at $t=0$),
and $\ddot{c}_{m}\left( 0\right) ~,\dddot{c}_{m}\left( 0\right) $ expressed
in terms of the initial data $z_{n}\left( 0\right) $, $\dot{z}_{n}\left(
0\right) $ and $w_{n}\left( 0\right) $, $\dot{w}_{n}\left( 0\right) $ by (%
\ref{c2zw}) and (\ref{c3zw}) (at $t=0$).

\textbf{Remark 2.1}. Above and hereafter we assume for simplicity that the $%
4N$ \textit{complex} numbers $\lambda _{m,k}=-a_{m}^{\left( k\right) }+\mathbf{i}%
~\omega _{m}^{\left( k\right) }$ (see (\ref{cmt})) are \textit{all different
among themselves}; otherwise some appropriate limit should be taken in (\ref%
{cmt}) and some of the statements made in the following \textbf{Remark 2.2}
would require additional restrictions.\textbf{\ }$\blacksquare $

\textbf{Remark 2.2}. The following properties of various subcases of the
many-body problem characterized by the Newtonian equations of motion (\ref%
{ManyBodyNewt}) are obviously implied by its \textit{general solution}, as
detailed above.

(i) If the $4N$ \textit{real} parameters $a_{m}^{\left( k\right) }$ are 
\textit{all nonnegative}, $a_{m}^{\left( k\right) }\geq 0,$ then \textit{all}
solutions of this many-body problem are, for \textit{all future time},
confined to a \textit{finite} region---the dimensions of which depend on the initial
data---of the complex $z$ and $w$ planes; and in particular if the $4N$ 
\textit{real} parameters $a_{m}^{\left( k\right) }$ are \textit{all positive}%
, $a_{m}^{\left( k\right) }>0$, \textit{all} solutions of this many-body
problem converge to the origin,%
\begin{equation}
\lim_{t\rightarrow \infty }\left[ z_{n}\left( t\right) \right]
=0~,~~~\lim_{t\rightarrow \infty }\left[ w_{n}\left( t\right) \right] =0~;
\end{equation}%
while if only \textit{some} of the $4N$ \textit{real} parameters $%
a_{m}^{\left( k\right) }$ are \textit{positive} and \textit{all} others 
\textit{vanish}, then this many-body problem is \textit{asymptotically
multiply periodic}; and if in addition the \textit{real} parameters $\omega
_{m}^{\left( k\right) }$, such that the corresponding parameter $%
a_{m}^{\left( k\right) }$ vanishes, are \textit{all integer multiples} of a
common (nonvanishing) \textit{real} factor $\omega \neq 0$, i. e. if for 
\textit{some} values of the indices $m$ and $k$ the parameters $%
a_{m}^{\left( k\right) }$ are positive, $a_{m}^{\left( k\right) }>0,$ while
for \textit{all other }values of the indices $m$ and $k$ 
\begin{equation}
a_{m}^{\left( k\right) }=0~,~~~\omega _{m}^{\left( k\right) }=p_{mk}~\omega 
\end{equation}%
with these parameters $p_{mk}$ being \textit{all integers} (positive,
negative or vanishing, but all different among themselves), then this
many-body system is \textit{asymptotically isochronous.} \cite{CG2008}

(ii) If the $4N$ \textit{real} parameters $a_{m}^{\left( k\right) }$ \textit{%
all} vanish, $a_{m}^{\left( k\right) }=0$, and the $4N$ \textit{real}
parameters $\omega _{m}^{\left( k\right) }$ are \textit{all integer multiples%
} of a common (nonvanishing) \textit{real} factor $\omega \neq 0$, $\omega
_{m}^{\left( k\right) }=p_{mk}~\omega $ with the $4N$ parameters $p_{mk}$ 
\textit{all integers} (positive, negative or vanishing, and of course all
different among themselves), then this many-body system is \textit{%
isochronous} \cite{C2008}.

(iii) If some or even all of the $4N$ \textit{real} parameters $%
a_{m}^{\left( k\right) }$ are \textit{negative}, then the solutions of this
many-body problem need not be confined, indeed they generally describe 
\textit{scattering} phenomena: for a detailed analysis of such behaviors see
Appendix~G (``Asymptotic behavior of the zeros of a polynomial whose
coefficients diverge exponentially'') of the book \cite{C2001}. $\blacksquare 
$

\textbf{Example 1.} If $N=2$, system~(\ref{ManyBodyNewt}) reduces to
\begin{subequations}
\begin{eqnarray}
&&\!\!\!\!  \!\!\!\! \ddot{z}_1=w_1,\;\; \ddot{z}_2=w_2,\notag\\
&&\!\!\!\!  \!\!\!\! \ddot{w}_1=G(\vec{z}, \dot{\vec{z}}, \vec{w}, \dot{\vec{w}})+\frac{1}{z_1-z_2}\left[ z_1 F_1(\vec{z}, \dot{\vec{z}}, \vec{w}, \dot{\vec{w}})- F_2(\vec{z}, \dot{\vec{z}}, \vec{w}, \dot{\vec{w}})\right],\notag\\
&& \!\!\!\!  \!\!\!\! \ddot{w}_2=-G(\vec{z}, \dot{\vec{z}}, \vec{w}, \dot{\vec{w}})+\frac{1}{z_1-z_2}\left[- z_2 F_1(\vec{z}, \dot{\vec{z}}, \vec{w}, \dot{\vec{w}})+F_2(\vec{z}, \dot{\vec{z}}, \vec{w}, \dot{\vec{w}})\right],
\end{eqnarray}
where
\begin{eqnarray}
&& G(\vec{z}, \dot{\vec{z}}, \vec{w}, \dot{\vec{w}})=\frac{4 \dot{w}_1 \dot{z}_2+ 4 \dot{w}_2 \dot{z}_1+6 w_1 w_2}{z_1-z_2},\\
&& F_1(\vec{z}, \dot{\vec{z}}, \vec{w}, \dot{\vec{w}})=\alpha_1(\dot{w}_1+\dot{w}_2)+\beta_1(w_1+w_2)\notag \\
&&+\gamma_1(\dot{z}_1+\dot{z}_2)+\delta_1(z_1+z_2),\\
&&F_2(\vec{z}, \dot{\vec{z}}, \vec{w}, \dot{\vec{w}})=\alpha_2(\dot{w}_1 z_2+3 w_1\dot{z}_2 +3 \dot{z}_1 w_2+z_1 \dot{w}_2)\notag\\
&&+\beta_2(w_1 z_2+2 \dot{z}_1 \dot{z}_2+z_1 w_2)
+\gamma_2(\dot{z}_1 z_2+z_1 \dot{z}_2)+\delta_2 z_1 z_2.
\end{eqnarray}
\label{SystemzwN2}
\end{subequations}

In Appendix~B we provide plots of the solutions of this system~(\ref{SystemzwN2}) with the following values of the parameters $\alpha_m, \beta_m, \gamma_m, \delta_m$, $m=1,2$, 
\begin{subequations}
\begin{equation}
\alpha_m=5 \mathbf{i}, \beta_m=5, \gamma_m=5 \mathbf{i}, \delta_m=6, \mbox{ for } m=1,2,
\label{N2Ex1Parameters}
\end{equation}
and the initial conditions
\begin{eqnarray}
z_1(0)=1+\mathbf{i}, \dot{z}_1(0)=1, z_2(0)=5+\mathbf{i}, \dot{z}_2(0)=1,\notag\\
w_1(0)=1, \dot{w}_1(0)=\mathbf{i}, w_2(0)=-\mathbf{i}, \dot{w}_2(0)=1.
\label{N2Ex1InitCond}
\end{eqnarray}
\label{N2Ex1}
\end{subequations}
For system~(\ref{SystemzwN2}), (\ref{N2Ex1Parameters}), each characteristic equation (\ref{Eqlanda}) has the four roots $-\mathbf{i},\mathbf{i}, 2 \mathbf{i}, 3 \mathbf{i}$. Therefore, by (ii) of \textbf{Remark 2.2,} system~(\ref{SystemzwN2}), (\ref{N2Ex1Parameters}) is \textit{isochronous}, see Figures~\ref{F11}, \ref{F12}, \ref{F13}, \ref{F14}, \ref{F15}, \ref{F16}, \ref{F17}, \ref{F18}  in Appendix~B.

Next,we provide plots of the solutions of system~(\ref{SystemzwN2}) with the following values of the parameters $\alpha_m, \beta_m, \gamma_m, \delta_m$, $m=1,2,$ \begin{subequations}
\begin{equation}
\alpha_m=-3 , \beta_m=-3, \gamma_m=-3 , \delta_m=-2, m=1,2,
\label{N2Ex2Parameters}
\end{equation}
and the initial conditions
\begin{eqnarray}
z_1(0)=-2-\mathbf{i}, \dot{z}_1(0)=1, z_2(0)=2+\mathbf{i}, \dot{z}_2(0)=1,\notag\\
w_1(0)=1, \dot{w}_1(0)=\mathbf{i}, w_2(0)=-\mathbf{i}, \dot{w}_2(0)=1.
\label{N2Ex2InitCond}
\end{eqnarray}
\label{N2Ex2}
\end{subequations}
For the initial value problem~(\ref{SystemzwN2}),~(\ref{N2Ex2Parameters}), each characteristic equation (\ref{Eqlanda}) has the four roots $-\mathbf{i},\mathbf{i}, -1,-2$. Therefore, by (i) of \textbf{Remark 2.2,} system~(\ref{SystemzwN2}),~(\ref{N2Ex2Parameters}) is \textit{asymptotically  isochronous}, see Figures~\ref{F21}, \ref{F22}, \ref{F23}, \ref{F24}, \ref{F25}, \ref{F26}, \ref{F27}, \ref{F28} in Appendix~B.

Next,we provide plots of the solutions of system~(\ref{SystemzwN2}) with the following values of the parameters $\alpha_m, \beta_m, \gamma_m, \delta_m$, $m=1,2$,
\begin{subequations}
\begin{eqnarray}
\alpha_m=-3+(\pi-1)\mathbf{i} , \beta_m=-(\pi+2)+3(\pi-1)\mathbf{i},\notag\\
 \gamma_m=-3 \pi+2(\pi-1)\mathbf{i} , \delta_m=-2 \pi, m=1,2,
\label{N2Ex3Parameters}
\end{eqnarray}
and the initial conditions
\begin{eqnarray}
z_1(0)=-2-\mathbf{i}, \dot{z}_1(0)=1, z_2(0)=2+\mathbf{i}, \dot{z}_2(0)=-1,\notag\\
w_1(0)=\mathbf{i}, \dot{w}_1(0)=1, w_2(0)=-\mathbf{i}, \dot{w}_2(0)=-1.
\label{N2Ex3InitCond}
\end{eqnarray}
\label{N2Ex3}
\end{subequations}
For system~(\ref{SystemzwN2}), (\ref{N2Ex3Parameters}), each characteristic equation (\ref{Eqlanda}) has the four roots $-\mathbf{i},\pi \mathbf{i}, -1,-2$. Therefore, by (i) of \textbf{Remark 2.2,} system~(\ref{SystemzwN2}), (\ref{N2Ex3Parameters}) is \textit{asymptotically multiply periodic}, see Figures~\ref{F31}, \ref{F32}, \ref{F33}, \ref{F34}, \ref{F35}, \ref{F36}, \ref{F37}, \ref{F38} in Appendix~B.

Next,we provide plots of the solutions of system~(\ref{SystemzwN2}) with the following values of the parameters $\alpha_m, \beta_m, \gamma_m, \delta_m$, $m=1,2$,\begin{subequations}
\begin{eqnarray}
\alpha_1=0.222+1.4 \mathbf{i}, \beta_1=0.41208-0.2208 \mathbf{i}, \notag\\
\gamma_1=-0.038436-0.018968 \mathbf{i} , \delta_1=0.000866464+0.0010224 \mathbf{i}, \notag\\
\alpha_2=0.172+1.1 \mathbf{i}, \beta_2=0.06952-0.1512 \mathbf{i}, \notag\\
\gamma_2=-0.006696+0.026376 \mathbf{i} , \delta_2=0.000104896-0.00047584 \mathbf{i},
\label{N2Ex4Parameters}
\end{eqnarray}
and the initial conditions
\begin{eqnarray}
z_1(0)=-2+3\mathbf{i}, \dot{z}_1(0)=7, z_2(0)=3+2 \mathbf{i}, \dot{z}_2(0)=-5,\notag\\
w_1(0)=2+4.2 \mathbf{i}, \dot{w}_1(0)=4.5, w_2(0)=3.1 \mathbf{i}, \dot{w}_2(0)=2.4.
\label{N2Ex4InitCond}
\end{eqnarray}
\label{N2Ex4}
\end{subequations}
For system~(\ref{SystemzwN2}), (\ref{N2Ex4Parameters}), 
 the characteristic equation (\ref{Eqlanda}) for $m=1$ has the four roots $0.04, 0.062+\mathbf{i},0.08+0.3\mathbf{i},0.04+0.1\mathbf{i}$ and the characteristic equation (\ref{Eqlanda}) for $m=2$ has the four roots $0.02, 0.032+\mathbf{i}, 0.06-0.1\mathbf{i}, 0.06+0.2\mathbf{i}$.
In agreement with (iii) of \textbf{Remark 2.2,} the components $z_1$ and $w_1$ of the solution of system~(\ref{SystemzwN2}), (\ref{N2Ex4Parameters}) exhibit \textit{scattering} phenomena, see Figures~\ref{F41}, \ref{F42}, \ref{F43}, \ref{F44}, \ref{F45}, \ref{F46}, \ref{F47}, \ref{F48} in Appendix~B. From these figures, it is clear that $z_1(t)$ diverges exponentially as $t \to \infty$ (and of course $w_1(t)$ features the same behavior), while $z_2(t)$ and $w_2(t)$ converge to zero as $t \to \infty$, which is consistent with the behavior of the zeros of polynomials whose coefficients depend on $t$ exponentially, as reported in Appendix~G of~\cite{C2001}.

\textbf{Example 2.}  If $N=3$, system~(\ref{ManyBodyNewt}) reduces to
\begin{subequations}
\begin{eqnarray}
&& \ddot{z}_1=w_1, \ddot{z}_2=w_2, \ddot{z}_3=w_3,\notag\\
&& \ddot{w}_1=\frac{4 \dot{w}_1 \dot{z}_2+4 \dot{w}_2 \dot{z}_1+6 w_1 w_2}{z_1-z_2}+\frac{4 \dot{w}_1 \dot{z}_3+4 \dot{w}_3 \dot{z}_1+6 w_1 w_3}{z_1-z_3}\notag\\
&&-\frac{1}{(z_1-z_2)(z_1-z_3)}\Bigg\{
12\left[ w_1\dot{z}_2 \dot{z}_3+\dot{z}_1(w_2 \dot{z}_3+w_3 \dot{z}_2) \right]+z_1^2 K_1(\vec{z}, \dot{\vec{z}}, \vec{w}, \dot{\vec{w}})\notag\\
&&+ z_1 K_2(\vec{z}, \dot{\vec{z}}, \vec{w}, \dot{\vec{w}})+ K_3(\vec{z}, \dot{\vec{z}}, \vec{w}, \dot{\vec{w}})
\Bigg\},\notag\\
&& \ddot{w}_2=\frac{4 \dot{w}_2 \dot{z}_1+4 \dot{w}_1 \dot{z}_2+6 w_1 w_2}{z_2-z_1}+\frac{4 \dot{w}_2 \dot{z}_3+4 \dot{w}_3 \dot{z}_2+6 w_2 w_3}{z_2-z_3}\notag\\
&&-\frac{1}{(z_2-z_1)(z_2-z_3)}\Bigg\{
12\left[ w_2\dot{z}_1 \dot{z}_3+\dot{z}_2(w_1 \dot{z}_3+w_3 \dot{z}_1) \right]+z_2^2 K_1(\vec{z}, \dot{\vec{z}}, \vec{w}, \dot{\vec{w}})\notag\\
&&+ z_2 K_2(\vec{z}, \dot{\vec{z}}, \vec{w}, \dot{\vec{w}})+ K_3(\vec{z}, \dot{\vec{z}}, \vec{w}, \dot{\vec{w}})
\Bigg\},\notag\\
&& \ddot{w}_3=\frac{4 \dot{w}_3 \dot{z}_1+4 \dot{w}_1 \dot{z}_3+6 w_1 w_3}{z_3-z_1}+\frac{4 \dot{w}_3 \dot{z}_2+4 \dot{w}_2 \dot{z}_3+6 w_2 w_3}{z_3-z_2}\notag\\
&&-\frac{1}{(z_3-z_1)(z_3-z_2)}\Bigg\{
12\left[ w_3\dot{z}_1 \dot{z}_2+\dot{z}_3(w_1 \dot{z}_2+w_2 \dot{z}_1) \right]+z_3^2 K_1(\vec{z}, \dot{\vec{z}}, \vec{w}, \dot{\vec{w}})\notag\\
&&+ z_3 K_2(\vec{z}, \dot{\vec{z}}, \vec{w}, \dot{\vec{w}})+ K_3(\vec{z}, \dot{\vec{z}}, \vec{w}, \dot{\vec{w}})
\Bigg\},\notag\\
\end{eqnarray}
where
\begin{eqnarray}
&& K_1(\vec{z}, \dot{\vec{z}}, \vec{w}, \dot{\vec{w}})=-\Big[ 
\alpha_1(\dot{w}_1+\dot{w}_2+\dot{w}_3)+\beta_1(w_1+w_2+w_3)\notag\\
&&+\gamma_1(\dot{z}_1+\dot{z}_2+\dot{z}_3)+\delta_1(z_1+z_2+z_3),
\end{eqnarray}
\begin{eqnarray}
 && K_2(\vec{z}, \dot{\vec{z}}, \vec{w}, \dot{\vec{w}})=
\alpha_2\Big[
\dot{w}_1(z_2+z_3)+\dot{w}_2(z_1+z_3)+\dot{w}_3(z_1+z_2)+2 w_1(\dot{z}_2+\dot{z}_3)\notag\\
&&+2 w_2(\dot{z}_1+\dot{z}_3)+2 w_3(\dot{z}_1+\dot{z}_2)+\dot{z}_1(w_2+w_3)+\dot{z}_2(w_1+w_3)+\dot{z}_3(w_1+w_2)\Big]\notag\\
&&+\beta_2\Big[
w_1(z_2+z_3)+w_2(z_1+z_3)+w_3(z_1+z_2)+\dot{z}_1(\dot{z}_2+\dot{z}_3)+\dot{z}_2(\dot{z}_1+\dot{z}_3)\notag\\
&&+\dot{z}_3(\dot{z}_1+\dot{z}_2)
\Big]+\gamma_2\Big[ 
\dot{z}_1(z_2+z_3)+\dot{z}_2(z_1+z_3)+\dot{z}_3(z_1+z_2)
\Big]\notag\\
&&+\delta_2(z_1 z_2+z_1 z_3+z_2 z_3),
\end{eqnarray}
\begin{eqnarray}
&& K_3(\vec{z}, \dot{\vec{z}}, \vec{w}, \dot{\vec{w}})=-\Bigg\{
\alpha_3\Big[
\dot{w}_1 z_2 z_3+\dot{w}_2 z_1 z_3+\dot{w}_3 z_1 z_2+ 2w_1(\dot{z}_2 z_3 + z_2 \dot{z}_3) \notag\\
&&+2w_2(\dot{z}_1 z_3 + z_1 \dot{z}_3)+ 2w_3(\dot{z}_1 z_2 + z_1 \dot{z}_2)
+\dot{z}_1(w_2 z_3+2 \dot{z}_2 \dot{z}_3 + z_2 w_3)\notag\\
&&+\dot{z}_2(w_1 z_3+2 \dot{z}_1 \dot{z}_3 + z_1 w_3)
+\dot{z}_3(w_1 z_2+2 \dot{z}_1 \dot{z}_2 + z_1 w_2)
\Big]\notag\\
&&+\beta_3\Big[
w_1 z_2 z_3 +w_2 z_1 z_3 +w_3 z_1 z_2 +\dot{z}_1(\dot{z}_2 z_3+ z_2 \dot{z}_3)+\dot{z}_2(\dot{z}_1 z_3 + z_1 \dot{z}_3)\notag\\
&&+\dot{z}_3 (\dot{z}_1 z_2+ z_1 \dot{z}_2)
\Big]
+\gamma_3\Big[
\dot{z}_1 z_2 z_3+ z_1 \dot{z}_2 z_3+ z_1 z_2 \dot{z}_3
\Big]+\delta_3 z_1 z_2 z_3
\Bigg\}.
\end{eqnarray}
\label{SystemzwN3}
\end{subequations}

In Appendix~B we provide plots of the solutions of system~(\ref{SystemzwN3}) with the following values of the parameters $\alpha_m, \beta_m, \gamma_m, \delta_m$, $m=1,2,3,$
\begin{subequations}
\begin{eqnarray}
&&\alpha_1=5 \mathbf{i}, \beta_1=5, \gamma_1=5 \mathbf{i}, \delta_1=6, \notag\\
&&\alpha_2=4 \mathbf{i}, \beta_2=-1, \gamma_2=16 \mathbf{i}, \delta_2=12, \notag\\
&&\alpha_3=0, \beta_3=-5, \gamma_3=0, \delta_3=-4, \notag\\
\label{N3Ex1Parameters}
\end{eqnarray}
 and the initial conditions
\begin{eqnarray}
&&z_1(0)=-1.45+1.1 \mathbf{i}, \dot{z}_1(0)=0.9,  \notag\\
&&z_2(0)=5.1+0.8 \mathbf{i}, \dot{z}_2(0)=1.2,  \notag\\
&& z_3(0)=2.5-0.2\mathbf{i}, \dot{z}_3(0)=-1.04,\notag\\
&&w_1(0)=1.23, \dot{w}_1(0)=0.84\mathbf{i}, \notag\\
&&w_2(0)=-2.26\mathbf{i}, \dot{w}_2(0)=2.16, \notag\\
&&w_3(0)=1.32 \mathbf{i}, \dot{w}_3(0)=-1.12.
\label{N3Ex1InitCond}
\end{eqnarray}
\label{N3Ex1}
\end{subequations}
For system~(\ref{SystemzwN3}), (\ref{N3Ex1Parameters}), the characteristic equation (\ref{Eqlanda}) for $m=1$ has the four roots $-\mathbf{i},\mathbf{i}, 2 \mathbf{i}, 3 \mathbf{i}$, the characteristic equation (\ref{Eqlanda}) for $m=2$ has the four roots $-2\mathbf{i},\mathbf{i}, 2 \mathbf{i}, 3 \mathbf{i}$ and the characteristic equation (\ref{Eqlanda}) for $m=3$ has the four roots $-2\mathbf{i},-\mathbf{i},  \mathbf{i}, 2 \mathbf{i}$,. Therefore, by (ii) of \textbf{Remark 2.2,} system~(\ref{SystemzwN3}), (\ref{N3Ex1Parameters}) is \textit{isochronous}, see Figures~\ref{N3F11}, \ref{N3F12}, \ref{N3F13}, \ref{N3F14}, \ref{N3F15}, \ref{N3F16}, \ref{N3F17}, \ref{N3F18}, \ref{N3F19}, \ref{N3F110} in Appendix~B.

\bigskip

\section{New \textit{solvable} dynamical systems and their solutions}

In this section we indicate how to identify endless sequences of \textit{%
solvable} dynamical systems describing the motion in the complex $z$-plane
of point-particles interacting among themselves with certain forces
depending on their positions and velocities. Let us reiterate that a
many-body model is considered \textit{solvable} if the configuration of the
system at any arbitrary time $t$ can be obtained---from any given \textit{%
initial} data: the \textit{initial} positions and velocities of the $N$
particles in the complex $z$-plane---by \textit{algebraic} operations, such
as finding the zeros of an \textit{explicitly} known time-dependent
polynomial.

\textbf{Remark 3.1}. Note however that knowledge of the configuration of the
many-body system at time $t$, with the (generally complex) values of its
coordinates given as the \textit{unordered} set of the zeros of a known
polynomial, does \textit{not} allow to identify the specific coordinate that
has evolved over time from the assignment of its specific initial position
and velocity; this additional information can only be gained by following
over time the evolution of the system, either by integrating numerically the
equations of motion, or by identifying the configurations of the system at a
sequence of time intervals sufficiently close to each other so as to
guarantee the identification by \textit{continuity} of the trajectory of each
particle (or at least of the specific particle under consideration). But
these additional operations need not be performed with great accuracy, even
when one wishes the final configuration---including the identity of each
particle---to be known with much greater accuracy.

Likewise---in the case of systems which have been identified as \textit{%
isochronous} because their solution is provided by the zeros of a
time-dependent polynomial which is itself periodic in time with period, say, 
$T$---an analogous procedure must be followed to ascertain whether the
period of the time evolution of a specific particle is $T$, or $pT$ (with $p$
a \textit{positive integer}), due to the possibility of a $T$-periodic
exchange of the correspondence between the zeros of the polynomial and the
particle identities (for a general discussion of this possibility in a
specific context see \cite{GS2005}). $\blacksquare $

The key formulas for the following developments are the identities (\ref%
{Iden1}), relating the time evolution of the \textit{zeros} $z_{n}\left(
t\right) $ of a time-dependent (monic) polynomial to that of the \textit{%
coefficients} $c_{m}\left( t\right) $ of the same polynomial, as well as the
relations (\ref{cm}) respectively (\ref{cmdot}) expressing the coefficients $%
c_{m}\left( t\right) $ of a monic polynomial respectively their time
derivatives in terms of the \textit{zeros} of the same polynomial and their
time derivatives.

In this paper we restrict for simplicity attention to the case of a \textit{%
linear decoupled} evolution of the coefficients $c_{m}\left( t\right) $,
namely we assume that these $N$ coefficients of the time-dependent
polynomial (\ref{Pol}) evolve in time according to the following system of
ODEs, 
\begin{equation}
\ddddot{c}_{m}=\alpha _{m}~\dddot{c}_{m}+\beta _{m}~\ddot{c}_{m}+\gamma _{m}~%
\dot{c}_{m}+\delta _{m}~,  \label{Systcm}
\end{equation}%
where the parameters $\alpha _{m},~\beta _{m},~\gamma _{m},~\delta _{m}$ are 
$4N$ \textit{generic} complex numbers such that for each $m$, the characteristic equation
\begin{equation}
\left( \lambda _{m}\right) ^{4}=\alpha _{m}~\left( \lambda _{m}\right)
^{3}+\beta _{m}~\left( \lambda _{m}\right) ^{2}+\gamma _{m}~\lambda
_{m}+\delta _{m}~,  \label{Eqlanda}
\end{equation}
has four \textit{distinct} roots $\lambda_{m,k}, k=1,2,3,4$ (see below).
 It is then plain that the 
\textit{general solution} of this system reads as follows : 
\begin{subequations}
\begin{equation}
c_{m}\left( t\right) =\sum_{k=1}^{4}\left[ b_{m}^{\left( k\right) }~\exp
\left( \lambda _{m,k}~t\right) \right] ~.  \label{GenSol}
\end{equation}%
The $4N$ numbers $\lambda _{m,k}$, labeled by the $4$
values of the index $k$, are denoted as follows:%
\begin{equation}
\lambda _{m,k}=-a_{m}^{\left( k\right) }+\mathbf{i}~\omega _{m}^{\left(
k\right) }~,~~~k=1,2,3,4~,  \label{landaaomega}
\end{equation}%
introducing thereby the $8N$ \textit{real} parameters $a_{m}^{\left(
1\right) }$, $a_{m}^{\left( 2\right) }$, $a_{m}^{\left( 3\right) }$, $%
a_{m}^{\left( 4\right) }$, $\omega _{m}^{\left( 1\right) }$, $\omega
_{m}^{\left( 2\right) }$, $\omega _{m}^{\left( 3\right) }$, $\omega
_{m}^{\left( 4\right) }$, implying that the general solution (\ref{GenSol})
can be equivalently written as (\ref{cmt}).

\textbf{Remark 3.2}. The fourth-degree algebraic equations (\ref{Eqlanda})
could be \textit{explicitly solved}, but the formulas expressing, for every
value of the index $m$, the $4$ exponents $\lambda _{m,k}$ in terms of the $4
$ parameters $\alpha _{m}$,$~\beta _{m}$, $\gamma _{m}$,$~\delta _{m}$ are
too complicated to be of much use. The converse formulas, expressing, for
every value of the index $m$, the $4$ parameters $\alpha _{m}$,$~\beta _{m}$%
, $\gamma _{m}$,$~\delta _{m}$ in terms of the $4$ exponents $\lambda _{m,k}$%
, or rather their real and imaginary parts, see (\ref{landaaomega}), are
instead rather neat, see (\ref{abcd}). $\blacksquare $

As for the $4N$ numbers $b_{m}^{\left( k\right) }$ in (\ref{GenSol}), they
are \textit{a priori} arbitrary; but can of course be determined in terms of
the initial data (thereby solving the \textit{initial value problem} of the
dynamical system (\ref{Systcm})) by solving, for each of the $N$ values of
the index $m$, the following system of $4$ \textit{linear algebraic}
equations,%
\begin{equation}
\sum_{k=1}^{4}\left[ b_{m}^{\left( k\right) }~\left( \lambda _{m,k}\right)
^{s}\right] =\left. \frac{d^{s}c_{m}\left( t\right) }{dt^{s}}\right\vert
_{t=0}~,~~~s=0,1,2,3~.  \label{Eqbmk}
\end{equation}

\textbf{Remark 3.3}. It is plain that one could have considered, instead of
the system of $N$ \textit{linear} \textit{decoupled} ODEs (\ref{Systcm}),
the more general system of $N$ \textit{linear coupled} ODEs 
\end{subequations}
\begin{equation}
\ddddot{c}_{m}=\sum_{n=1}^{N}\left( A_{mn}~\dddot{c}_{n}+B_{mn}~\ddot{c}%
_{n}+C_{mn}~\dot{c}_{n}+D_{mn}~c_{n}\right) ~,
\end{equation}%
which is of course also \textit{solvable} by algebraic operations, while
featuring more arbitrary constants ($4N^{2}$ instead than $4N$). $%
\blacksquare $

The \textit{solvable} character of the dynamical system characterized by the
following $N$ \textit{coupled nonlinear} ODEs to be satisfied by the $N$
dependent variables $z_{n}\equiv z_{n}\left( t\right) $ is then clearly
implied by the formula (\ref{zndot4}):%
\begin{eqnarray}
&&\ddddot{z}_{n}=\sum_{\ell =1}^{N}{}^{\prime }\left( \frac{4~\dddot{z}_{n}~%
\dot{z}_{\ell }+4~\dddot{z}_{\ell }~\dot{z}_{n}+6~\ddot{z}_{n}~\ddot{z}%
_{\ell }}{z_{n}-z_{\ell }}\right)   \notag \\
&&-6~\sum_{\ell _{1},\ell _{2}=1}^{N}{}^{\prime }\left[ \frac{\ddot{z}_{n}~%
\dot{z}_{\ell _{1}}~\dot{z}_{\ell _{2}}+2~\ddot{z}_{\ell _{1}}~\dot{z}_{n}~%
\dot{z}_{\ell _{2}}}{\left( z_{n}-z_{\ell _{1}}\right) ~\left( z_{n}-z_{\ell
_{2}}\right) }\right]   \notag \\
&&+4~\sum_{\ell _{1},\ell _{2},~\ell _{3}=1}^{N}{}^{\prime }\left[ \frac{%
\dot{z}_{n}~\dot{z}_{\ell _{1}}~\dot{z}_{\ell _{2}}~\dot{z}_{\ell _{3}}}{%
\left( z_{n}-z_{\ell _{1}}\right) ~\left( z_{n}-z_{\ell _{2}}\right) ~\left(
z_{n}-z_{\ell _{3}}\right) }\right] -\left[ \prod\limits_{\ell =1,~\ell \neq
n}^{N}\left( z_{n}-z_{\ell }\right) ^{-1}\right] \cdot   \notag \\
&&\cdot \sum_{m=1}^{N}\left[ \left( \alpha _{m}~\dddot{c}_{m}+\beta _{m}~%
\ddot{c}_{m}+\gamma _{m}~\dot{c}_{m}+\delta _{m}~c_{m}\right) ~\left(
z_{n}\right) ^{N-m}\right] ~.  \label{zn4dot}
\end{eqnarray}%
In the last term the $4$ quantities $\dddot{c}_{m}$, $\ddot{c}_{m}$, $\dot{c}%
_{m}$ and $c_{m}$ must of course be expressed in terms of the dependent
variables $z_{n}$ and their time-derivatives by the formulas (\ref{cmdot})
and (\ref{cm}) (of course with (\ref{sigma})). Indeed the solution of this
dynamical system---(\ref{zn4dot}) with (\ref{cmdot}) and (\ref{cm})---is
clearly provided by the $N$ zeros of the monic polynomial (\ref{Polcm})
where the coefficients $c_{m}\left( t\right) $ are given by the formulas (%
\ref{cmt})---with the coefficients $b_{m}^{\left( k\right) }$ appearing
there expressed, as indicated above after eq. (\ref{Eqbmk}), in terms of the
initial data $c_{m}\left( 0\right) ,$ $\dot{c}_{m}\left( 0\right) ,$ $\ddot{c%
}_{m}\left( 0\right) ,$ $\dddot{c}_{m}\left( 0\right) ,$ themselves
expressed in terms of the initial data $z_{n}\left( 0\right) ,$ $\dot{z}%
_{n}\left( 0\right) ,$ $\ddot{z}_{n}\left( 0\right) ,$ $\dddot{z}_{n}\left(
0\right) $ via the formulas (\ref{cm}) and (\ref{cmdot}) at $t=0$.

\textbf{Remark 3.4}. If in the (last term in the) right-hand side of (\ref%
{zn4dot}) any one of the parameters $\alpha _{m},~\beta _{m},~\gamma
_{m},~\delta _{m}$ is independent of the index $m,$ say $\beta _{m}=\beta $,
then the corresponding term can be replaced by a simpler expression via the
appropriate identity (\ref{Iden1}), implying, say,%
\begin{equation}
\left[ \prod\limits_{\ell =1,~\ell \neq n}^{N}\left( z_{n}-z_{\ell }\right)
^{-1}\right] ~\sum_{m=1}^{N}\left[ \beta ~\ddot{c}_{m}~\left( z_{n}\right)
^{N-m}\right] =\beta ~\left[ \ddot{z}_{n}-\sum_{\ell =1}^{N}{}^{\prime
}\left( \frac{2~\dot{z}_{n}~\dot{z}_{\ell }}{z_{n}-z_{\ell }}\right) \right]
~,
\end{equation}%
see (\ref{zndot2}). $\blacksquare $

\bigskip

\section{Outlook}

The findings reported in this paper suggest further developments, which
ourselves or others might pursue and report in future publications.

One direction of future research is the exploration of the \textit{solvable}
dynamical systems and many-body problems of Newtonian type that are obtained by 
\textit{iterating} the type of approach described above, along the lines
discussed in \cite{BC2015b}.

It would also be of interest to obtain generalizations of the identities (%
\ref{Iden1}) to derivatives of order higher than 4, indeed hopefully of
arbitrary order.

And of course further explorations are appealing of the detailed behaviors
of the solutions of the dynamical systems obtained via this approach, as
well as---in the case of \textit{solvable} many-body problems of Newtonian
type allowing a Hamiltonian formulation---the exploration of their quantal
versions.

\bigskip
\appendix

\section{ Relations among the time derivatives of the zeros and
the coefficients of a time-dependent polynomial}

In this Appendix~A we tersely outline for the convenience of the reader the
proof of the $4$ identities (\ref{Iden1}) relating the time evolution of the 
$N$ zeros $z_{n}\left( t\right) $ of a time-dependent monic polynomial of
degree $N$ in the independent variable $z$ to the time-evolution of its $N$
coefficients $c_{m}\left( t\right) ,$ see (\ref{Pol}). A proof of the first $%
2$ of these $4$ identities was already provided in \cite{C2015a}, hence the
first part of the following treatment reports almost \textit{verbatim} that
presentation.

The starting point to prove the relation (\ref{zndot1}) are the two
relations 
\begin{subequations}
\begin{equation}
\psi _{t}\left( z;t\right) =\sum_{m=1}^{N}\left[ \dot{c}_{m}~z^{N-m}\right]
~,  \label{psict}
\end{equation}%
\begin{equation}
\psi _{t}\left( z;t\right) =-\sum_{m=1}^{N}\left[ \dot{z}_{m}\prod\limits_{%
\ell =1,~\ell \neq m}^{N}\left( z-z_{\ell }\right) \right] ~,  \label{psizt}
\end{equation}%
which clearly obtain time-differentiating (\ref{Polcm}) respectively (\ref%
{Polzn}). They imply the relation%
\begin{equation}
\sum_{m=1}^{N}\left[ \dot{z}_{m}\prod\limits_{\ell =1,~\ell \neq
m}^{N}\left( z-z_{\ell }\right) \right] =-\sum_{m=1}^{N}\left[ \dot{c}%
_{m}~z^{N-m}\right] ~,
\end{equation}%
and it is plain that, for $z=z_{n},$ this formula yields (\ref{zndot1}).

Likewise, an additional time-differentiation of (\ref{psict}) yields 
\end{subequations}
\begin{subequations}
\begin{equation}
\psi _{tt}\left( z;t\right) =\sum_{m=1}^{N}\left( \ddot{c}%
_{m}~z^{N-m}\right) ~,  \label{psictt}
\end{equation}%
while an additional time-differentiation of (\ref{psizt}) yields%
\begin{eqnarray}
&&\psi _{tt}\left( z;t\right) =-\sum_{m=1}^{N}\left\{ \ddot{z}_{m}\left[
\prod\limits_{\ell =1,~\ell \neq m}^{N}\left( z-z_{\ell }\right) \right]
\right\}  \notag \\
&&+\sum_{\ell _{1},\ell _{2}=1,~\ell _{1}\neq \ell _{2}}^{N}\left\{ \dot{z}%
_{\ell _{1}}~\dot{z}_{\ell _{2}}~\left[ \prod\limits_{\ell ^{\prime
}=1,~\ell ^{\prime }\neq \ell _{1},\ell _{2}}^{N}\left( z-z_{\ell ^{\prime
}}\right) \right] \right\}  \notag \\
&=&\sum_{m=1}^{N}\left( \ddot{c}_{m}~z^{N-m}\right) ~,
\end{eqnarray}%
where the second equality is implied by (\ref{psictt}). It is then again
plain that, for $z=z_{n},$ one obtains (\ref{zndot2}).

To obtain (\ref{zndot3}) and (\ref{zndot4}) we proceeded in an analogous
manner: the calculations involved in the additional two
time-differentiations of (\ref{Polcm}) are clearly trivial, while the
successive time-differentiations of (\ref{Polzn}) become progressively more
complicated; but the two yielding (\ref{zndot3}) and (\ref{zndot4}) are
still quite manageable by hand, so that their detailed treatment can be left
to the diligent reader.

\textbf{Remark A.1}. It is plain from a perusal of the $4$ formulas (\ref%
{Iden1})---as well as from their derivation outlined above---how to guess
the structure of the \textit{general} formula of this type, featuring in its
right-hand side the term $\left( -\right) ^{k}~\sum_{m=1}^{N}\left(
c_{m}^{\left( k\right) }~z^{N-m}\right) $ with $c_{m}^{\left( k\right)
}\equiv c_{m}^{\left( k\right) }\left( t\right) =d^{k}c_{m}\left( t\right)
/dt^{k},$ for $k$ an \textit{arbitrary} positive integer; but the numerical
coefficients appearing in the left-hand side of this formula are---to the
best of our knowledge---not known, so that their computation (for \textit{%
arbitrary} values of the two positive integers $N$ and $k$) remains an open
problem. Of course for \textit{small} values of either one of these two
parameters it is an easy task, as shown for small $k$ and arbitrary $N$ by
the $4$ formulas (\ref{Iden1}) and, for small $N$ and arbitrary $k$ by the
last statement in \textbf{Notation 1.1}:\textbf{\ }for instance for $N=2$%
\end{subequations}
\begin{eqnarray}
\frac{d^{k}z_{n}}{dt^{k}}=\left( z_{n}-z_{n+1}\right) ^{-1}~\left[ \frac{%
d^{k}}{dt^{k}}\left( z_{n}~z_{n+1}\right) -\left( \frac{d^{k}z_{n}}{dt^{k}}%
\right) ~z_{n+1}-z_{n}~\left( \frac{d^{k}z_{n+1}}{dt^{k}}\right) \right.  &&
\notag \\
\left. +\left( -1\right) ^{k}~\left( z_{n}~\frac{d^{k}c_{1}}{dt^{k}}+\frac{%
d^{k}c_{2}}{dt^{k}}\right) \right] ~,~~~n=1,2~\operatorname{mod}(2),~k=1,~2,~3~,...~%
\text{.~}\blacksquare  &&
\end{eqnarray}

\bigskip

\section{Plots of the Solutions of the Initial Value Problems\\ in Examples 1 and 2 of Section 2}

\begin{minipage}{\linewidth}
      \centering
      \begin{minipage}{0.45\linewidth}
          \begin{figure}[H]
              \includegraphics[width=\linewidth]{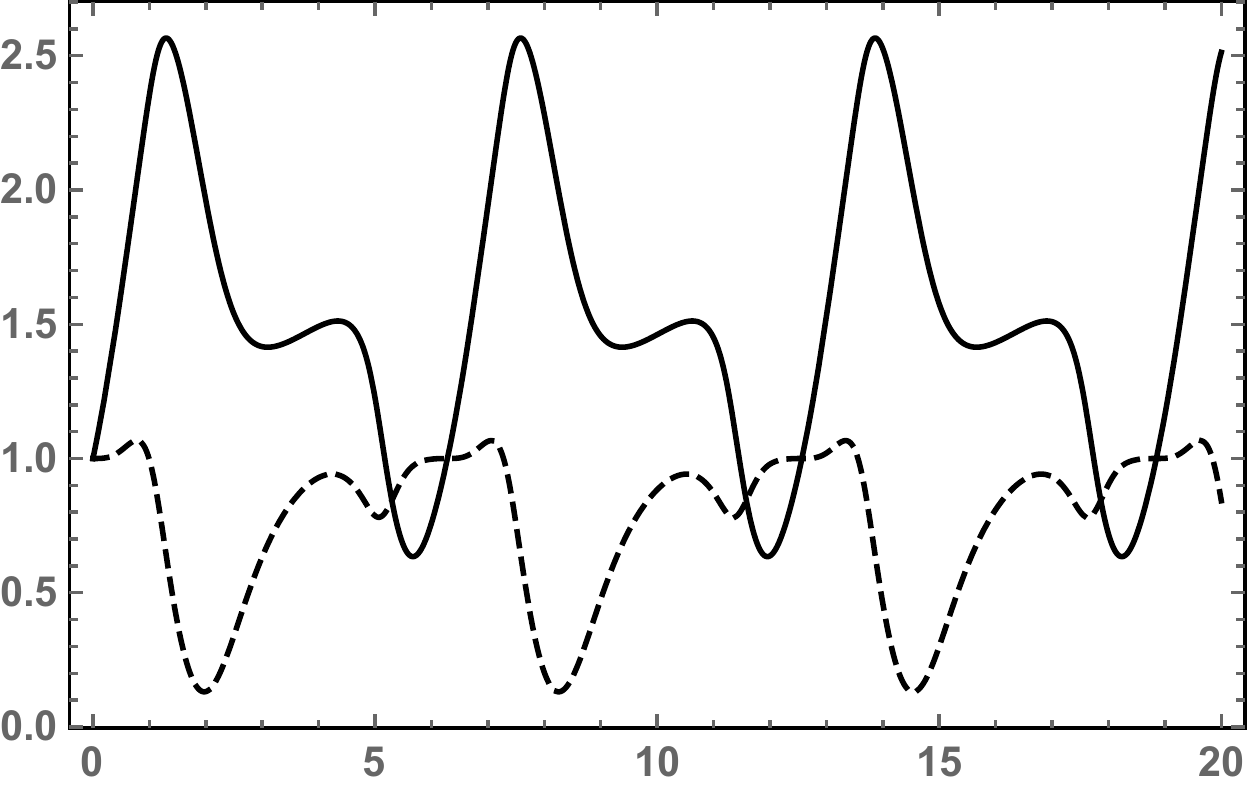}
              \caption{Initial value problem~(\ref{SystemzwN2}),~(\ref{N2Ex1}). Graphs of the real (bold curve) and imaginary 
              (dashed curve) parts of the coordinate $z_1(t)$; period $2\pi$.}
              \label{F11}
          \end{figure}
      \end{minipage}
      \hspace{0.05\linewidth}
      \begin{minipage}{0.45\linewidth}
          \begin{figure}[H]
              \includegraphics[width=\linewidth]{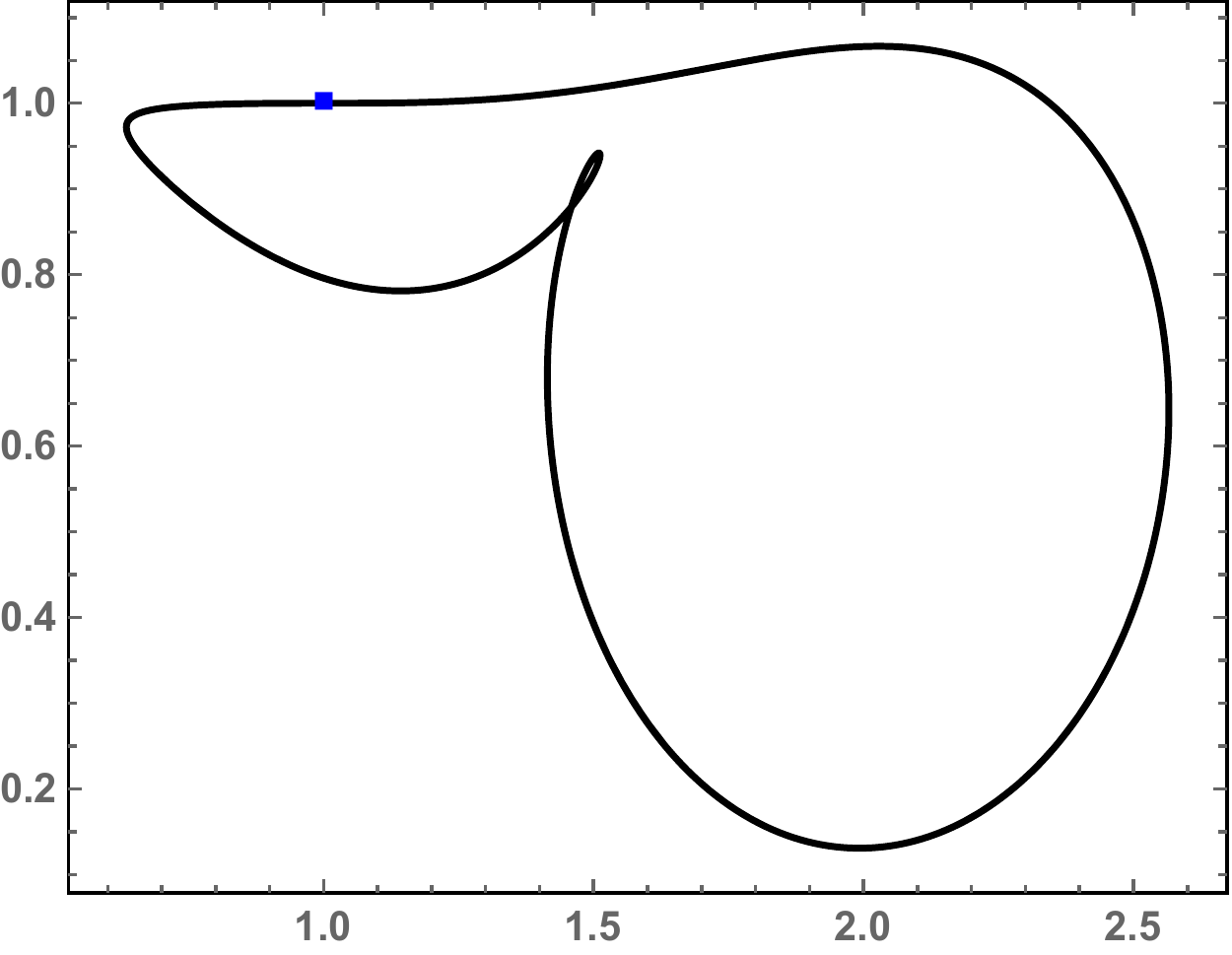}
              \caption{Initial value problem~(\ref{SystemzwN2}),~(\ref{N2Ex1}). Trajectory, in the complex $z$-plane, of  $z_1(t)$; 
              period $2 \pi$. The   square indicates the initial condition $z_1(0)=1+\mathbf{i}$.}
              \label{F12}
          \end{figure}
      \end{minipage}
  \end{minipage}

\begin{minipage}{\linewidth}
      \centering
      \begin{minipage}{0.45\linewidth}
          \begin{figure}[H]
             \includegraphics[width=\linewidth]{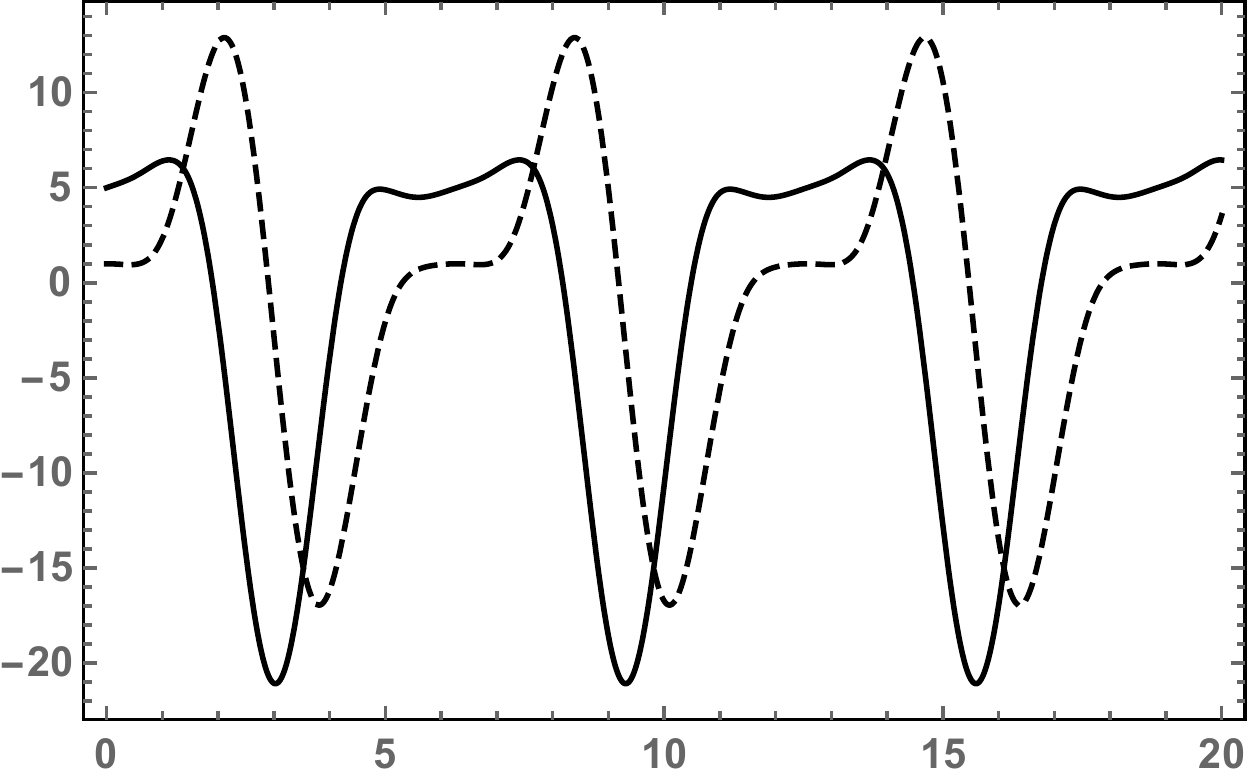}
             \caption{ Initial value problem~(\ref{SystemzwN2}),~(\ref{N2Ex1}). Graphs of the real (bold curve) and imaginary 
             (dashed curve) parts of the coordinate $z_2(t)$; period $2\pi$.}
             \label{F13}
          \end{figure}
      \end{minipage}
      \hspace{0.05\linewidth}
      \begin{minipage}{0.45\linewidth}
          \begin{figure}[H]
             \includegraphics[width=\linewidth]{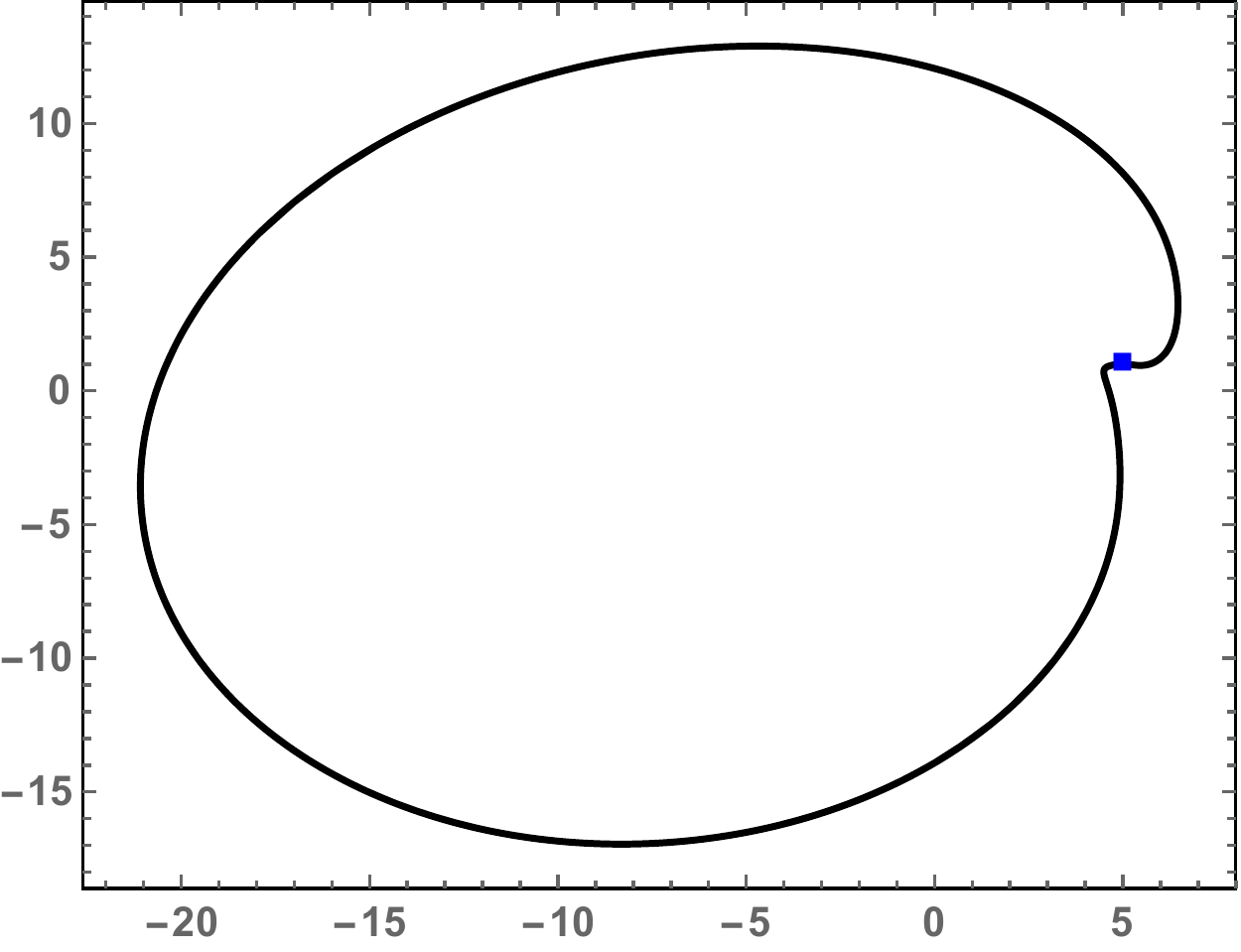}
             \caption{Initial value problem~(\ref{SystemzwN2}),~(\ref{N2Ex1}). Trajectory, in the complex $z$-plane, 
             of  $z_2(t)$; period $2 \pi$. The   square indicates the initial condition $z_2(0)=5+\mathbf{i}$.}
             \label{F14}          
             \end{figure}
      \end{minipage}
  \end{minipage}

\begin{minipage}{\linewidth}
      \centering
      \begin{minipage}{0.45\linewidth}
          \begin{figure}[H]
            \includegraphics[width=\linewidth]{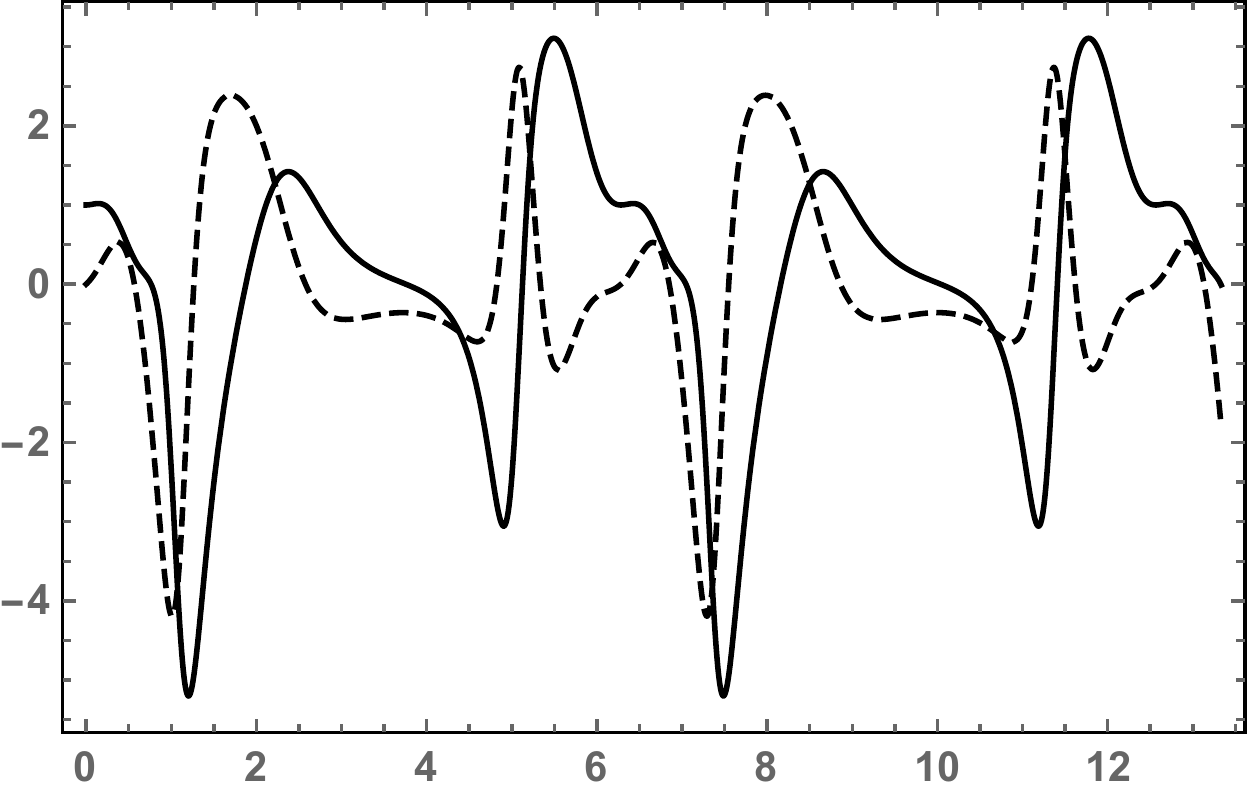}
\caption{ Initial value problem~(\ref{SystemzwN2}),~(\ref{N2Ex1}). Graphs of the real (bold curve) and imaginary (dashed curve) parts of the coordinate $w_1(t)$; period $2\pi$.}
\label{F15}
          \end{figure}
      \end{minipage}
      \hspace{0.05\linewidth}
      \begin{minipage}{0.45\linewidth}
          \begin{figure}[H]
           \includegraphics[width=\linewidth]{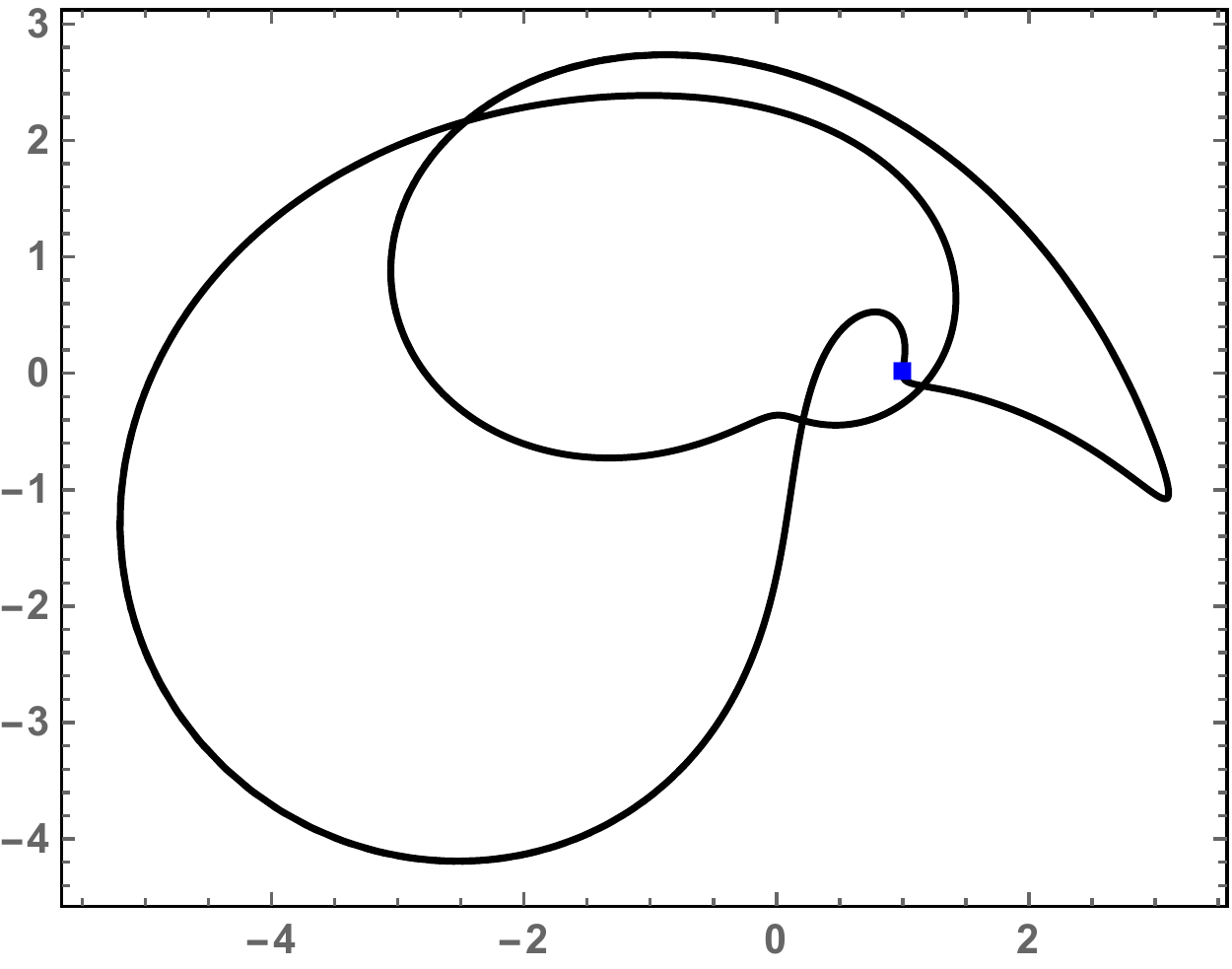}
\caption{Initial value problem~(\ref{SystemzwN2}),~(\ref{N2Ex1}). Trajectory, in the complex $z$-plane, of  $w_1(t)$; period $2 \pi$. The   square indicates the initial condition $w_1(0)=1$.}
\label{F16}
             \end{figure}
      \end{minipage}
  \end{minipage}
  
\begin{minipage}{\linewidth}
      \centering
      \begin{minipage}{0.45\linewidth}
          \begin{figure}[H]
		\includegraphics[width=\linewidth]{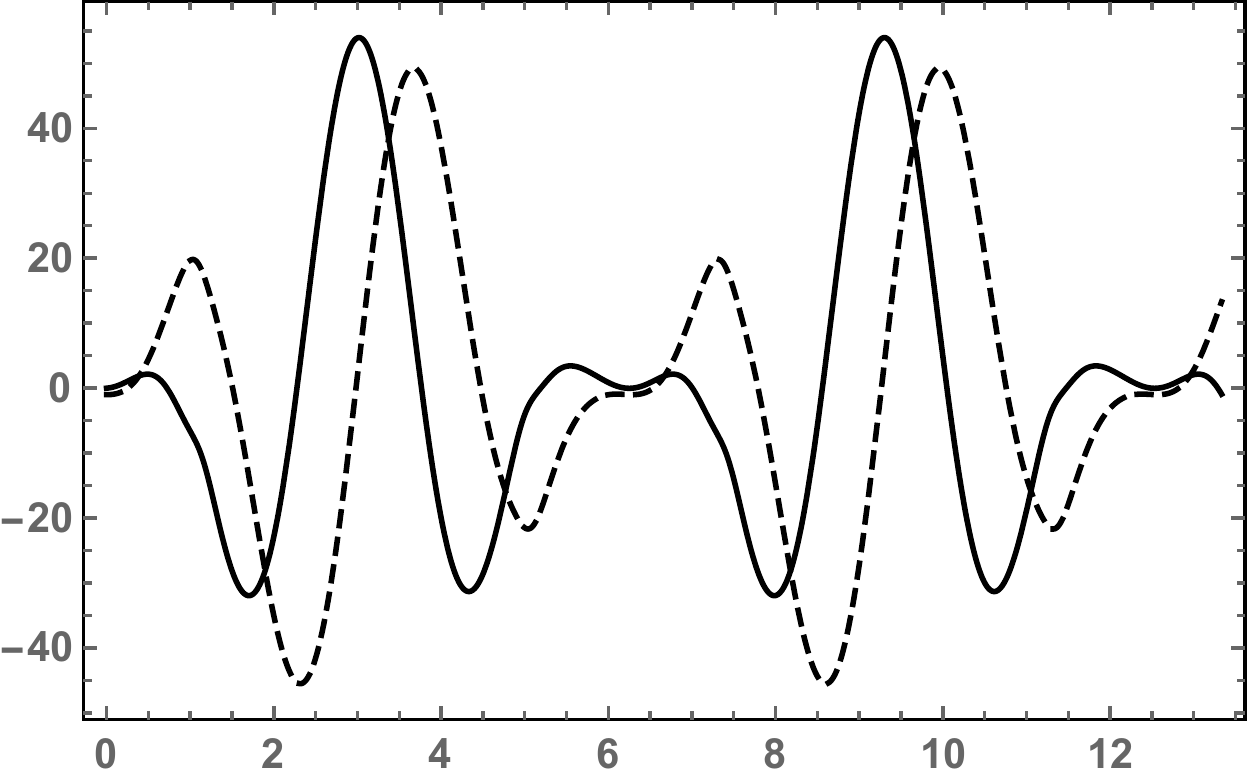}
		\caption{ Initial value problem~(\ref{SystemzwN2}),~(\ref{N2Ex1}). Graphs of the real (bold curve) and imaginary (dashed curve) parts of the coordinate $w_2(t)$; period 				$2\pi$.}
		\label{F17}
          \end{figure}
      \end{minipage}
      \hspace{0.05\linewidth}
      \begin{minipage}{0.45\linewidth}
          \begin{figure}[H]
            \includegraphics[width=\linewidth]{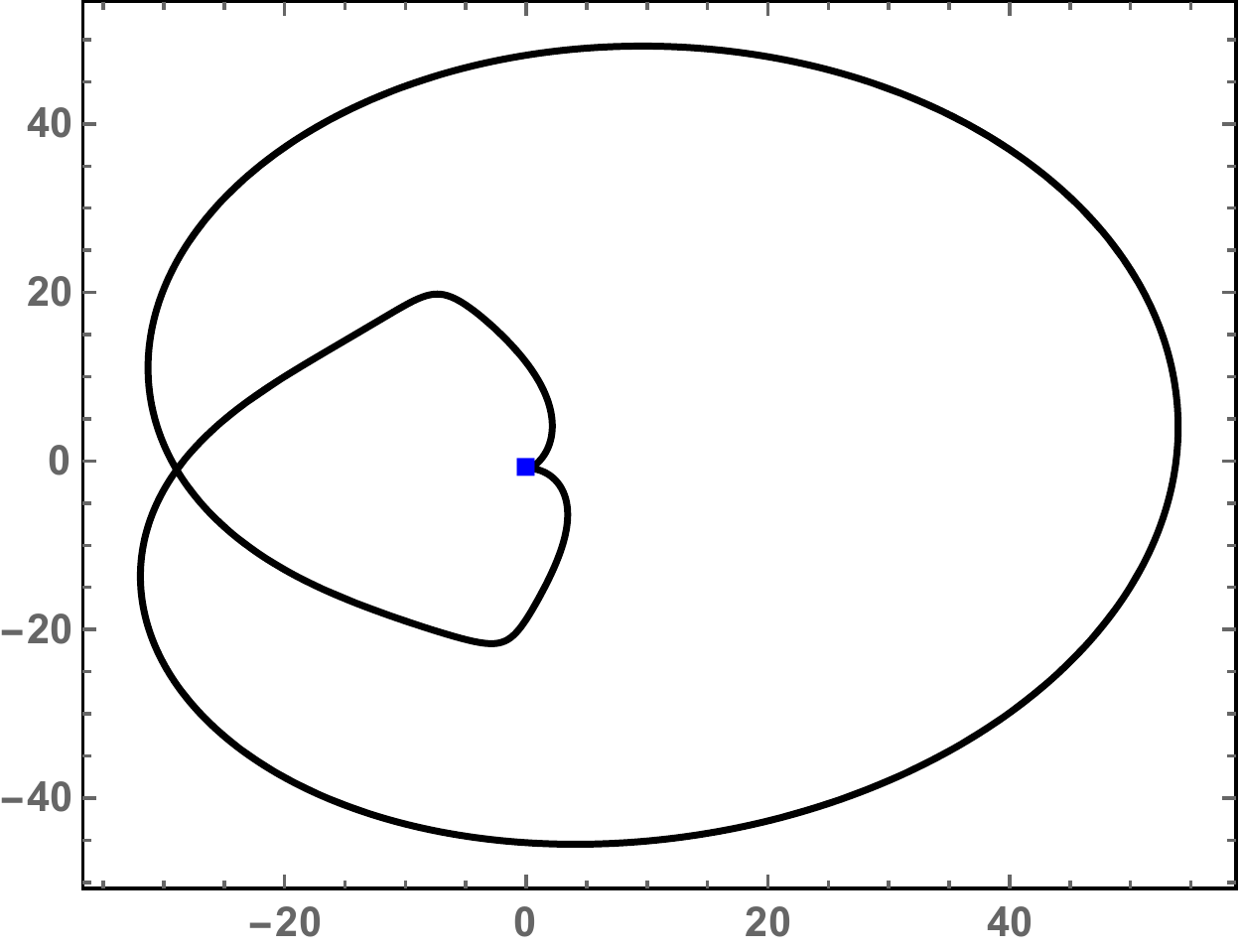}
\caption{Initial value problem~(\ref{SystemzwN2}),~(\ref{N2Ex1}). Trajectory, in the complex $z$-plane, of  $w_2(t)$; period $2 \pi$. The   square indicates the initial condition $w_2(0)=-\mathbf{i}$.}
\label{F18}     
             \end{figure}
      \end{minipage}
  \end{minipage}

\clearpage

\begin{minipage}{\linewidth}
      \centering
      \begin{minipage}{0.45\linewidth}
          \begin{figure}[H]
		\includegraphics[width=\linewidth]{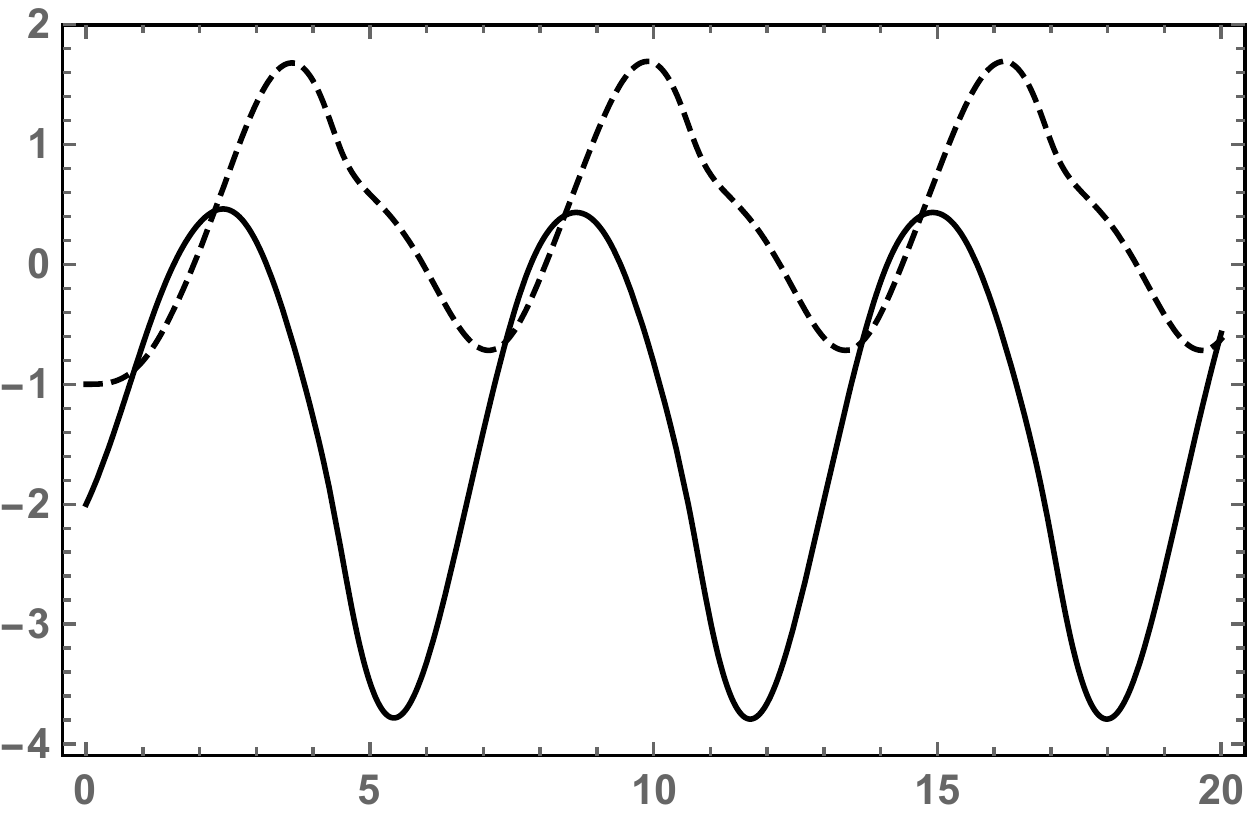}
\caption{ Initial value problem~(\ref{SystemzwN2}),~(\ref{N2Ex2}). Graphs of the real (bold curve) and imaginary (dashed curve) parts of the coordinate $z_1(t)$.}
\label{F21}
          \end{figure}
      \end{minipage}
      \hspace{0.05\linewidth}
      \begin{minipage}{0.45\linewidth}
          \begin{figure}[H]
            \includegraphics[width=\linewidth]{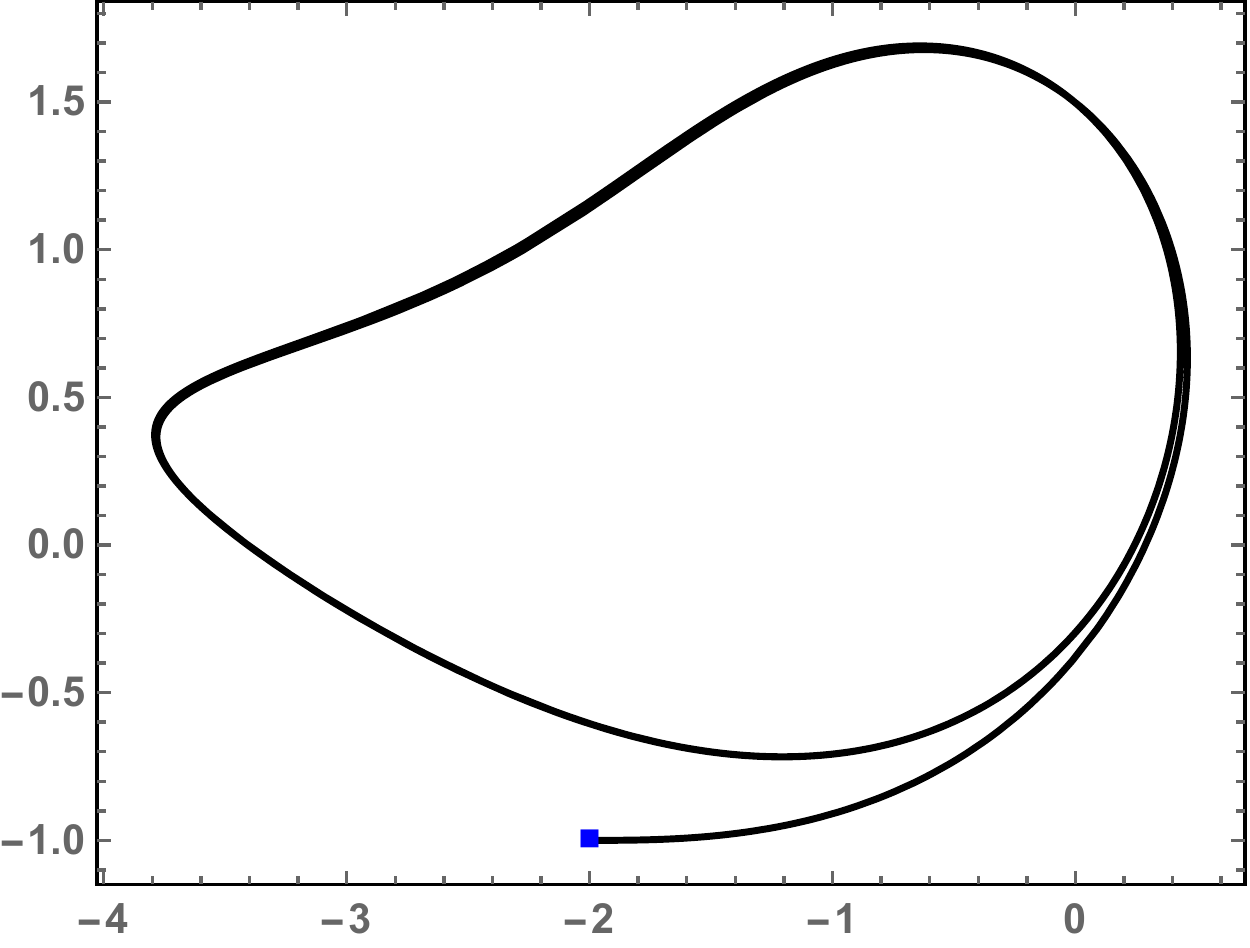}
\caption{Initial value problem~(\ref{SystemzwN2}),~(\ref{N2Ex2}). Trajectory, in the complex $z$-plane, of  $z_1(t)$. The   square indicates the initial condition $z_1(0)=-2-\mathbf{i}$.}
\label{F22}
             \end{figure}
      \end{minipage}
  \end{minipage}

\begin{minipage}{\linewidth}
      \centering
      \begin{minipage}{0.45\linewidth}
          \begin{figure}[H]
		\includegraphics[width=\linewidth]{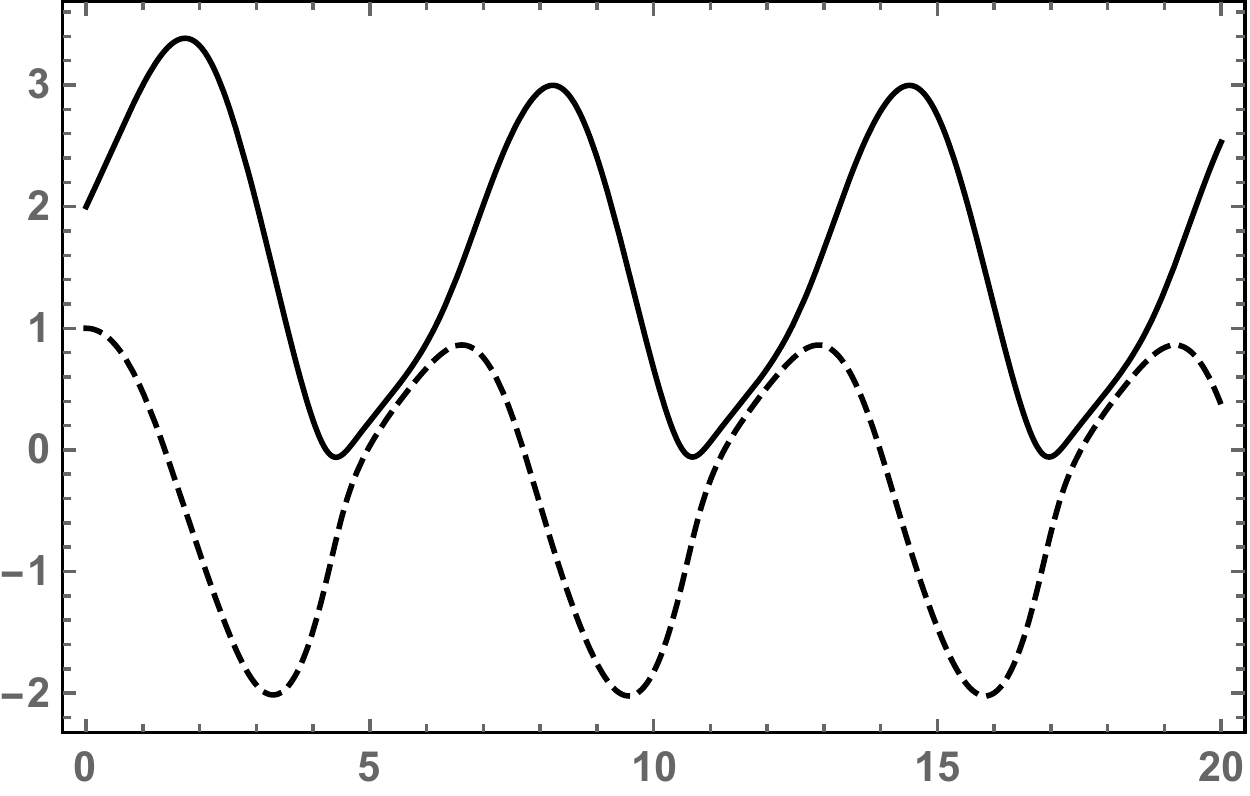}
\caption{ Initial value problem~(\ref{SystemzwN2}),~(\ref{N2Ex2}). Graphs of the real (bold curve) and imaginary (dashed curve) parts of the coordinate $z_2(t)$.}
\label{F23}
          \end{figure}
      \end{minipage}
      \hspace{0.05\linewidth}
      \begin{minipage}{0.45\linewidth}
          \begin{figure}[H]
           \includegraphics[width=\linewidth]{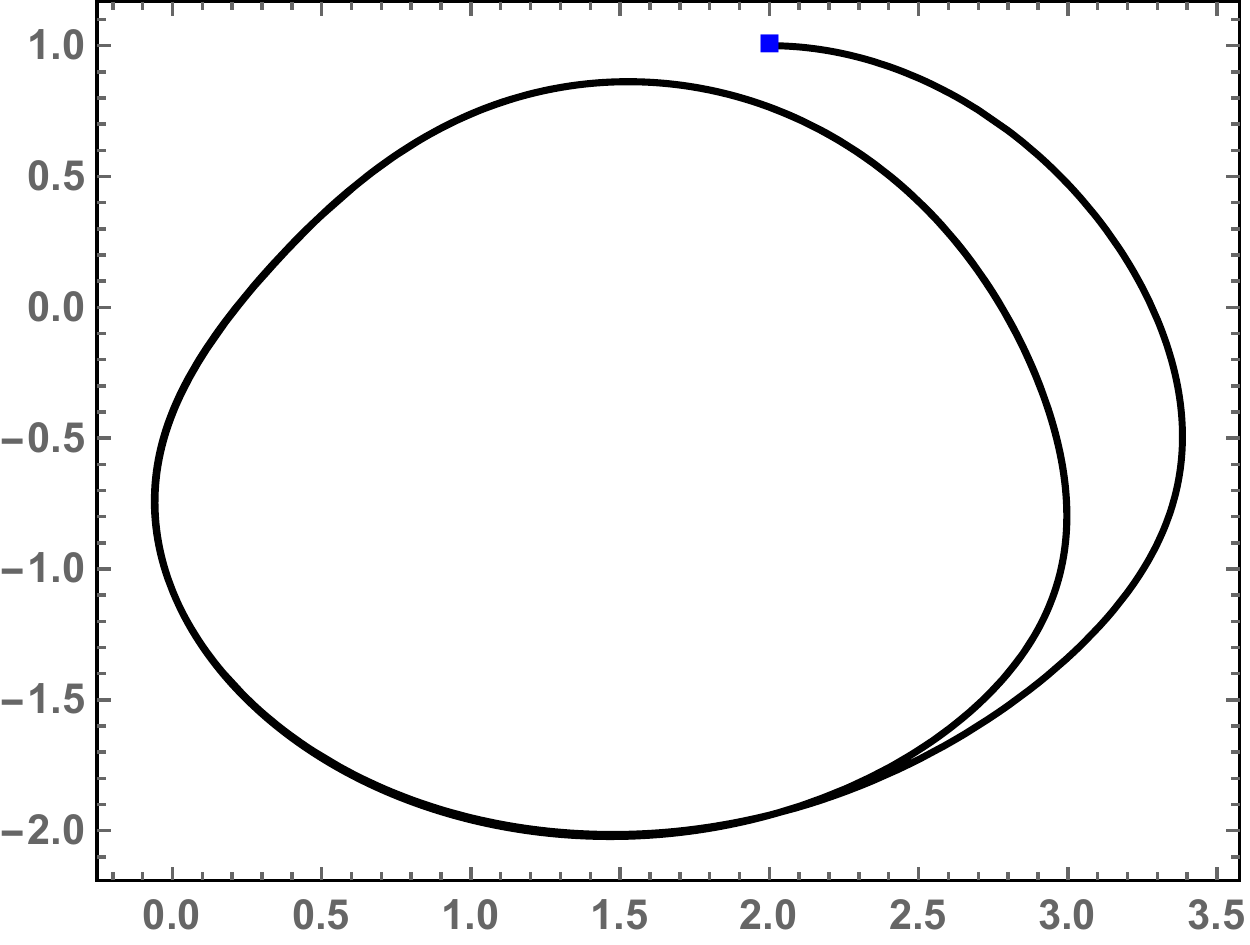}
\caption{Initial value problem~(\ref{SystemzwN2}),~(\ref{N2Ex2}). Trajectory, in the complex $z$-plane, of  $z_2(t)$. The   square indicates the initial condition $z_2(0)=2+\mathbf{i}$.}
\label{F24}
             \end{figure}
      \end{minipage}
  \end{minipage}

\begin{minipage}{\linewidth}
      \centering
      \begin{minipage}{0.45\linewidth}
          \begin{figure}[H]
		\includegraphics[width=\linewidth]{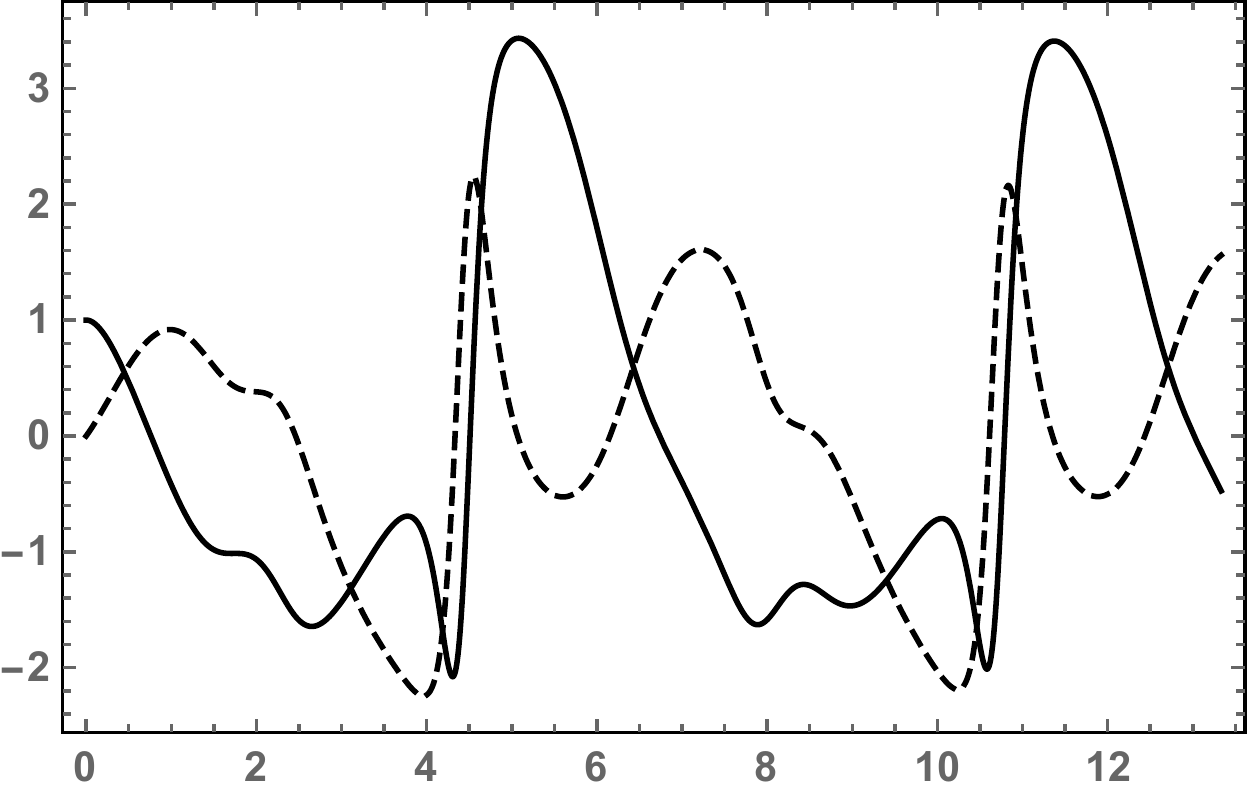}
\caption{ Initial value problem~(\ref{SystemzwN2}),~(\ref{N2Ex2}). Graphs of the real (bold curve) and imaginary (dashed curve) parts of the coordinate $w_1(t)$.}
\label{F25}
          \end{figure}
      \end{minipage}
      \hspace{0.05\linewidth}
      \begin{minipage}{0.45\linewidth}
          \begin{figure}[H]
\includegraphics[width=\linewidth]{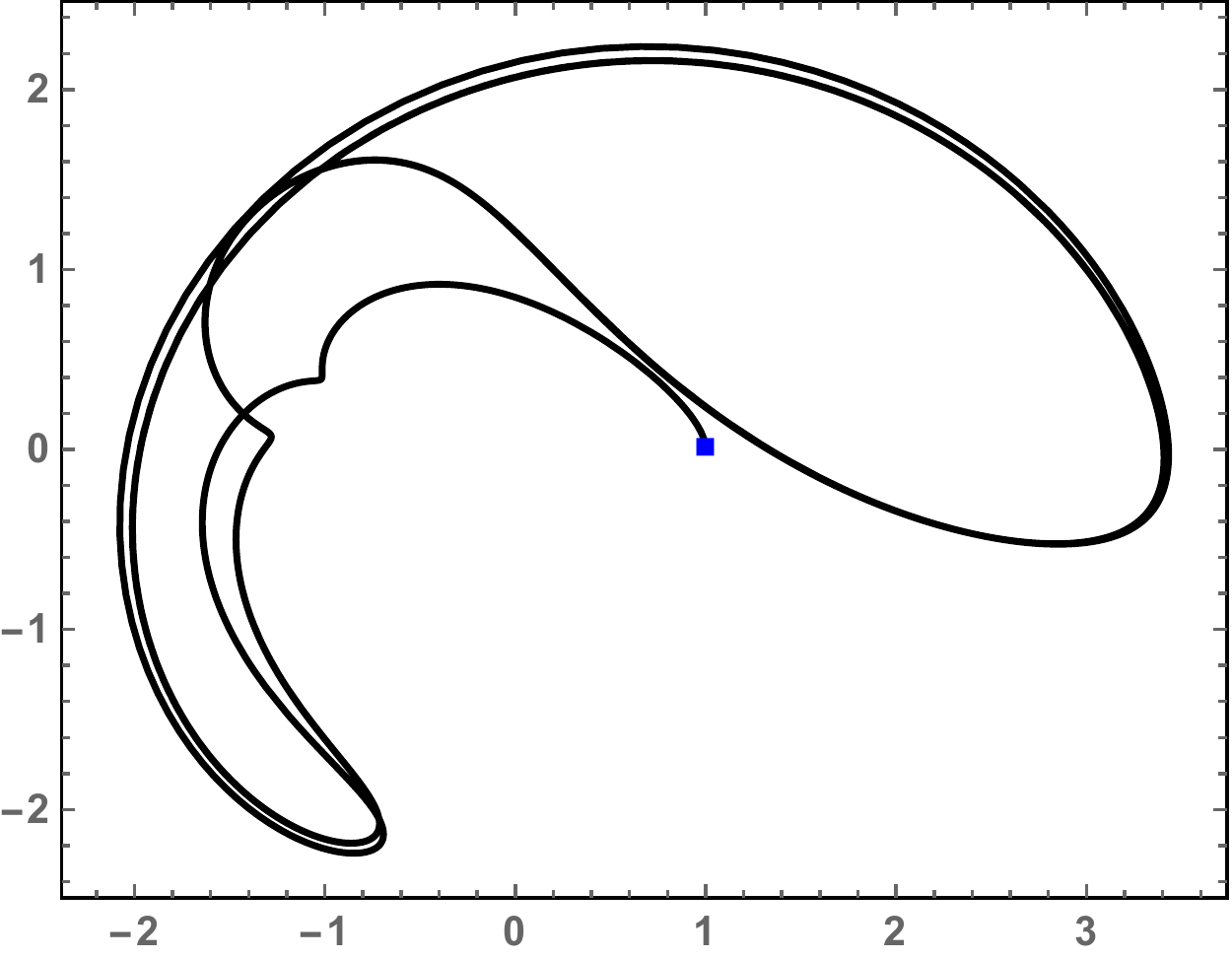}
\caption{Initial value problem~(\ref{SystemzwN2}),~(\ref{N2Ex2}). Trajectory, in the complex $z$-plane, of  $w_1(t)$. The   square indicates the initial condition $w_1(0)=1$.}
\label{F26}
             \end{figure}
      \end{minipage}
  \end{minipage}

\begin{minipage}{\linewidth}
      \centering
      \begin{minipage}{0.45\linewidth}
          \begin{figure}[H]
\includegraphics[width=\linewidth]{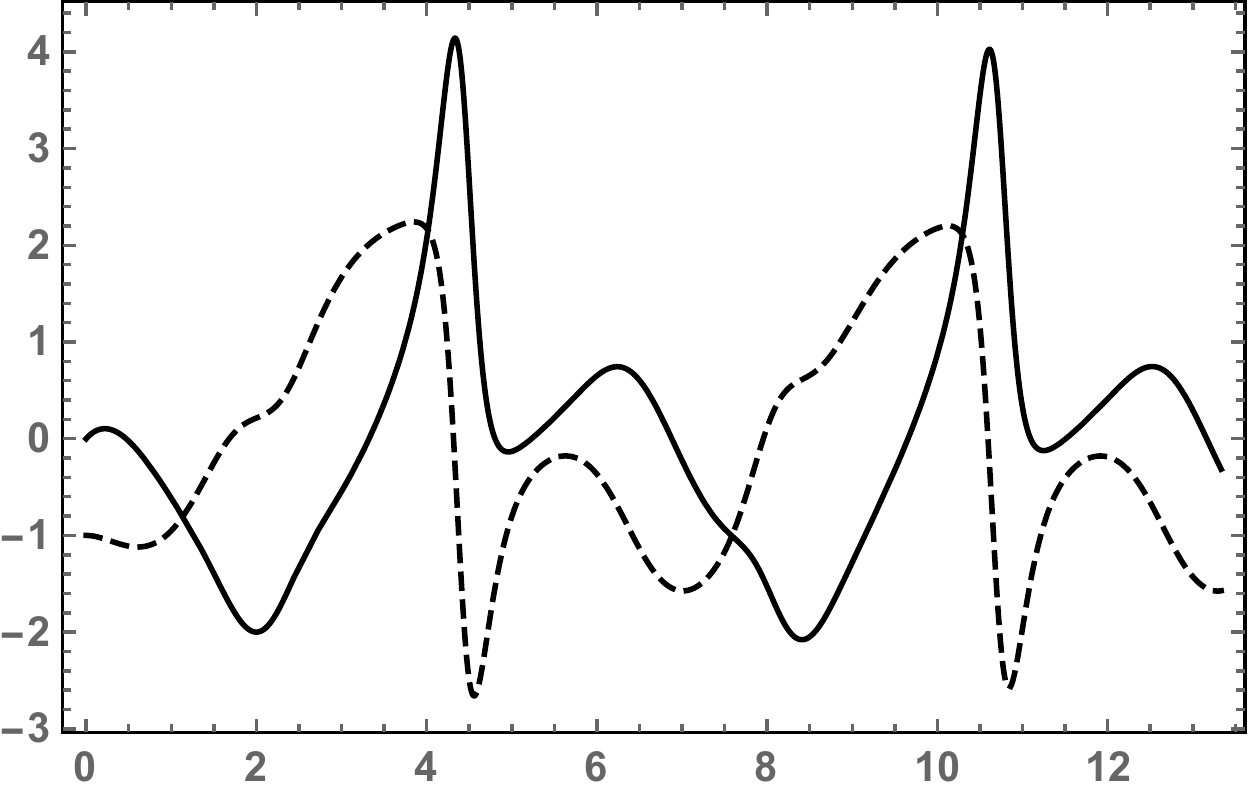}
\caption{ Initial value problem~(\ref{SystemzwN2}),~(\ref{N2Ex2}). Graphs of the real (bold curve) and imaginary (dashed curve) parts of the coordinate $w_2(t)$.}
\label{F27}
          \end{figure}
      \end{minipage}
      \hspace{0.05\linewidth}
      \begin{minipage}{0.45\linewidth}
          \begin{figure}[H]
\includegraphics[width=\linewidth]{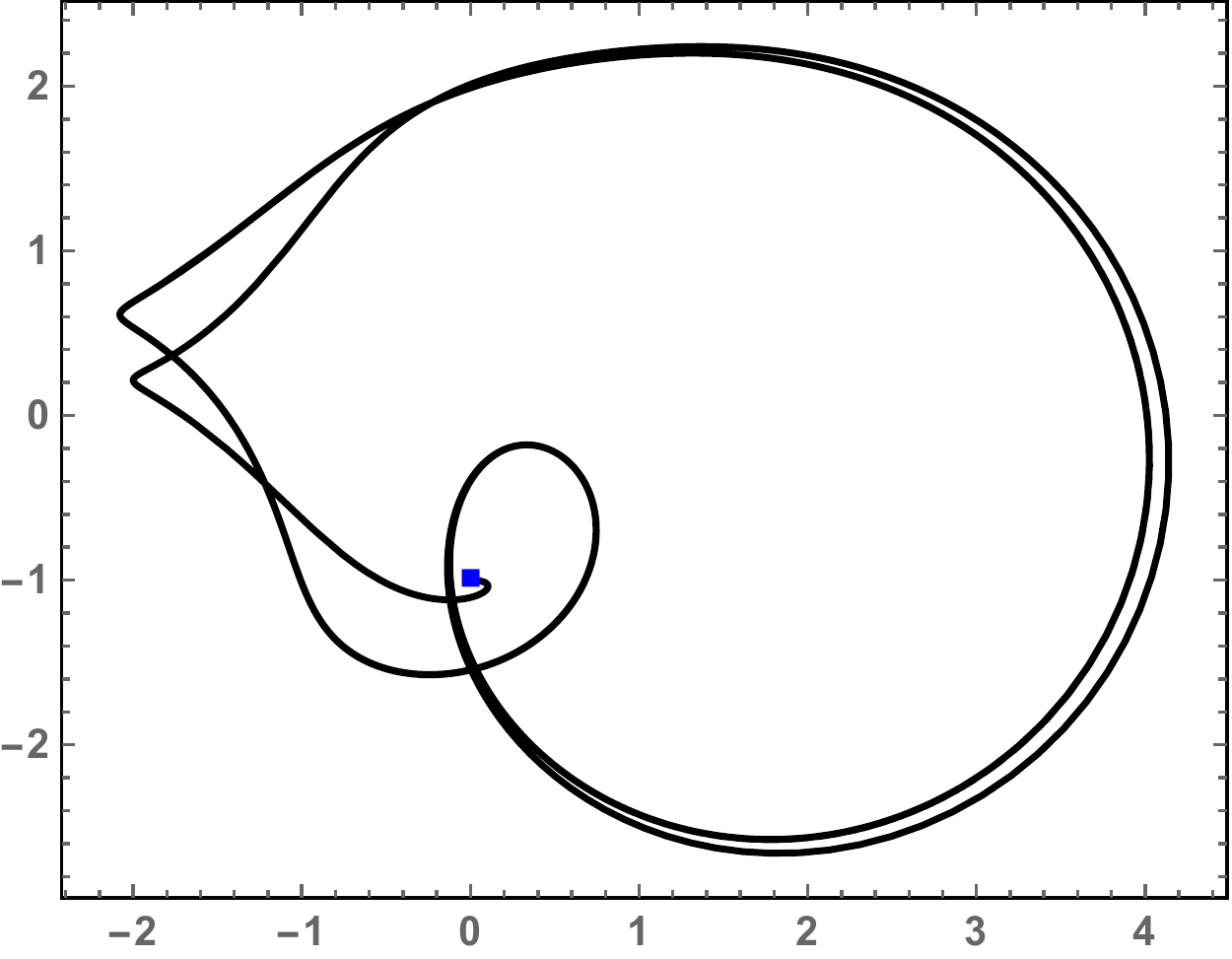}
\caption{Initial value problem~(\ref{SystemzwN2}),~(\ref{N2Ex2}). Trajectory, in the complex $z$-plane, of  $w_2(t)$. The   square indicates the initial condition $w_2(0)=-\mathbf{i}$.}
\label{F28}
             \end{figure}
      \end{minipage}
  \end{minipage}

\clearpage

\begin{minipage}{\linewidth}
      \centering
      \begin{minipage}{0.45\linewidth}
          \begin{figure}[H]
\includegraphics[width=\linewidth]{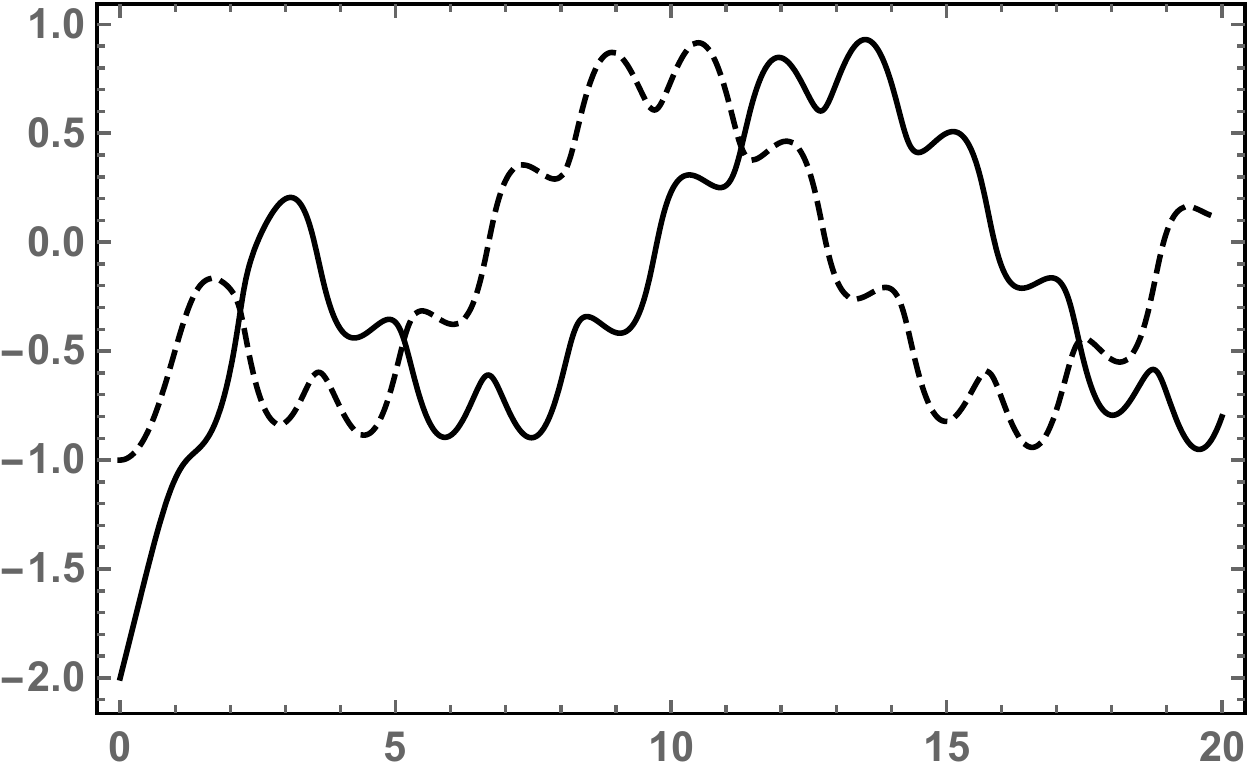}
\caption{ Initial value problem~(\ref{SystemzwN2}),~(\ref{N2Ex3}). Graphs of the real (bold curve) and imaginary (dashed curve) parts of the coordinate $z_1(t)$.}
\label{F31}
          \end{figure}
      \end{minipage}
      \hspace{0.05\linewidth}
      \begin{minipage}{0.45\linewidth}
          \begin{figure}[H]
\includegraphics[width=\linewidth]{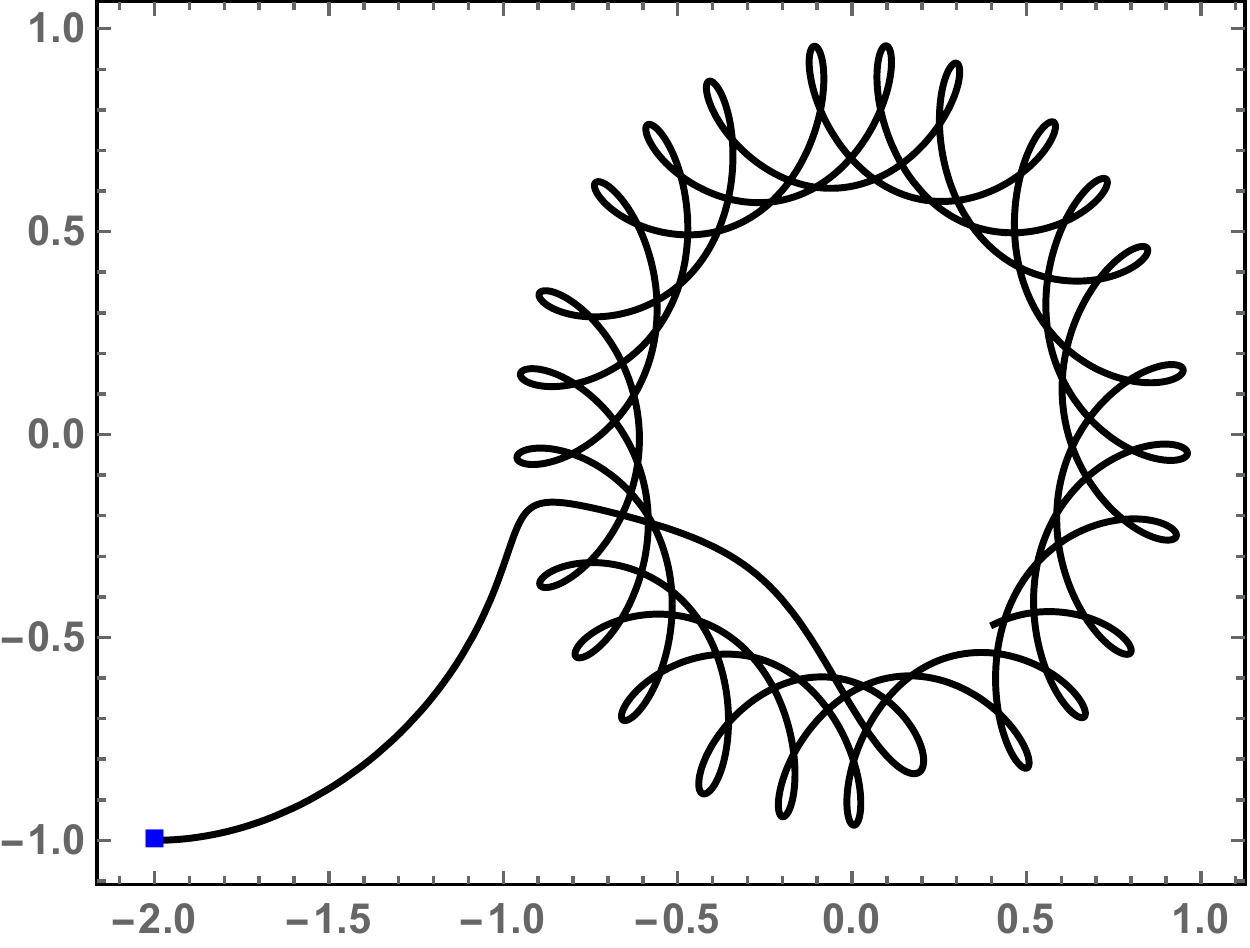}
\caption{Initial value problem~(\ref{SystemzwN2}),~(\ref{N2Ex3}). Trajectory, in the complex $z$-plane, of  $z_1(t)$. The   square indicates the initial condition $z_1(0)=-2-\mathbf{i}$.}
\label{F32}
             \end{figure}
      \end{minipage}
  \end{minipage}

\begin{minipage}{\linewidth}
      \centering
      \begin{minipage}{0.45\linewidth}
          \begin{figure}[H]
\includegraphics[width=\linewidth]{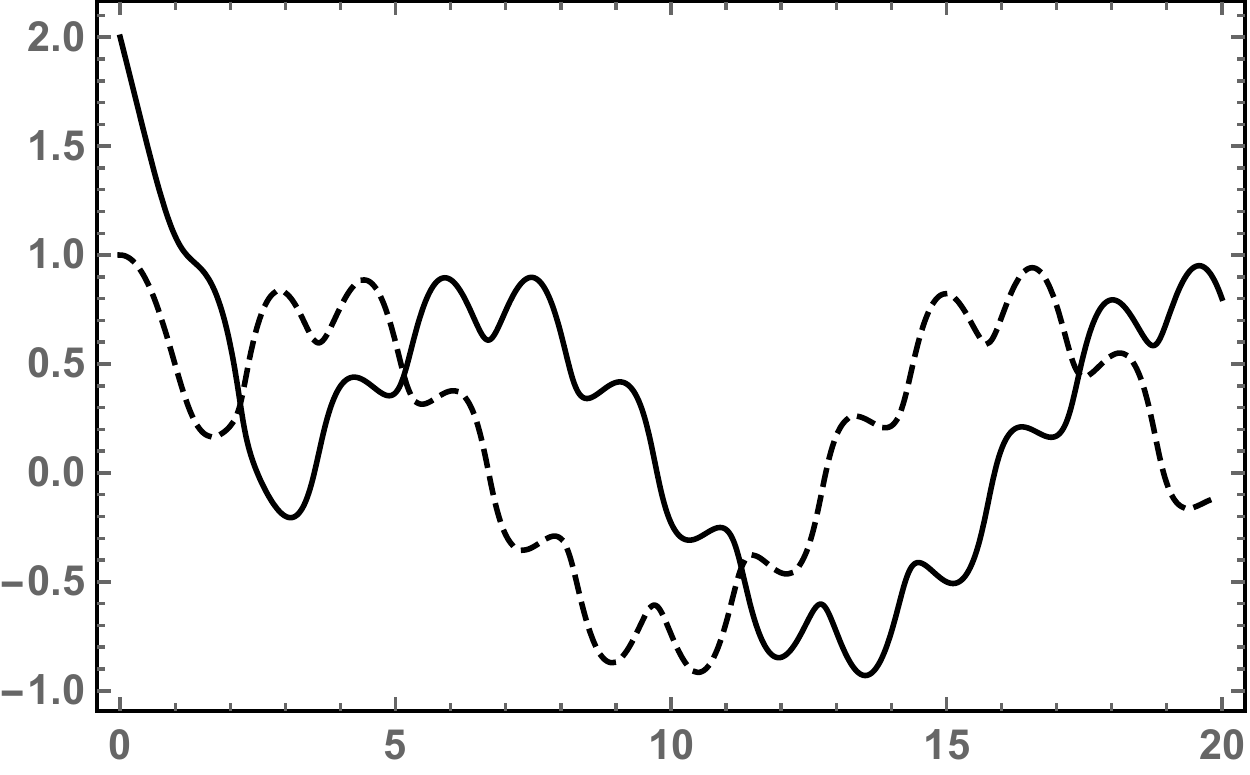}
\caption{ Initial value problem~(\ref{SystemzwN2}),~(\ref{N2Ex3}). Graphs of the real (bold curve) and imaginary (dashed curve) parts of the coordinate $z_2(t)$.}
\label{F33}
          \end{figure}
      \end{minipage}
      \hspace{0.05\linewidth}
      \begin{minipage}{0.45\linewidth}
          \begin{figure}[H]
\includegraphics[width=\linewidth]{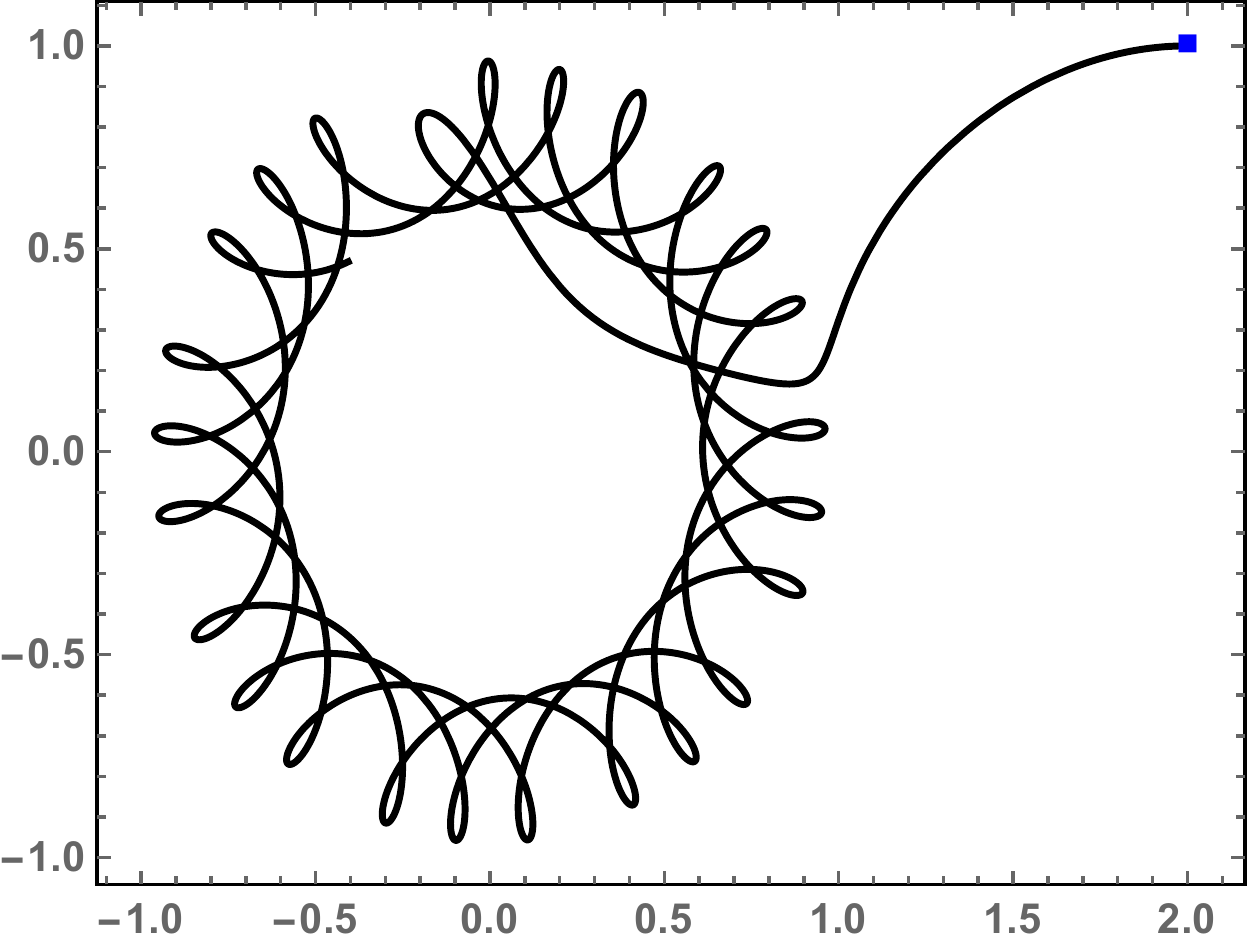}
\caption{Initial value problem~(\ref{SystemzwN2}),~(\ref{N2Ex3}). Trajectory, in the complex $z$-plane, of  $z_2(t)$. The   square indicates the initial condition $z_2(0)=2+\mathbf{i}$.}
\label{F34}
             \end{figure}
      \end{minipage}
  \end{minipage}
 
 \begin{minipage}{\linewidth}
      \centering
      \begin{minipage}{0.45\linewidth}
          \begin{figure}[H]
\includegraphics[width=\linewidth]{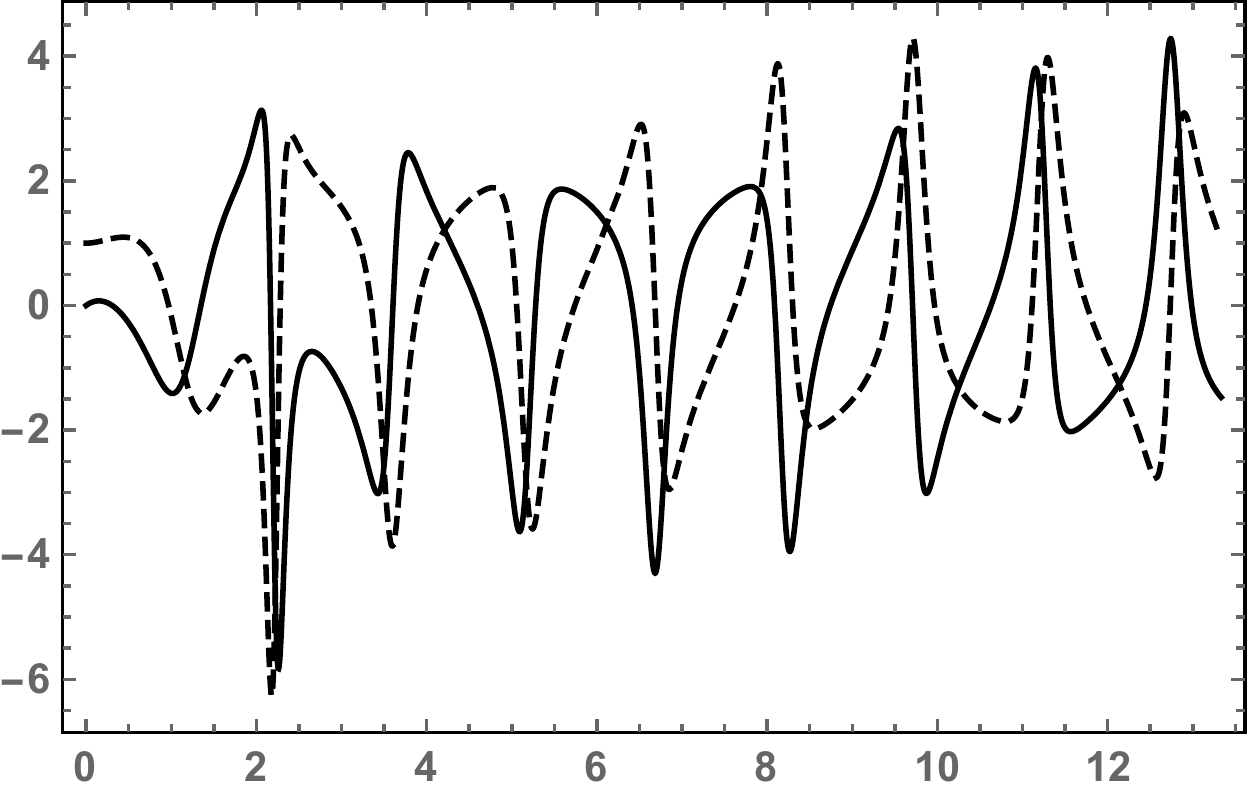}
\caption{ Initial value problem~(\ref{SystemzwN2}),~(\ref{N2Ex3}). Graphs of the real (bold curve) and imaginary (dashed curve) parts of the coordinate $w_1(t)$.}
\label{F35}
          \end{figure}
      \end{minipage}
      \hspace{0.05\linewidth}
      \begin{minipage}{0.45\linewidth}
          \begin{figure}[H]
\includegraphics[width=\linewidth]{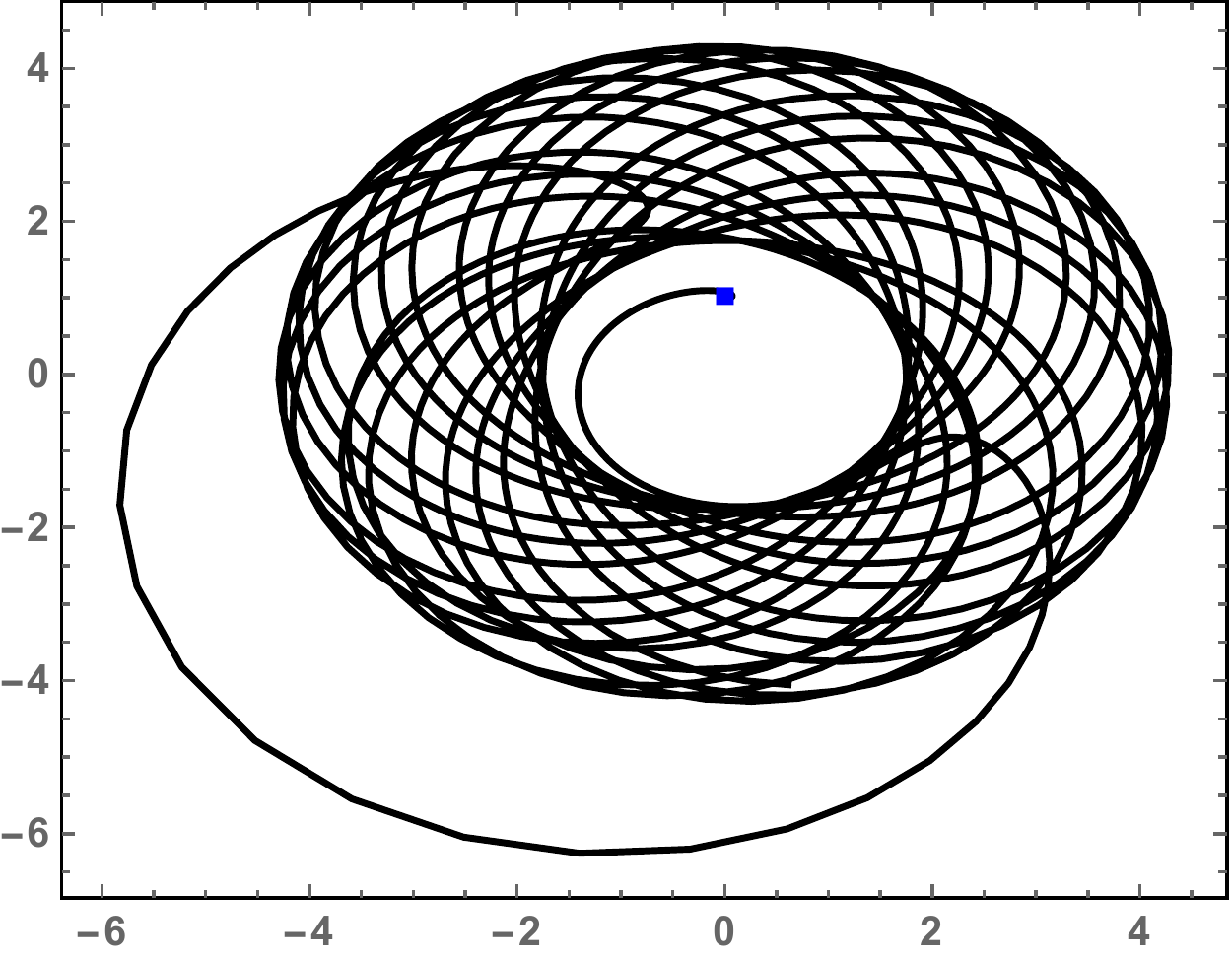}
\caption{Initial value problem~(\ref{SystemzwN2}),~(\ref{N2Ex3}). Trajectory, in the complex $z$-plane, of  $w_1(t)$. The   square indicates the initial condition $w_1(0)=\mathbf{i}$.}
\label{F36}
             \end{figure}
      \end{minipage}
  \end{minipage} 

 \begin{minipage}{\linewidth}
      \centering
      \begin{minipage}{0.45\linewidth}
            \begin{figure}[H]
         \includegraphics[width=\linewidth]{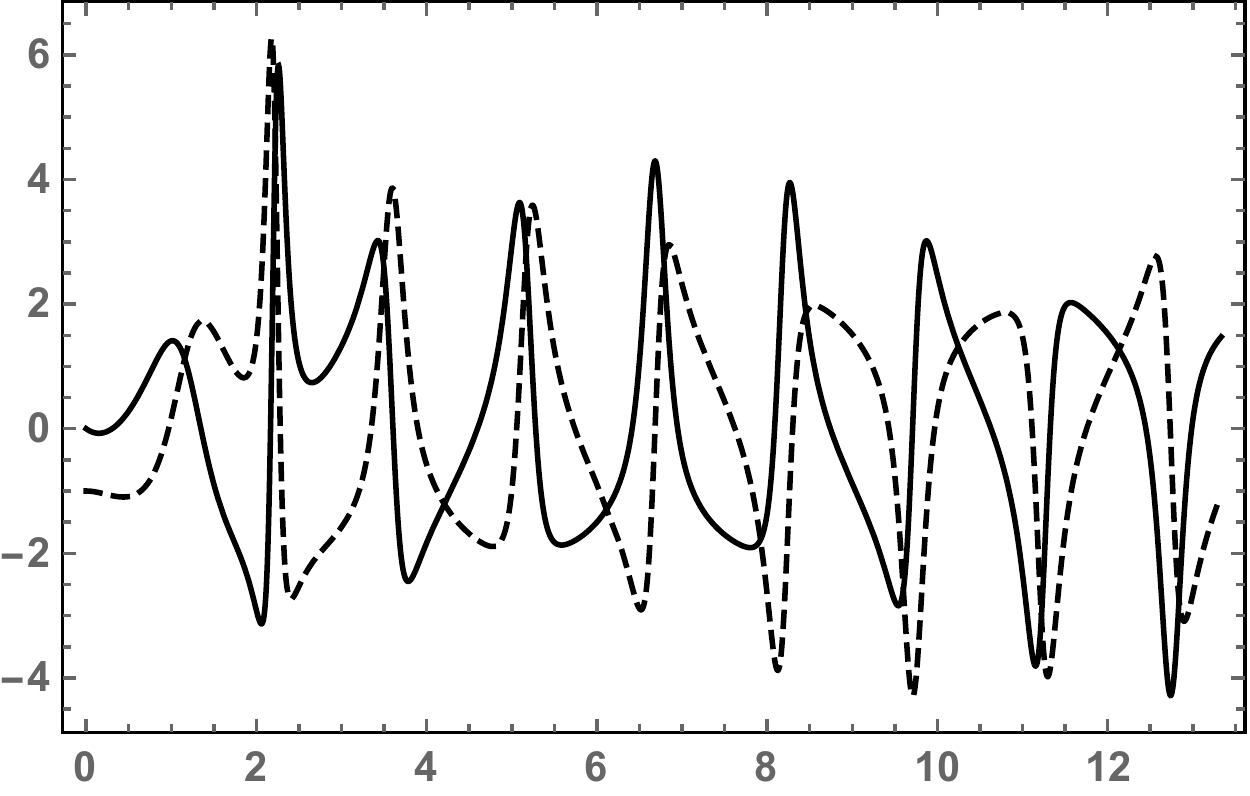}
\caption{ Initial value problem~(\ref{SystemzwN2}),~(\ref{N2Ex3}). Graphs of the real (bold curve) and imaginary (dashed curve) parts of the coordinate $w_2(t)$.}
\label{F37}
          \end{figure}
      \end{minipage}
      \hspace{0.05\linewidth}
      \begin{minipage}{0.45\linewidth}
          \begin{figure}[H]
\includegraphics[width=\linewidth]{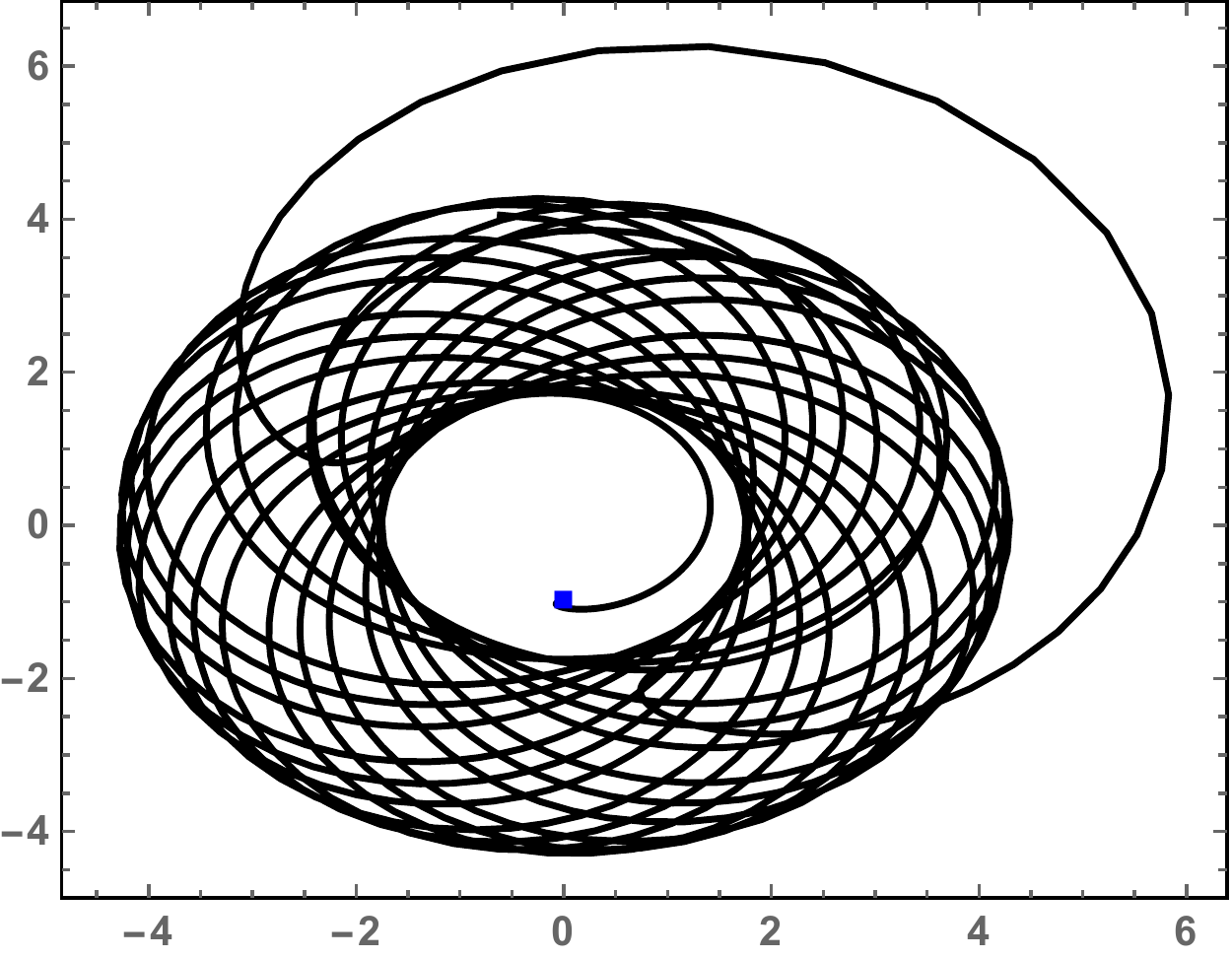}
\caption{Initial value problem~(\ref{SystemzwN2}),~(\ref{N2Ex3}). Trajectory, in the complex $z$-plane, of  $w_2(t)$. The   square indicates the initial condition $w_2(0)=-\mathbf{i}$.}
\label{F38}
             \end{figure}
      \end{minipage}
  \end{minipage}

\clearpage 

 \begin{minipage}{\linewidth}
      \centering
      \begin{minipage}{0.45\linewidth}
            \begin{figure}[H]
\includegraphics[width=\linewidth]{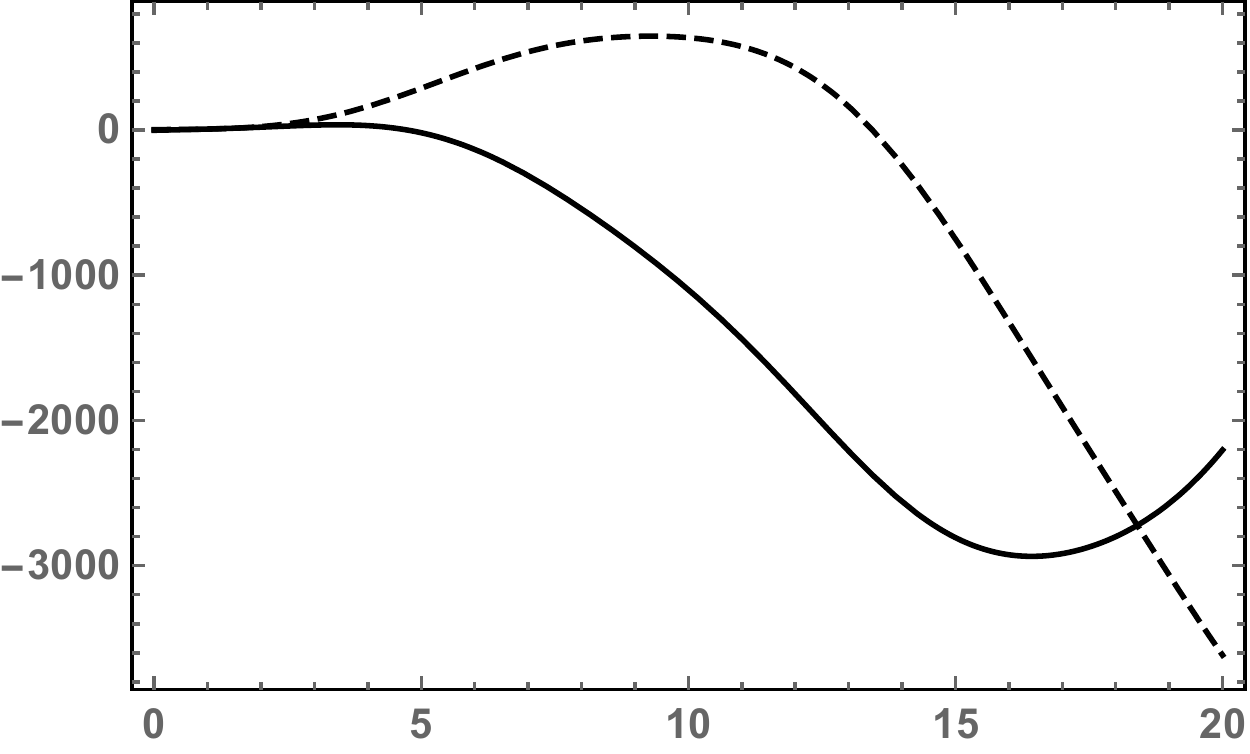}
\caption{ Initial value problem~(\ref{SystemzwN2}),~(\ref{N2Ex4}). Graphs of the real (bold curve) and imaginary (dashed curve) parts of the coordinate $z_1(t)$.}
\label{F41}
          \end{figure}
      \end{minipage}
      \hspace{0.05\linewidth}
      \begin{minipage}{0.45\linewidth}
          \begin{figure}[H]
\includegraphics[width=\linewidth]{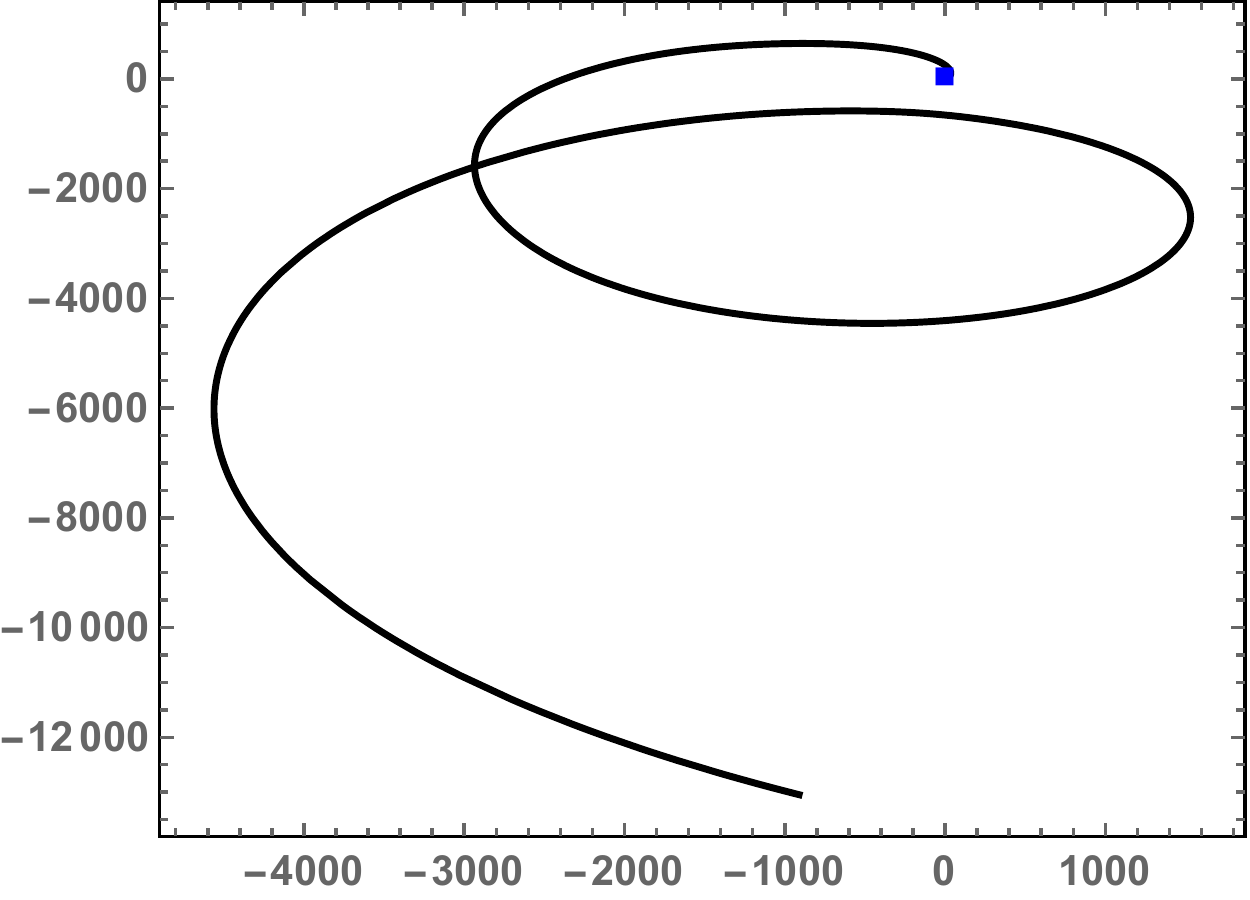}
\caption{Initial value problem~(\ref{SystemzwN2}),~(\ref{N2Ex4}). Trajectory, in the complex $z$-plane, of  $z_1(t)$. The   square indicates the initial condition $z_1(0)=-2+3\mathbf{i}$.}
\label{F42}
             \end{figure}
      \end{minipage}
  \end{minipage}

 \begin{minipage}{\linewidth}
      \centering
      \begin{minipage}{0.45\linewidth}
            \begin{figure}[H]
\includegraphics[width=\linewidth]{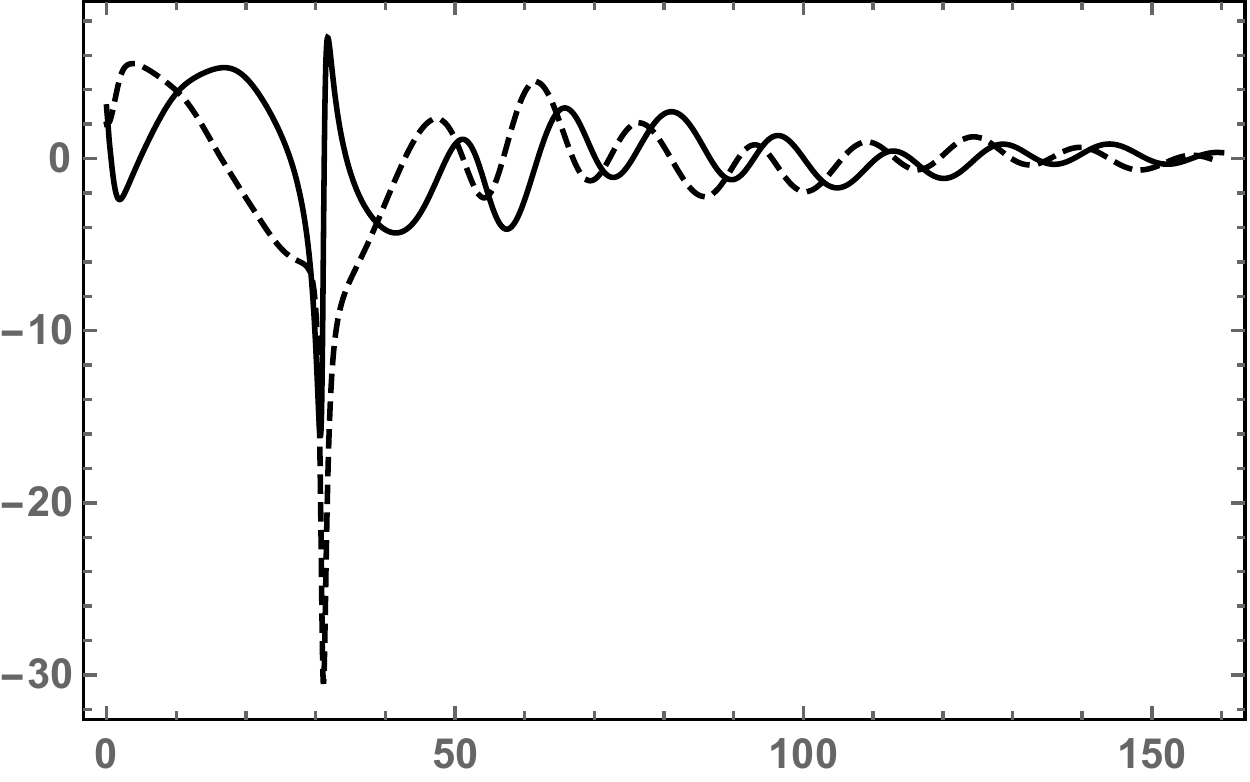}
\caption{ Initial value problem~(\ref{SystemzwN2}),~(\ref{N2Ex4}). Graphs of the real (bold curve) and imaginary (dashed curve) parts of the coordinate $z_2(t)$.}
\label{F43}
          \end{figure}
      \end{minipage}
      \hspace{0.05\linewidth}
      \begin{minipage}{0.45\linewidth}
          \begin{figure}[H]
\includegraphics[width=\linewidth]{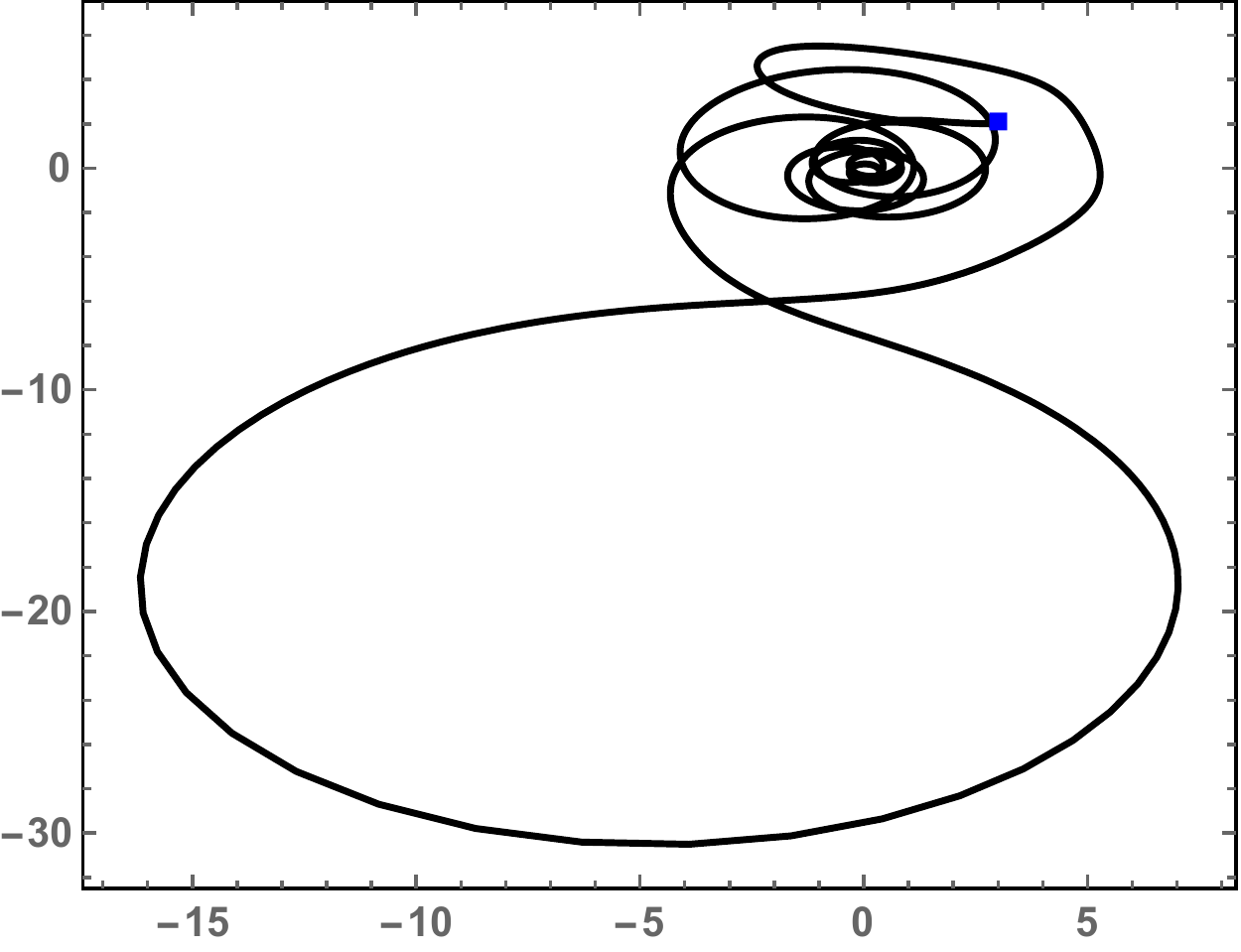}
\caption{Initial value problem~(\ref{SystemzwN2}),~(\ref{N2Ex4}). Trajectory, in the complex $z$-plane, of  $z_2(t)$. The   square indicates the initial condition $z_2(0)=3+2 \mathbf{i}$.}
\label{F44}
             \end{figure}
      \end{minipage}
  \end{minipage} 
  
  \pagebreak
  
\begin{minipage}{\linewidth}
      \centering
      \begin{minipage}{0.45\linewidth}
            \begin{figure}[H]
\includegraphics[width=\linewidth]{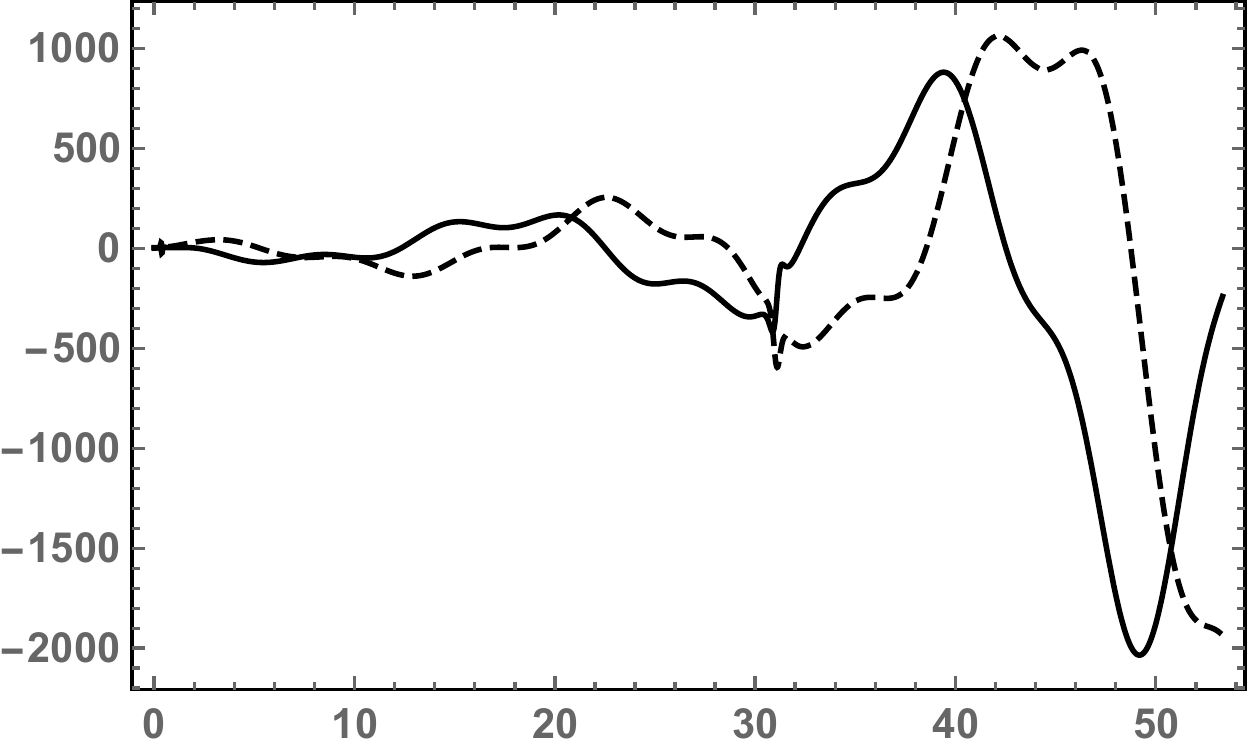}
\caption{ Initial value problem~(\ref{SystemzwN2}),~(\ref{N2Ex4}). Graphs of the real (bold curve) and imaginary (dashed curve) parts of the coordinate $w_1(t)$.}
\label{F45}
          \end{figure}
      \end{minipage}
      \hspace{0.05\linewidth}
      \begin{minipage}{0.45\linewidth}
          \begin{figure}[H]
\includegraphics[width=\linewidth]{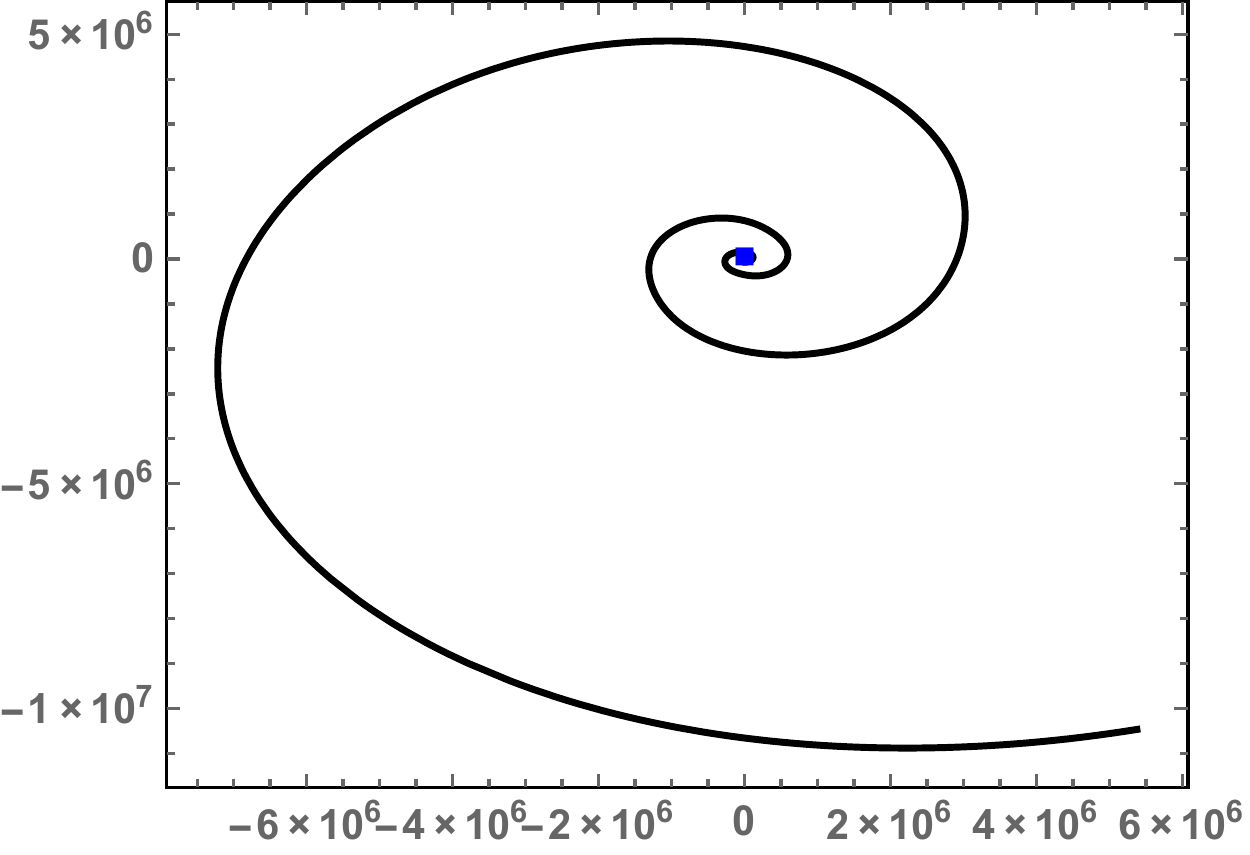}
\caption{Initial value problem~(\ref{SystemzwN2}),~(\ref{N2Ex4}). Trajectory, in the complex $z$-plane, of  $w_1(t)$. The   square indicates the initial condition $w_1(0)=2+4.2 \mathbf{i}$.}
\label{F46}
             \end{figure}
      \end{minipage}
  \end{minipage} 
  
\begin{minipage}{\linewidth}
      \centering
      \begin{minipage}{0.45\linewidth}
            \begin{figure}[H]
\includegraphics[width=\linewidth]{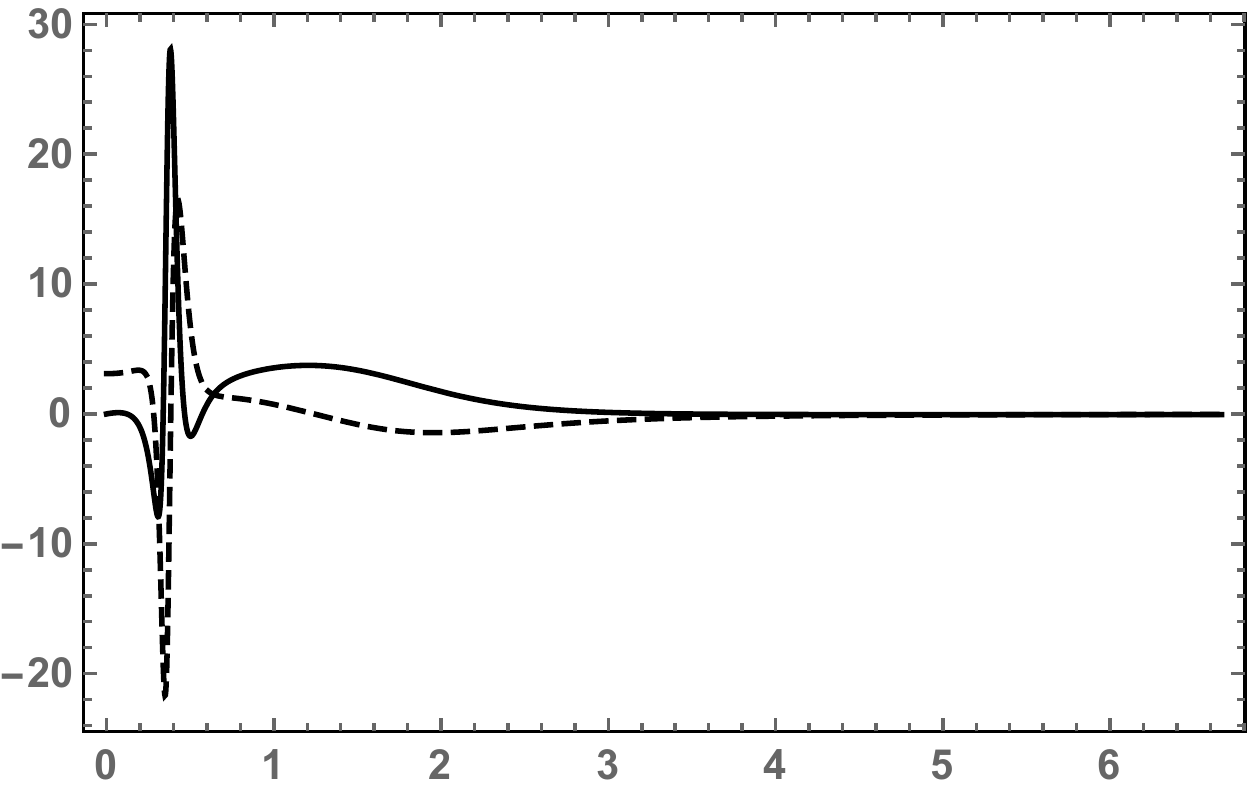}
\caption{ Initial value problem~(\ref{SystemzwN2}),~(\ref{N2Ex4}). Graphs of the real (bold curve) and imaginary (dashed curve) parts of the coordinate $w_2(t)$.}
\label{F47}
          \end{figure}
      \end{minipage}
      \hspace{0.05\linewidth}
      \begin{minipage}{0.45\linewidth}
          \begin{figure}[H]
\includegraphics[width=\linewidth]{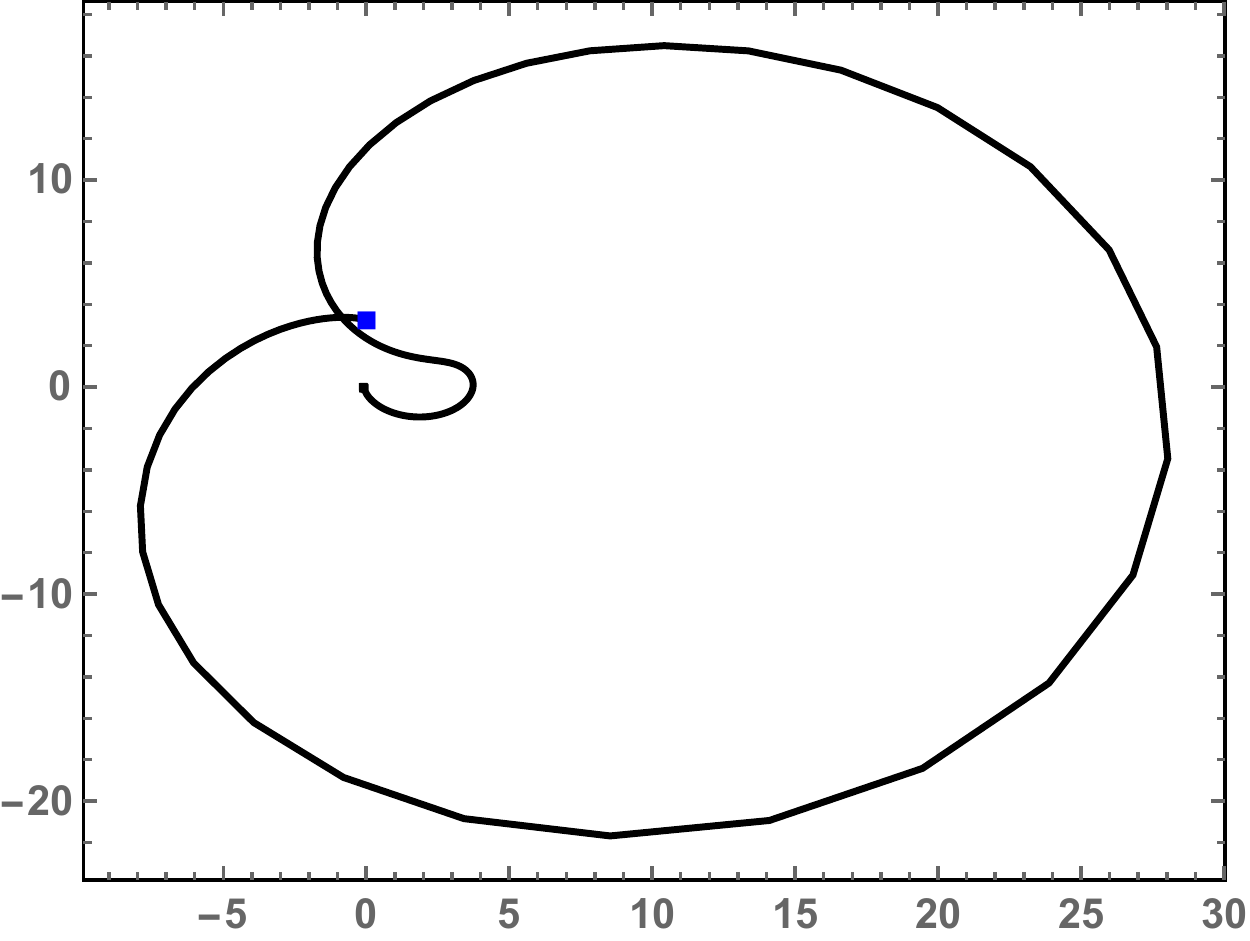}
\caption{Initial value problem~(\ref{SystemzwN2}),~(\ref{N2Ex4}). Trajectory, in the complex $z$-plane, of  $w_2(t)$. The   square indicates the initial condition $w_2(0)=3.1 \mathbf{i}$.}
\label{F48}
             \end{figure}
      \end{minipage}
  \end{minipage} 

\clearpage 

\begin{minipage}{\linewidth}
      \centering
      \begin{minipage}{0.45\linewidth}
            \begin{figure}[H]
\includegraphics[width=\linewidth]{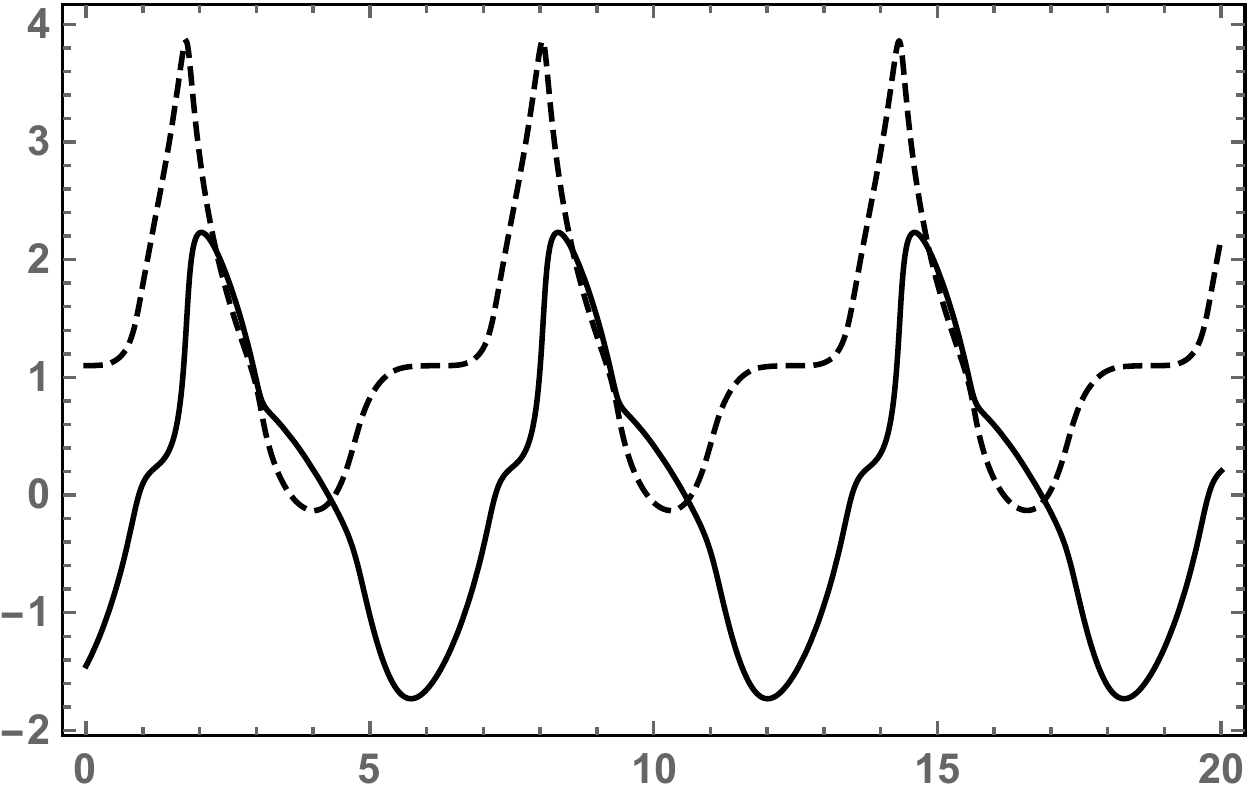}
\caption{ Initial value problem~(\ref{SystemzwN3}),~(\ref{N3Ex1}). Graphs of the real (bold curve) and imaginary (dashed curve) parts of the coordinate $z_1(t)$.}
\label{N3F11}
          \end{figure}
      \end{minipage}
      \hspace{0.05\linewidth}
      \begin{minipage}{0.45\linewidth}
          \begin{figure}[H]
\includegraphics[width=\linewidth]{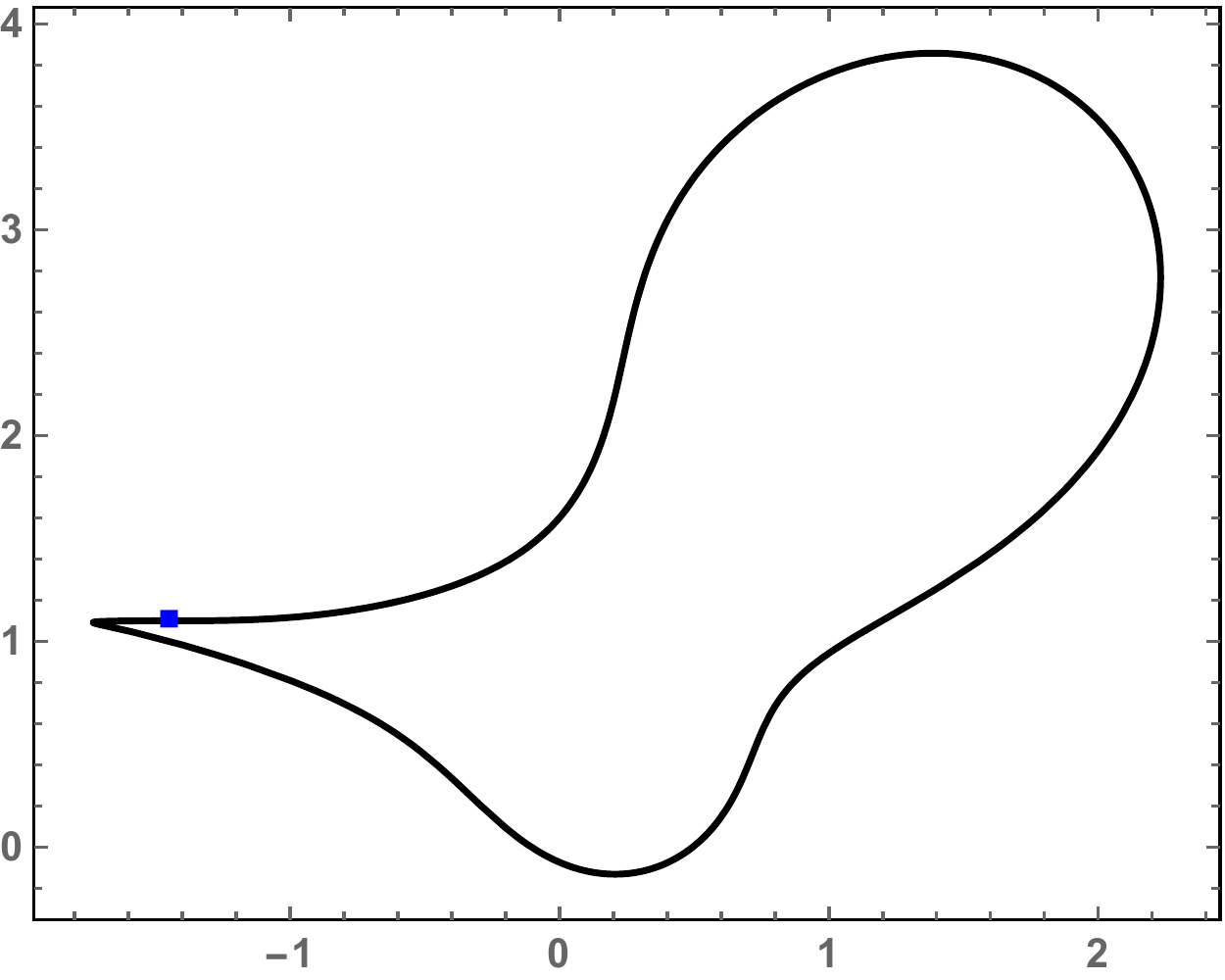}
\caption{Initial value problem~(\ref{SystemzwN3}),~(\ref{N3Ex1}). Trajectory, in the complex $z$-plane, of  $z_1(t)$; period $2\pi$. The   square indicates the initial condition $z_1(0)=-1.45+1.1 \mathbf{i}$.}
\label{N3F12}
             \end{figure}
      \end{minipage}
  \end{minipage} 

\begin{minipage}{\linewidth}
      \centering
      \begin{minipage}{0.45\linewidth}
            \begin{figure}[H]
\includegraphics[width=\linewidth]{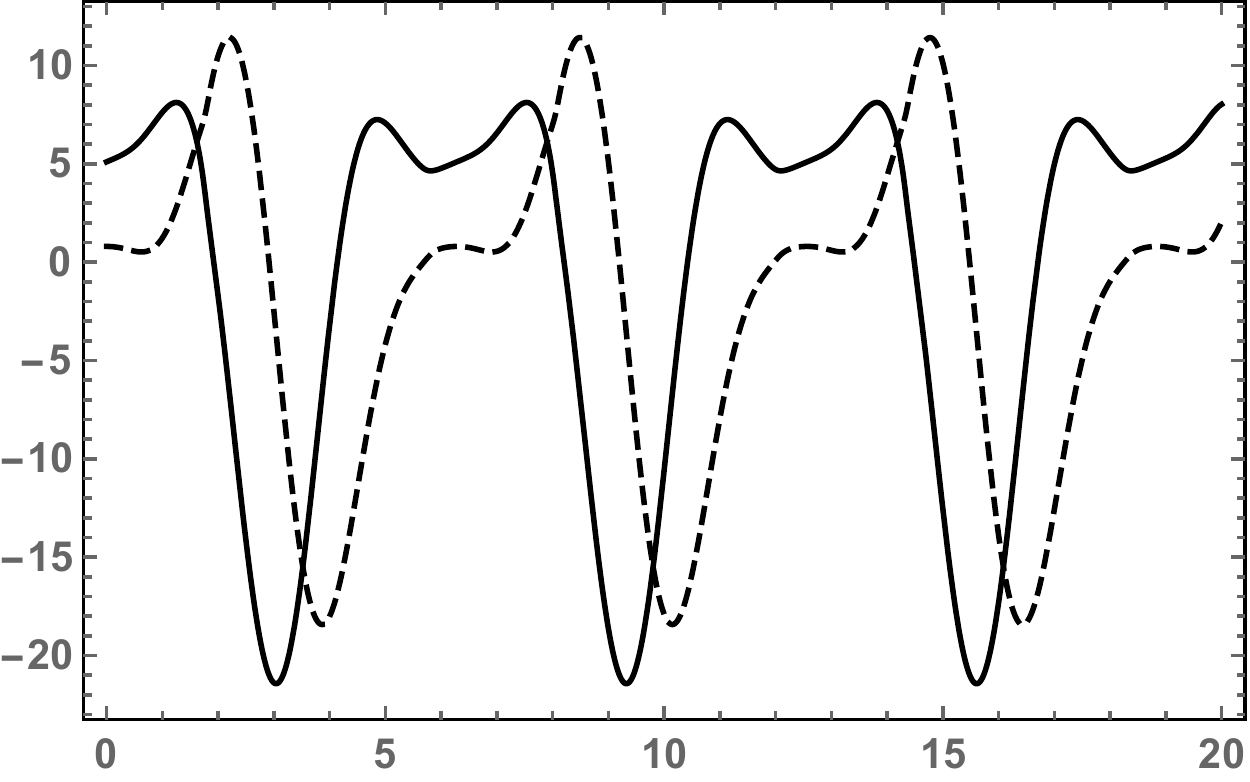}
\caption{ Initial value problem~(\ref{SystemzwN3}),~(\ref{N3Ex1}). Graphs of the real (bold curve) and imaginary (dashed curve) parts of the coordinate $z_2(t)$.}
\label{N3F13}
          \end{figure}
      \end{minipage}
      \hspace{0.05\linewidth}
      \begin{minipage}{0.45\linewidth}
          \begin{figure}[H]
\includegraphics[width=\linewidth]{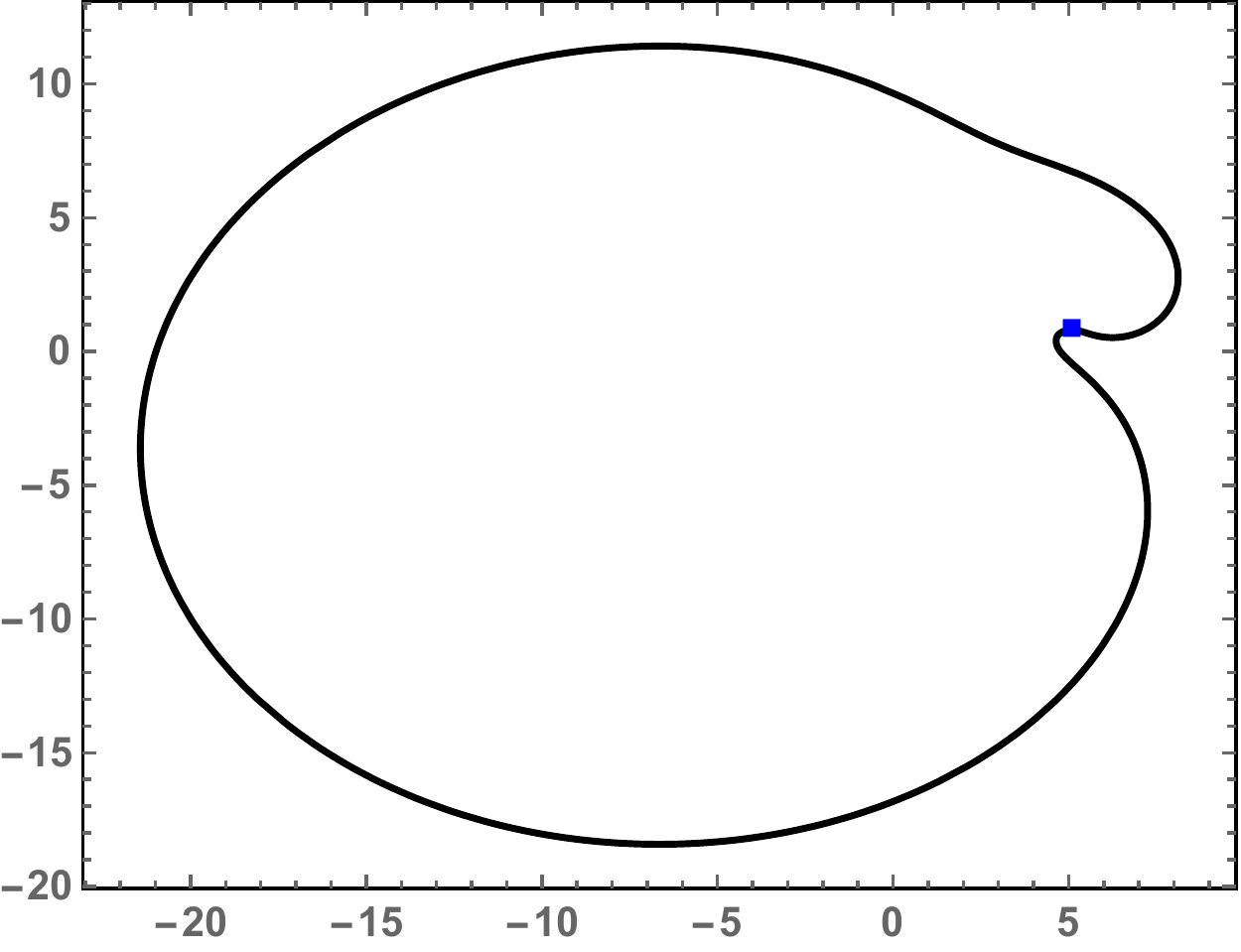}
\caption{Initial value problem~(\ref{SystemzwN3}),~(\ref{N3Ex1}). Trajectory, in the complex $z$-plane, of  $z_2(t)$; period $4\pi$. The   square indicates the initial condition $z_2(0)=5.1+0.8 \mathbf{i}$.}
\label{N3F14}
             \end{figure}
      \end{minipage}
  \end{minipage} 

\begin{minipage}{\linewidth}
      \centering
      \begin{minipage}{0.45\linewidth}
            \begin{figure}[H]
\includegraphics[width=\linewidth]{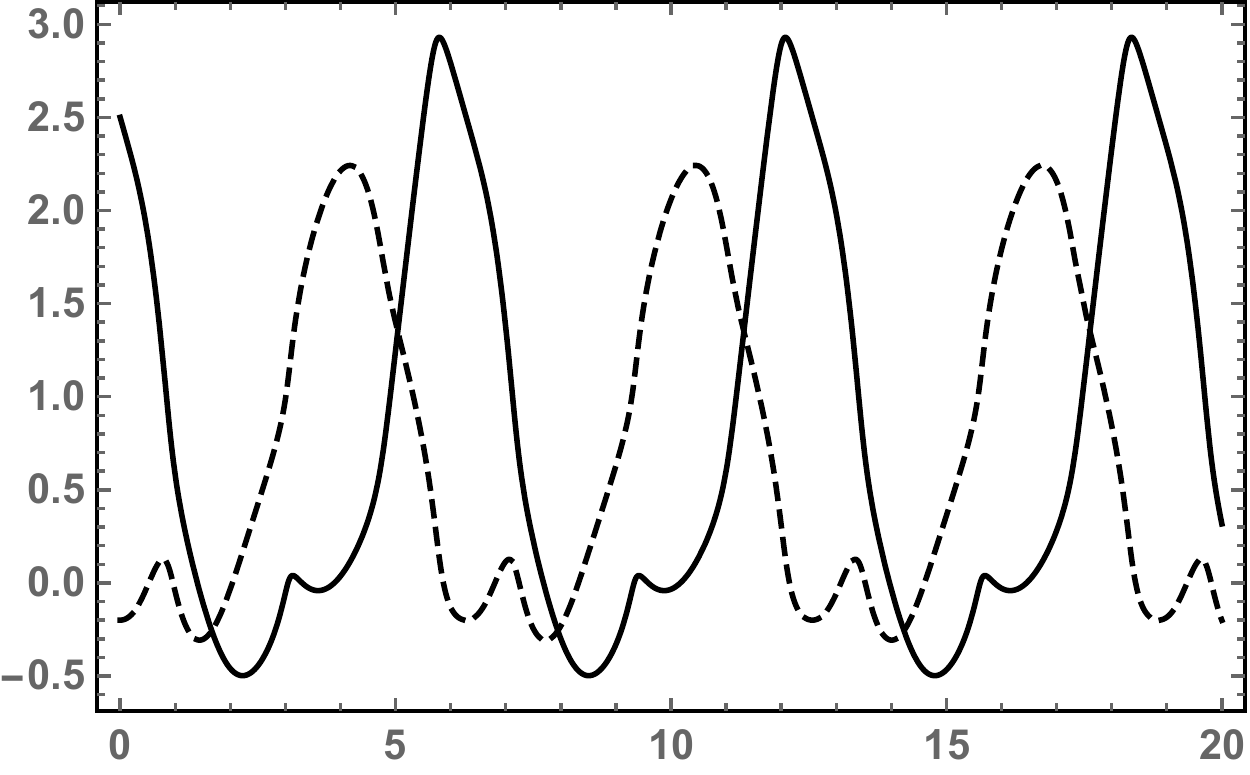}
\caption{ Initial value problem~(\ref{SystemzwN3}),~(\ref{N3Ex1}). Graphs of the real (bold curve) and imaginary (dashed curve) parts of the coordinate $z_3(t)$.}
\label{N3F15}
          \end{figure}
      \end{minipage}
      \hspace{0.05\linewidth}
      \begin{minipage}{0.45\linewidth}
          \begin{figure}[H]
\includegraphics[width=\linewidth]{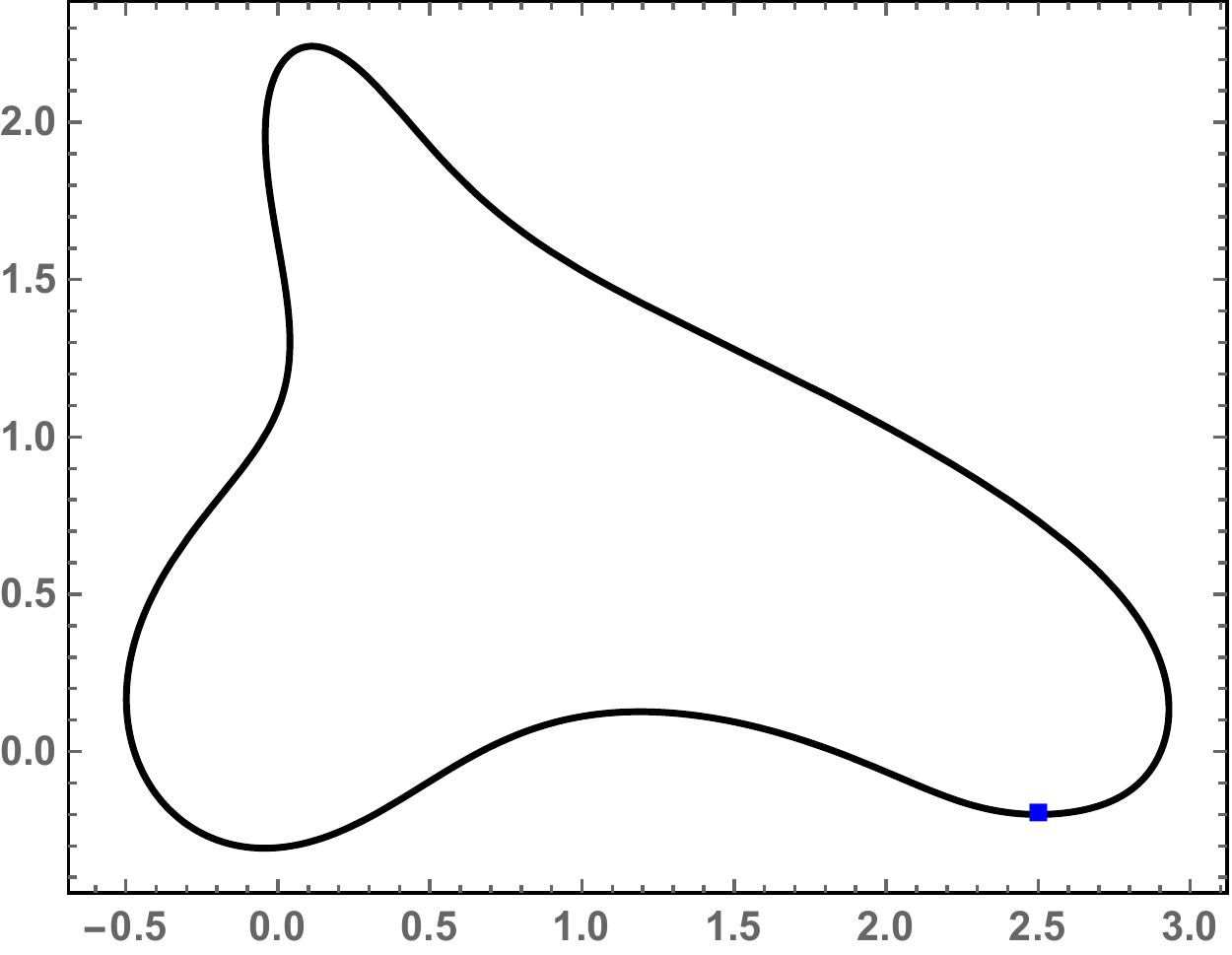}
\caption{Initial value problem~(\ref{SystemzwN3}),~(\ref{N3Ex1}). Trajectory, in the complex $z$-plane, of  $z_3(t)$; period $4\pi$.  The   square indicates the initial condition $z_3(0)=2.5-0.2 \mathbf{i}$.}
\label{N3F16}
             \end{figure}
      \end{minipage}
  \end{minipage} 
  
\begin{minipage}{\linewidth}
      \centering
      \begin{minipage}{0.45\linewidth}
            \begin{figure}[H]
\includegraphics[width=\linewidth]{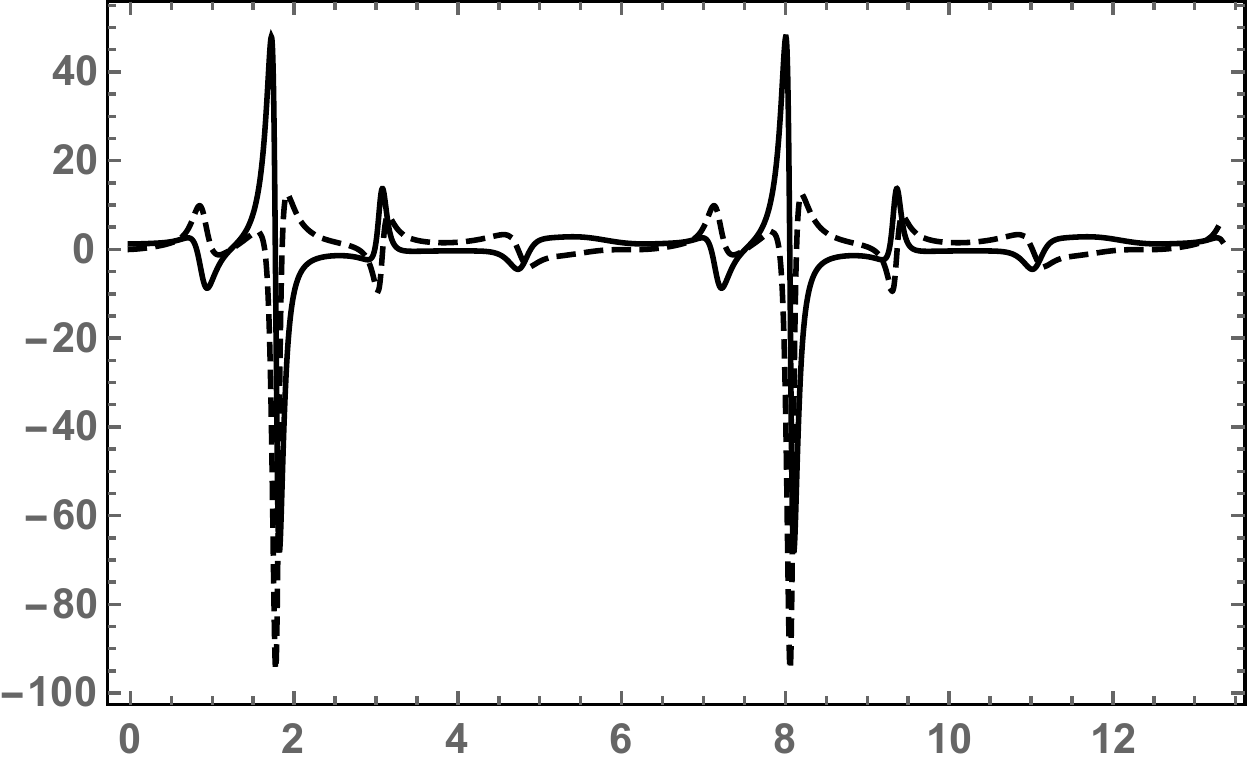}
\caption{ Initial value problem~(\ref{SystemzwN3}),~(\ref{N3Ex1}). Graphs of the real (bold curve) and imaginary (dashed curve) parts of the coordinate $w_1(t)$.}
\label{N3F17}
          \end{figure}
      \end{minipage}
      \hspace{0.05\linewidth}
      \begin{minipage}{0.45\linewidth}
          \begin{figure}[H]
\includegraphics[width=\linewidth]{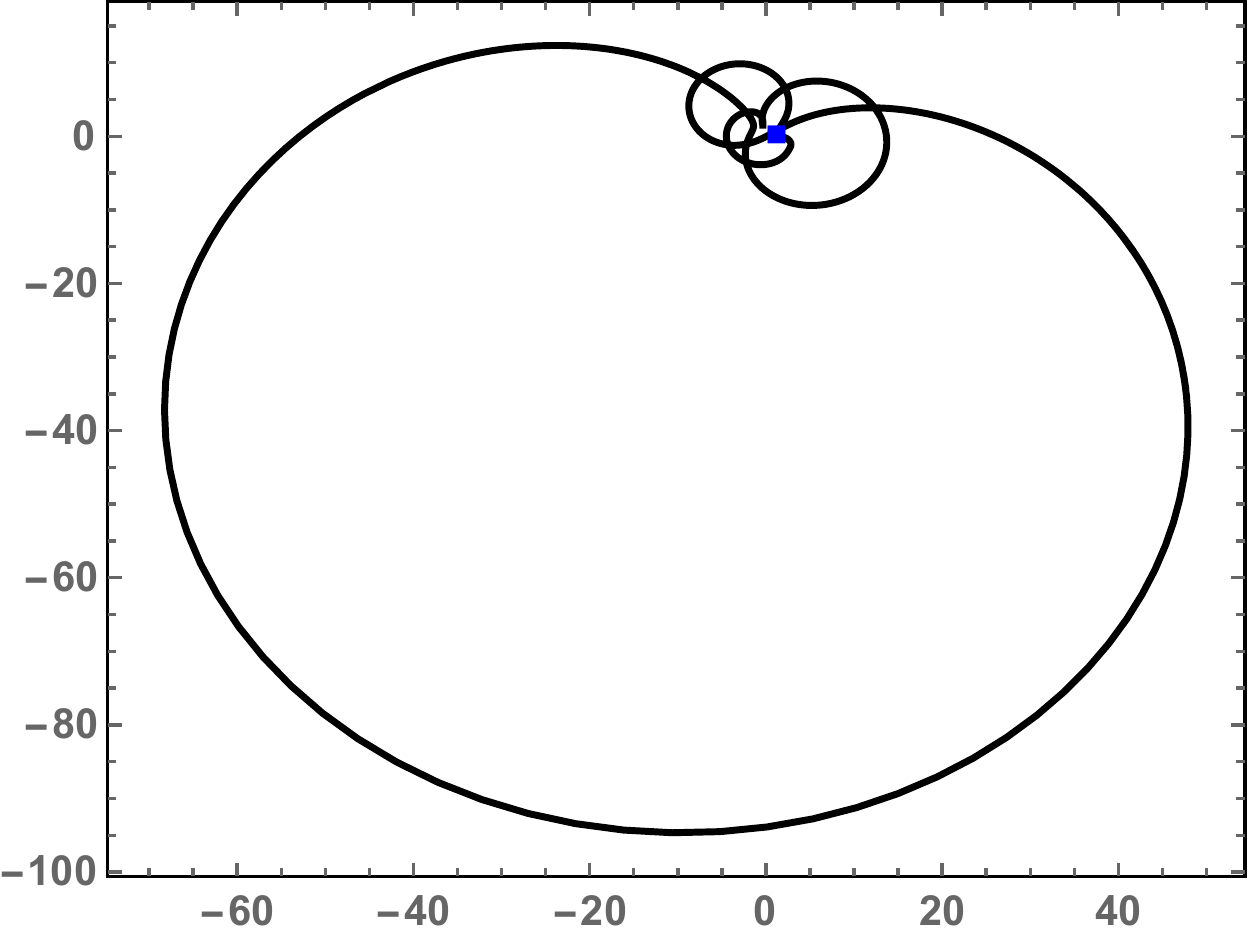}
\caption{Initial value problem~(\ref{SystemzwN3}),~(\ref{N3Ex1}). Trajectory, in the complex $z$-plane, of  $w_1(t)$; period $2\pi$. . The   square indicates the initial condition $w_1(0)=1.23$.}
\label{N3F18}
             \end{figure}
      \end{minipage}
  \end{minipage} 
  
\begin{minipage}{\linewidth}
      \centering
      \begin{minipage}{0.45\linewidth}
            \begin{figure}[H]
\includegraphics[width=\linewidth]{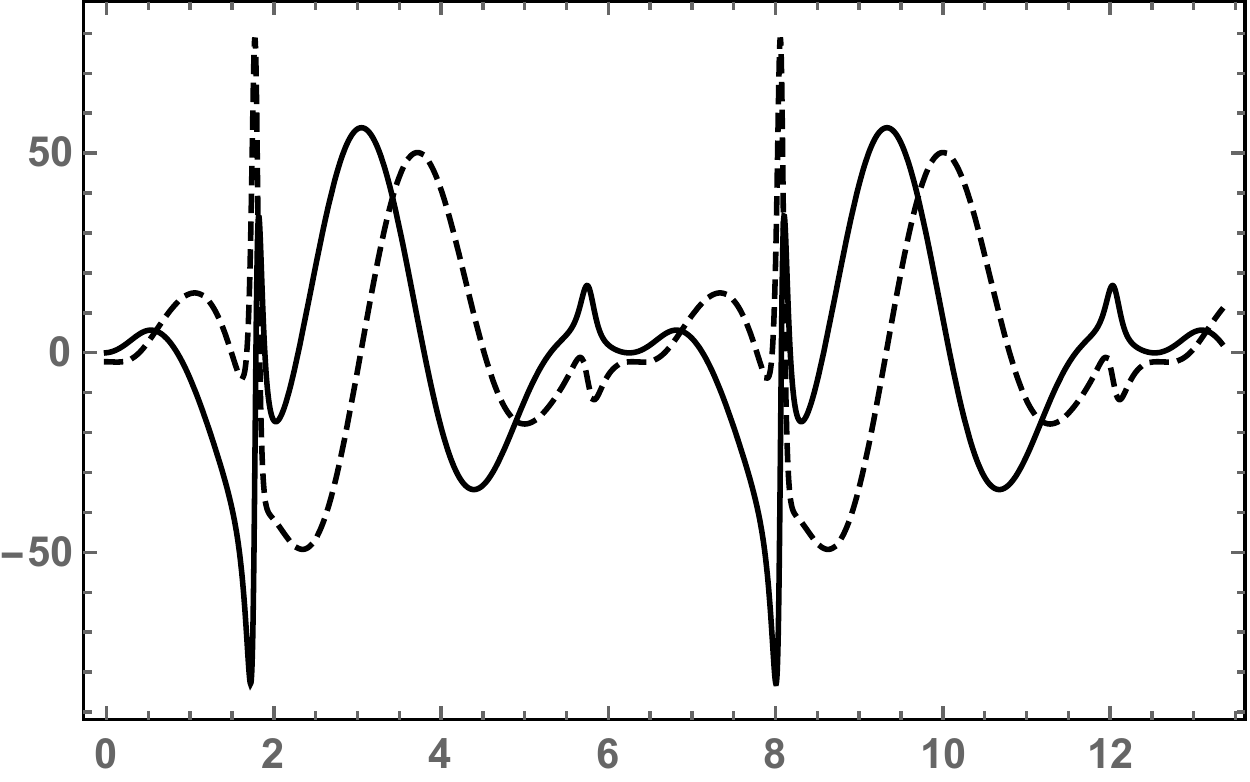}
\caption{ Initial value problem~(\ref{SystemzwN3}),~(\ref{N3Ex1}). Graphs of the real (bold curve) and imaginary (dashed curve) parts of the coordinate $w_2(t)$.}
\label{N3F19}
          \end{figure}
      \end{minipage}
      \hspace{0.05\linewidth}
      \begin{minipage}{0.45\linewidth}
          \begin{figure}[H]
\includegraphics[width=\linewidth]{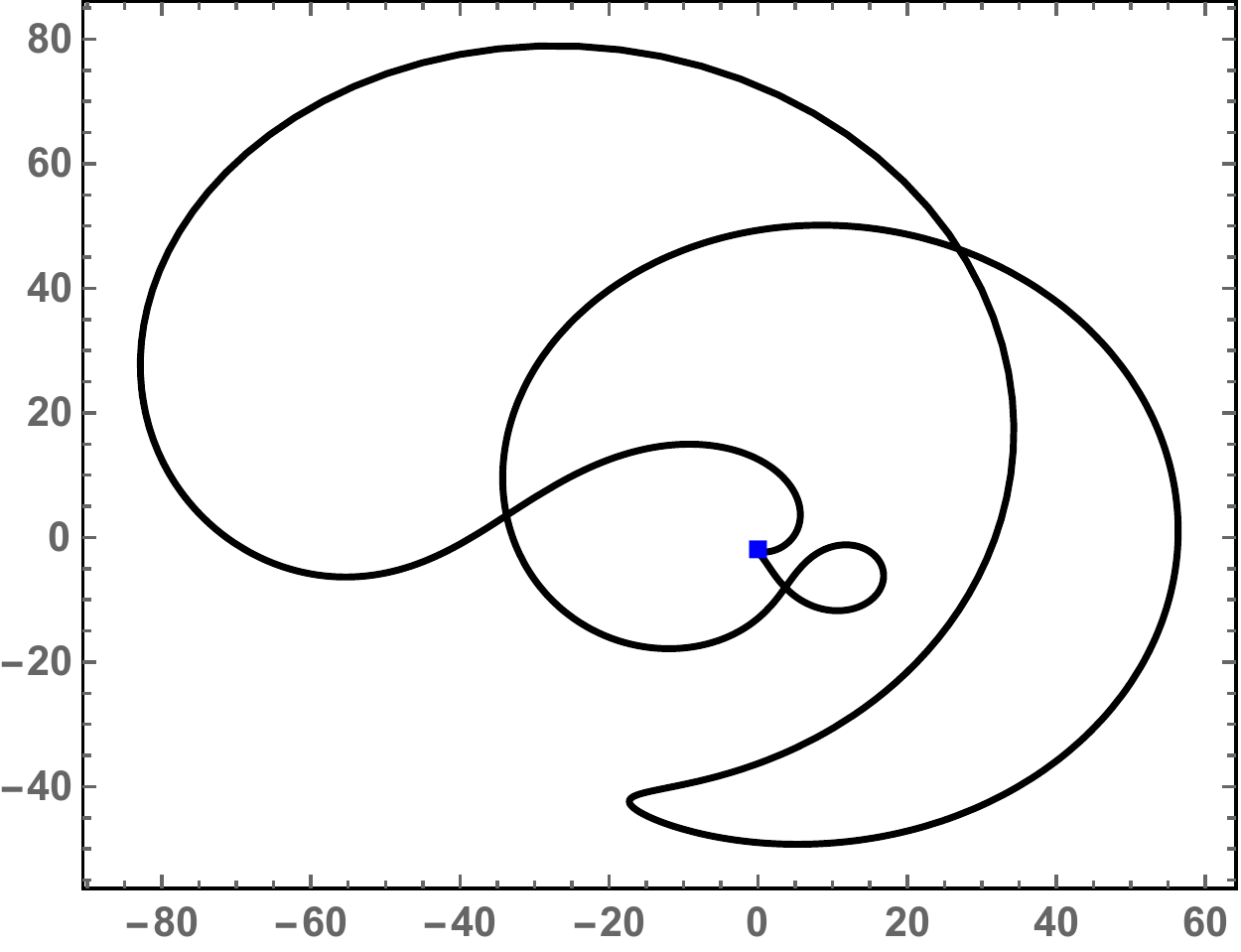}
\caption{Initial value problem~(\ref{SystemzwN3}),~(\ref{N3Ex1}). Trajectory, in the complex $z$-plane, of  $w_2(t)$; period $4\pi$. The   square indicates the initial condition $w_2(0)=-2.26 \mathbf{i}$.}
\label{N3F110}
             \end{figure}
      \end{minipage}
  \end{minipage} 

\begin{minipage}{\linewidth}
      \centering
      \begin{minipage}{0.45\linewidth}
            \begin{figure}[H]
\includegraphics[width=\linewidth]{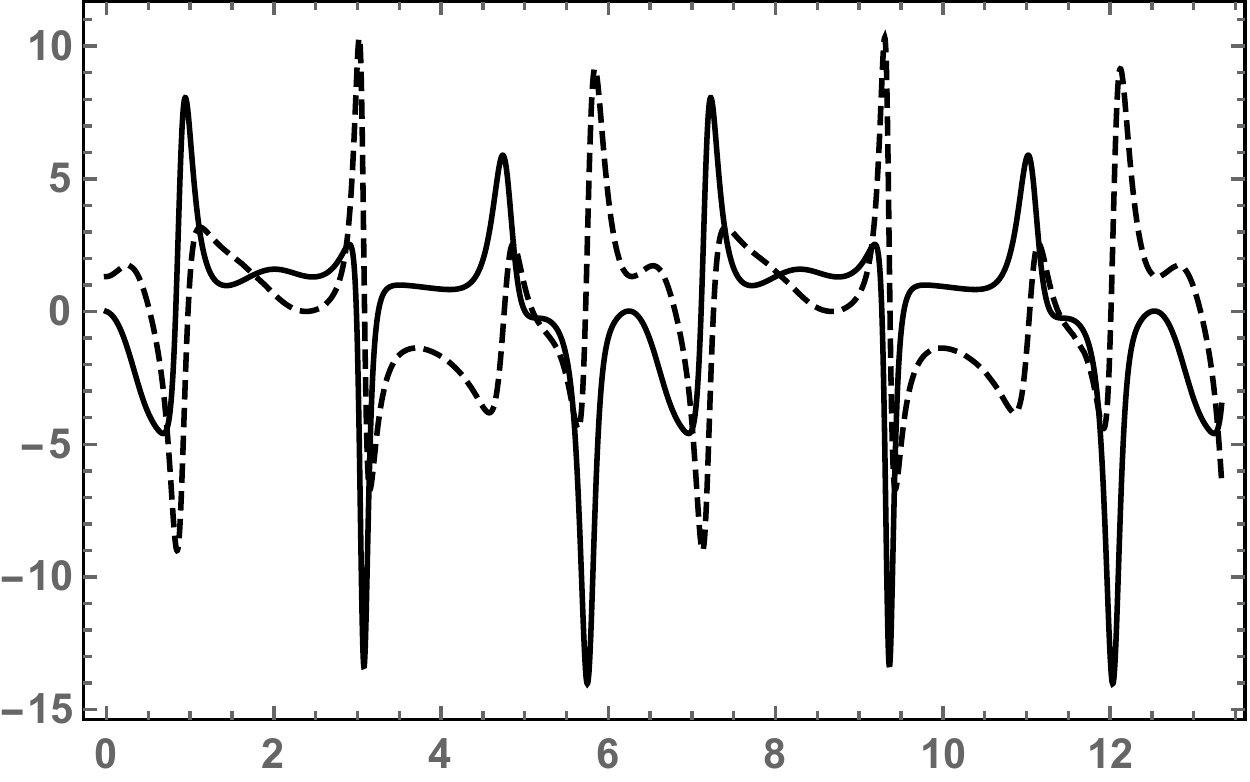}
\caption{ Initial value problem~(\ref{SystemzwN3}),~(\ref{N3Ex1}). Graphs of the real (bold curve) and imaginary (dashed curve) parts of the coordinate $w_3(t)$.}
\label{N3F111}
          \end{figure}
      \end{minipage}
      \hspace{0.05\linewidth}
      \begin{minipage}{0.45\linewidth}
          \begin{figure}[H]
\includegraphics[width=\linewidth]{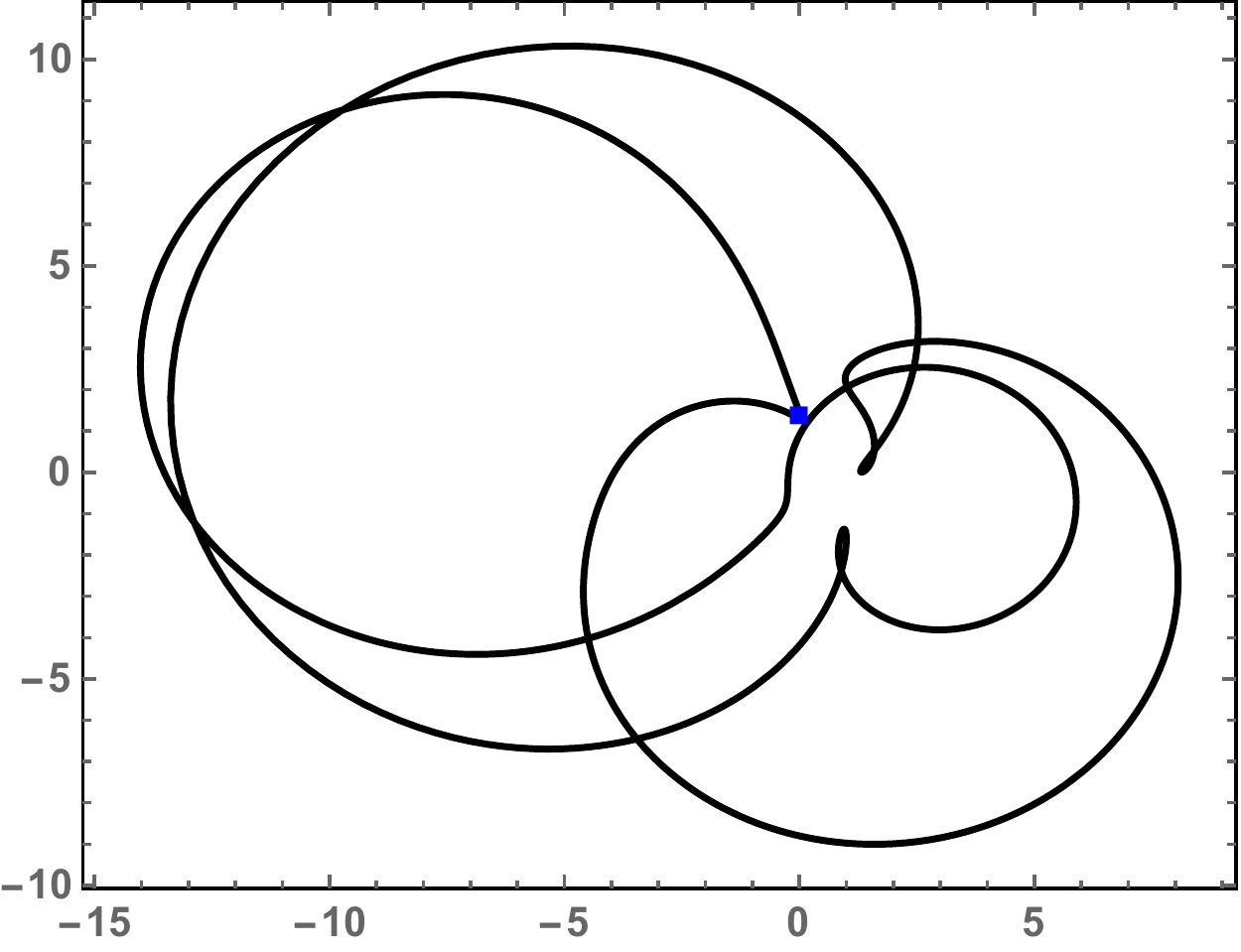}
\caption{Initial value problem~(\ref{SystemzwN3}),~(\ref{N3Ex1}). Trajectory, in the complex $z$-plane, of  $w_3(t)$; period $4\pi$. The   square indicates the initial condition $w_3(0)=1.32 \mathbf{i}$.}
\label{N3F112}
             \end{figure}
      \end{minipage}
  \end{minipage} 

\section{Mathematica Code}

In this appendix, we provide the Mathematica 10 code that was used for producing the plots in Appendix B. Let us recall that these plots illustrate Examples 1 and 2 of Section 2.



\section*{Supplementary material}

```mathematica
(* This is the Mathematica code used to produce the plots
 for the four examples within Example 1 in Section 2 of the paper
 "Novel Solvable Many-body Problems" by Oksana Bihun and Francesco Calogero*)
α1 = 1; α2 = 1; β1 = 1; β2 = 1; γ1 = 1; γ2 = 1; δ1 = 1; δ2 = 1;
N[Solve[λ^4 == α1 λ^3 + β1 λ^2 + γ1 λ + δ1, λ]]
N[Solve[λ^4 == α2 λ^3 + β2 λ^2 + γ2 λ + δ2, λ]]

(*Example 1*)
λ11 = I; λ12 = 2 I; λ13 = 3 I; λ14 = -I;
λ21 = 2 I; λ22 = 3 I; λ23 = -I; λ24 = I;
α1 = λ11 + λ12 + λ13 + λ14
β1 = -(λ11 λ12 + λ11 λ13 + λ11 λ14 + λ12 λ13 + λ12 λ14 + λ13 λ14)
γ1 = λ11 λ12 λ13 + λ11 λ12 λ14 + λ11 λ13 λ14 + λ12 λ13 λ14
δ1 = -λ11 λ12 λ13 λ14
α2 = λ21 + λ22 + λ23 + λ24
β2 = -(λ21 λ22 + λ21 λ23 + λ21 λ24 + λ22 λ23 + λ22 λ24 + λ23 λ24)
γ2 = λ21 λ22 λ23 + λ21 λ22 λ24 + λ21 λ23 λ24 + λ22 λ23 λ24
δ2 = -λ21 λ22 λ23 λ24
z10 = 1 + I; z1p0 = 1; z20 = 5 + I; z2p0 = 1;
w10 = 1;
w1p0 = I;
w20 = -I;
w2p0 = 1;

(*Example 2*)
λ11 = -1; λ12 = -2; λ13 = -I; λ14 = I;
λ21 = -1; λ22 = -2; λ23 = -I; λ24 = I;
α1 = λ11 + λ12 + λ13 + λ14
β1 = -(λ11 λ12 + λ11 λ13 + λ11 λ14 + λ12 λ13 + λ12 λ14 + λ13 λ14)
γ1 = λ11 λ12 λ13 + λ11 λ12 λ14 + λ11 λ13 λ14 + λ12 λ13 λ14
δ1 = -λ11 λ12 λ13 λ14
α2 = λ21 + λ22 + λ23 + λ24
β2 = -(λ21 λ22 + λ21 λ23 + λ21 λ24 + λ22 λ23 + λ22 λ24 + λ23 λ24)
γ2 = λ21 λ22 λ23 + λ21 λ22 λ24 + λ21 λ23 λ24 + λ22 λ23 λ24
δ2 = -λ21 λ22 λ23 λ24
z10 = -2 - I; z1p0 = 1; z20 = 2 + I; z2p0 = 1;
w10 = 1;
w1p0 = I;
w20 = -I;
w2p0 = 1;
```

```
(*Example 3*)
λ11 = -1; λ12 = -2; λ13 = -I; λ14 = Pi I;
λ21 = -1; λ22 = -2; λ23 = -I; λ24 = Pi I;
α1 = λ11 + λ12 + λ13 + λ14
β1 = - (λ11 λ12 + λ11 λ13 + λ11 λ14 + λ12 λ13 + λ12 λ14 + λ13 λ14)
γ1 = λ11 λ12 λ13 + λ11 λ12 λ14 + λ11 λ13 λ14 + λ12 λ13 λ14
δ1 = -λ11 λ12 λ13 λ14
α2 = λ21 + λ22 + λ23 + λ24
β2 = - (λ21 λ22 + λ21 λ23 + λ21 λ24 + λ22 λ23 + λ22 λ24 + λ23 λ24)
γ2 = λ21 λ22 λ23 + λ21 λ22 λ24 + λ21 λ23 λ24 + λ22 λ23 λ24
δ2 = -λ21 λ22 λ23 λ24
z10 = -2 - I; z1p0 = 1; z20 = 2 + I; z2p0 = -1;
w10 = I;
w1p0 = 1;
w20 = -I;
w2p0 = -1;

(*Example 4*)
λ11 = 0.04; λ12 = 0.062 + I; λ13 = 0.08 + 0.3 I; λ14 = 0.04 + 0.1 I;
λ21 = 0.02; λ22 = 0.032 + I; λ23 = 0.06 - 0.1 I; λ24 = 0.06 + 0.2 I;
α1 = λ11 + λ12 + λ13 + λ14
β1 = - (λ11 λ12 + λ11 λ13 + λ11 λ14 + λ12 λ13 + λ12 λ14 + λ13 λ14)
γ1 = λ11 λ12 λ13 + λ11 λ12 λ14 + λ11 λ13 λ14 + λ12 λ13 λ14
δ1 = -λ11 λ12 λ13 λ14
α2 = λ21 + λ22 + λ23 + λ24
β2 = - (λ21 λ22 + λ21 λ23 + λ21 λ24 + λ22 λ23 + λ22 λ24 + λ23 λ24)
γ2 = λ21 λ22 λ23 + λ21 λ22 λ24 + λ21 λ23 λ24 + λ22 λ23 λ24
δ2 = -λ21 λ22 λ23 λ24
z10 = -2 + 3 I; z1p0 = 7; z20 = 3 + 2 I; z2p0 = -5;
w10 = 2 + 4.2 I;
w1p0 = 4.5;
w20 = 3.1 I;
w2p0 = 2.4;
```

```
Tmax = 40;
s = NDSolve[{z1''[t] == w1[t], z2''[t] == w2[t],
    w1''[t] == (4 w1'[t] z2'[t] + 4 w2'[t] z1'[t] + 6 w1[t] w2[t]) / (z1[t] - z2[t]) + 1 / (z1[t] - z2[t])
       (z1[t] (α1 (w1'[t] + w2'[t]) + β1 (w1[t] + w2[t]) + γ1 (z1'[t] + z2'[t]) +
          δ1 (z1[t] + z2[t]))) - (α2 (w1'[t] z2[t] + 3 w1[t] z2'[t] + 3 z1'[t] w2[t] +
          z1[t] w2'[t]) + β2 (w1[t] z2[t] + 2 z1'[t] z2'[t] + z1[t] w2[t]) +
          γ2 (z1'[t] z2[t] + z1[t] z2'[t]) + δ2 (z1[t] z2[t]))),
    w2''[t] == - (4 w1'[t] z2'[t] + 4 w2'[t] z1'[t] + 6 w1[t] w2[t]) / (z1[t] - z2[t]) + 1 / (z1[t] - z2[t])
       (-z2[t] (α1 (w1'[t] + w2'[t]) + β1 (w1[t] + w2[t]) + γ1 (z1'[t] + z2'[t]) +
          δ1 (z1[t] + z2[t]))) + (α2 (w1'[t] z2[t] + 3 w1[t] z2'[t] + 3 z1'[t] w2[t] +
          z1[t] w2'[t]) + β2 (w1[t] z2[t] + 2 z1'[t] z2'[t] + z1[t] w2[t]) +
          γ2 (z1'[t] z2[t] + z1[t] z2'[t]) + δ2 (z1[t] z2[t]))),
    z1[0] == z10, z1'[0] == z1p0, z2[0] == z20, z2'[0] == z2p0, w1[0] == w10,
    w1'[0] == w1p0, w2[0] == w20, w2'[0] == w2p0}, {z1, z2, w1, w2}, {t, Tmax}];
```

```mathematica
Plot[{{Re[Evaluate[{z1[t]} /. s]][[1]][[1]]], Im[Evaluate[{z1[t]} /. s]][[1]][[1]]]},
 {t, 0, Tmax / 2}, PlotRange → All, Axes → False, Frame → True,
 FrameStyle → Directive[Bold, 12], PlotStyle → {Black, {Black, Dashed}, Thick}]
Graphz1 = ParametricPlot[{Re[Evaluate[{z1[t]} /. s]][[1]][[1]],
    Im[Evaluate[{z1[t]} /. s]][[1]][[1]]}, {t, 0, Tmax}, PlotRange → All,
   Axes → False, Frame → True, FrameStyle → Directive[Bold, 12],
   PlotStyle → {Black, Thick}, AspectRatio → 1 / 1.3];
Show[Graphz1, Graphics[
  {PointSize[Large], Text[Style[■, Medium, Blue], {Re[z10], Im[z10]}]
   }]]
PeriodGuess = N[2 Pi];
Plot[{{Re[Evaluate[{z1[t + PeriodGuess]} /. s]][[1]][[1]] -
     Re[Evaluate[{z1[t]} /. s]][[1]][[1]], PlotStyle → {Black, Thick}},
  Im[Evaluate[{z1[t + PeriodGuess]} /. s]][[1]][[1]] -
   Im[Evaluate[{z1[t]} /. s]][[1]][[1]]}, {t, 0, Tmax / 2}, PlotRange → All,
 Axes → True, AxesStyle → Directive[Bold, 12], PlotStyle → {Black, Thick}]


Plot[{{Re[Evaluate[{z2[t]} /. s]][[1]][[1]]], Im[Evaluate[{z2[t]} /. s]][[1]][[1]]]},
 {t, 0, Tmax}, PlotRange → All, Axes → False, Frame → True,
 FrameStyle → Directive[Bold, 12], PlotStyle → {Black, {Black, Dashed}, Thick}]
Graphz2 = ParametricPlot[{Re[Evaluate[{z2[t]} /. s]][[1]][[1]],
    Im[Evaluate[{z2[t]} /. s]][[1]][[1]]}, {t, 0, Tmax}, PlotRange → All,
   Axes → False, Frame → True, FrameStyle → Directive[Bold, 12],
   PlotStyle → {Black, Thick}, AspectRatio → 1 / 1.3];
Show[Graphz2, Graphics[
  {PointSize[Large], Text[Style[■, Medium, Blue], {Re[z20], Im[z20]}]
   }]]
PeriodGuess = N[2 Pi];
Plot[{{Re[Evaluate[{z2[t + PeriodGuess]} /. s]][[1]][[1]] -
     Re[Evaluate[{z2[t]} /. s]][[1]][[1]], PlotStyle → {Black, Thick}},
  Im[Evaluate[{z2[t + PeriodGuess]} /. s]][[1]][[1]] -
   Im[Evaluate[{z2[t]} /. s]][[1]][[1]]}, {t, 0, Tmax / 2}, PlotRange → All,
 Axes → True, AxesStyle → Directive[Bold, 12], PlotStyle → {Black, Thick}]


Plot[{{Re[Evaluate[{w1[t]} /. s]][[1]][[1]]], Im[Evaluate[{w1[t]} /. s]][[1]][[1]]]},
 {t, 0, Tmax / 3}, PlotRange → All, Axes → False, Frame → True,
 FrameStyle → Directive[Bold, 12], PlotStyle → {Black, {Black, Dashed}, Thick}]
Graphw1 = ParametricPlot[{Re[Evaluate[{w1[t]} /. s]][[1]][[1]],
    Im[Evaluate[{w1[t]} /. s]][[1]][[1]]}, {t, 0, Tmax}, PlotRange → All,
   Axes → False, Frame → True, FrameStyle → Directive[Bold, 12],
   PlotStyle → {Black, Thick}, AspectRatio → 1 / 1.3];
Show[Graphw1, Graphics[
  {PointSize[Large], Text[Style[■, Medium, Blue], {Re[w10], Im[w10]}]
   }]]
PeriodGuess = N[2 Pi];
Plot[{{Re[Evaluate[{w1[t + PeriodGuess]} /. s]][[1]][[1]] -
     Re[Evaluate[{w1[t]} /. s]][[1]][[1]], PlotStyle → {Black, Thick}},
  Im[Evaluate[{w1[t + PeriodGuess]} /. s]][[1]][[1]] -
   Im[Evaluate[{w1[t]} /. s]][[1]][[1]]}, {t, 0, Tmax / 4}, PlotRange → All,
 Axes → True, AxesStyle → Directive[Bold, 12], PlotStyle → {Black, Thick}]
```

```
Plot[{{Re[Evaluate[{w2[t]} /. s]][[1]][[1]]], Im[Evaluate[{w2[t]} /. s]][[1]][[1]]]},
 {t, 0, Tmax / 6}, PlotRange → All, Axes → False, Frame → True,
 FrameStyle → Directive[Bold, 12], PlotStyle → {Black, {Black, Dashed}, Thick}]
Graphw2 = ParametricPlot[{Re[Evaluate[{w2[t]} /. s]][[1]][[1]],
    Im[Evaluate[{w2[t]} /. s]][[1]][[1]]], {t, 0, Tmax / 2}, PlotRange → All,
   Axes → False, Frame → True, FrameStyle → Directive[Bold, 12],
   PlotStyle → {Black, Thick}, AspectRatio → 1 / 1.3];
Show[Graphw2, Graphics[
  {PointSize[Large], Text[Style[■, Medium, Blue], {Re[w20], Im[w20]}]
   }]]
PeriodGuess = N[2 Pi];
Plot[{{Re[Evaluate[{w2[t + PeriodGuess]} /. s]][[1]][[1]] -
     Re[Evaluate[{w2[t]} /. s]][[1]][[1]], PlotStyle → {Black, Thick}},
  Im[Evaluate[{w2[t + PeriodGuess]} /. s]][[1]][[1]] -
   Im[Evaluate[{w2[t]} /. s]][[1]][[1]]], {t, 0, Tmax / 2}, PlotRange → All,
 Axes → True, AxesStyle → Directive[Bold, 12], PlotStyle → {Black, Thick}]
```

```mathematica
(* This is the Mathematica code used to produce
 the plots for Example 2 in Section 2 of the paper
 "Novel Solvable Many-body Problems" by Oksana Bihun and Francesco Calogero*)

α1 = 1;  α2 = 1;  β1 = 1;  β2 = 1;  γ1 = 1;  γ2 = 1;  δ1 = 1;  δ2 = 1;
N[Solve[λ^4 == α1 λ^3 + β1 λ^2 + γ1 λ + δ1, λ]]
N[Solve[λ^4 == α2 λ^3 + β2 λ^2 + γ2 λ + δ2, λ]]

(*Example N3Ex1*)
λ11 = I; λ12 = 2 I; λ13 = 3 I; λ14 = -I;
λ21 = 2 I; λ22 = 3 I; λ23 = -2 I; λ24 = I;
λ31 = I; λ32 = 2 I; λ33 = -2 I; λ34 = -I;
α1 = λ11 + λ12 + λ13 + λ14
β1 = - (λ11 λ12 + λ11 λ13 + λ11 λ14 + λ12 λ13 + λ12 λ14 + λ13 λ14)
γ1 = λ11 λ12 λ13 + λ11 λ12 λ14 + λ11 λ13 λ14 + λ12 λ13 λ14
δ1 = -λ11 λ12 λ13 λ14
α2 = λ21 + λ22 + λ23 + λ24
β2 = - (λ21 λ22 + λ21 λ23 + λ21 λ24 + λ22 λ23 + λ22 λ24 + λ23 λ24)
γ2 = λ21 λ22 λ23 + λ21 λ22 λ24 + λ21 λ23 λ24 + λ22 λ23 λ24
δ2 = -λ21 λ22 λ23 λ24
α3 = λ31 + λ32 + λ33 + λ34
β3 = - (λ31 λ32 + λ31 λ33 + λ31 λ34 + λ32 λ33 + λ32 λ34 + λ33 λ34)
γ3 = λ31 λ32 λ33 + λ31 λ32 λ34 + λ31 λ33 λ34 + λ32 λ33 λ34
δ3 = -λ31 λ32 λ33 λ34
z10 = -1.45 + 1.1 I; z1p0 = 0.9;
z20 = 5.1 + 0.8 I; z2p0 = 1.2;
z30 = 2.5 - 0.2 I; z3p0 = -1.04;
w10 = 1.23;  w1p0 = 0.84 I; w20 = -2.26 I;   w2p0 = 2.16;
w30 = 1.32 I;  w3p0 = -1.12;

K1[z1_, z1p_, z2_, z2p_, z3_, z3p_, w1_, w1p_, w2_, w2p_, w3_, w3p_] =
 - (α1 (w1p + w2p + w3p) + β1 (w1 + w2 + w3) + γ1 (z1p + z2p + z3p) + δ1 (z1 + z2 + z3));


K2[z1_, z1p_, z2_, z2p_, z3_, z3p_, w1_, w1p_, w2_, w2p_, w3_, w3p_] = α2 (
      w1p (z2 + z3) + w2p (z1 + z3) + w3p (z1 + z2) + 2 w1 (z2p + z3p) +
      2 w2 (z1p + z3p) + 2 w3 (z1p + z2p) + z1p (w2 + w3) + z2p (w1 + w3) + z3p (w1 + w2)
    ) + β2 (
      w1 (z2 + z3) + w2 (z1 + z3) +
      w3 (z1 + z2) + z1p (z2p + z3p) + z2p (z1p + z3p) + z3p (z1p + z2p)
    ) + γ2 (
      z1p (z2 + z3) + z2p (z1 + z3) + z3p (z1 + z2)
    ) + δ2 (z1 z2 + z1 z3 + z2 z3);


K3[z1_, z1p_, z2_, z2p_, z3_, z3p_, w1_, w1p_, w2_, w2p_, w3_, w3p_] = - (
      α3 (w1p z2 z3 + w2p z1 z3 + w3p z1 z2 + 2 w1 (z2p z3 + z2 z3p) +
        2 w2 (z1p z3 + z1 z3p) + 2 w3 (z1p z2 + z1 z2p) + z1p (w2 z3 + 2 z2p z3p + z2 w3) +
        z2p (w1 z3 + 2 z1p z3p + z1 w3) + z3p (w1 z2 + 2 z1p z2p + z1 w2)
      )
      + β3 (
        w1 z2 z3 + w2 z1 z3 + w3 z1 z2 +
        z1p (z2p z3 + z2 z3p) + z2p (z1p z3 + z1 z3p) + z3p (z1p z2 + z1 z2p)
      ) + γ3 (
        z1p z2 z3 + z1 z2p z3 + z1 z2 z3p
      ) + δ3 z1 z2 z3
    );
```

```mathematica
Tmax = 20;
s = NDSolve[{z1''[t] == w1[t], z2''[t] == w2[t], z3''[t] == w3[t],
   w1''[t] == (4 w1'[t] z2'[t] + 4 w2'[t] z1'[t] + 6 w1[t] w2[t]) / (z1[t] - z2[t]) +
     (4 w1'[t] z3'[t] + 4 w3'[t] z1'[t] + 6 w1[t] w3[t]) / (z1[t] - z3[t]) - 1 / ((z1[t] - z2[t]) (z1[t] - z3[t]))
     (12 (w1[t] z2'[t] z3'[t] + z1'[t] (w2[t] z3'[t] + w3[t] z2'[t])) +
      z1[t]^2 K1[z1[t], z1'[t], z2[t], z2'[t], z3[t], z3'[t], w1[t], w1'[t],
       w2[t], w2'[t], w3[t], w3'[t]] + z1[t] K2[z1[t], z1'[t], z2[t], z2'[t],
       z3[t], z3'[t], w1[t], w1'[t], w2[t], w2'[t], w3[t], w3'[t]] + K3[z1[t],
       z1'[t], z2[t], z2'[t], z3[t], z3'[t], w1[t], w1'[t], w2[t], w2'[t], w3[
        t], w3'[t]]), w2''[t] == (4 w2'[t] z1'[t] + 4 w1'[t] z2'[t] + 6 w1[t] w2[t]) / (z2[t] - z1[t]) +
     (4 w2'[t] z3'[t] + 4 w3'[t] z2'[t] + 6 w2[t] w3[t]) / (z2[t] - z3[t]) - 1 / ((z2[t] - z1[t]) (z2[t] - z3[t]))
     (12 (w2[t] z1'[t] z3'[t] + z2'[t] (w1[t] z3'[t] + w3[t] z1'[t])) +
      z2[t]^2 K1[z1[t], z1'[t], z2[t], z2'[t], z3[t], z3'[t], w1[t], w1'[t],
       w2[t], w2'[t], w3[t], w3'[t]] + z2[t] K2[z1[t], z1'[t], z2[t], z2'[t],
       z3[t], z3'[t], w1[t], w1'[t], w2[t], w2'[t], w3[t], w3'[t]] + K3[z1[t],
       z1'[t], z2[t], z2'[t], z3[t], z3'[t], w1[t], w1'[t], w2[t], w2'[t], w3[
        t], w3'[t]]), w3''[t] == (4 w3'[t] z1'[t] + 4 w1'[t] z3'[t] + 6 w1[t] w3[t]) / (z3[t] - z1[t]) +
     (4 w3'[t] z2'[t] + 4 w2'[t] z3'[t] + 6 w2[t] w3[t]) / (z3[t] - z2[t]) - 1 / ((z3[t] - z1[t]) (z3[t] - z2[t]))
     (12 (w3[t] z1'[t] z2'[t] + z3'[t] (w1[t] z2'[t] + w2[t] z1'[t])) +
      z3[t]^2 K1[z1[t], z1'[t], z2[t], z2'[t], z3[t],
       z3'[t], w1[t], w1'[t], w2[t], w2'[t], w3[t], w3'[t]] +
      z3[t] K2[z1[t], z1'[t], z2[t], z2'[t], z3[t], z3'[t], w1[t], w1'[t],
       w2[t], w2'[t], w3[t], w3'[t]] + K3[z1[t], z1'[t], z2[t], z2'[t],
       z3[t], z3'[t], w1[t], w1'[t], w2[t], w2'[t], w3[t], w3'[t]]),
   z1[0] == z10, z1'[0] == z1p0, z2[0] == z20, z2'[0] == z2p0, z3[0] == z30,
   z3'[0] == z3p0, w1[0] == w10, w1'[0] == w1p0, w2[0] == w20, w2'[0] == w2p0,
   w3[0] == w30, w3'[0] == w3p0}, {z1, z2, z3, w1, w2, w3}, {t, Tmax}];
```

```
Plot[{{Re[Evaluate[{z1[t]} /. s]][[1]][[1]]], Im[Evaluate[{z1[t]} /. s]][[1]][[1]]]},
 {t, 0, Tmax}, PlotRange → All, Axes → False, Frame → True,
 FrameStyle → Directive[Bold, 12], PlotStyle → {Black, {Black, Dashed}, Thick}]
Graphz1 = ParametricPlot[{Re[Evaluate[{z1[t]} /. s]][[1]][[1]],
     Im[Evaluate[{z1[t]} /. s]][[1]][[1]]}, {t, 0, Tmax}, PlotRange → All,
    Axes → False, Frame → True, FrameStyle → Directive[Bold, 12],
    PlotStyle → {Black, Thick}, AspectRatio → 1 / 1.3];
Show[Graphz1, Graphics[
   {PointSize[Large], Text[Style[■, Medium, Blue], {Re[z10], Im[z10]}]
    }]]
PeriodGuess = N[2 Pi];
Plot[{{Re[Evaluate[{z1[t + PeriodGuess]} /. s]][[1]][[1]] -
     Re[Evaluate[{z1[t]} /. s]][[1]][[1]], PlotStyle → {Black, Thick}},
  Im[Evaluate[{z1[t + PeriodGuess]} /. s]][[1]][[1]] -
   Im[Evaluate[{z1[t]} /. s]][[1]][[1]]}, {t, 0, Tmax / 2}, PlotRange → All,
 Axes → True, AxesStyle → Directive[Bold, 12], PlotStyle → {Black, Thick}]

Plot[{{Re[Evaluate[{z2[t]} /. s]][[1]][[1]]], Im[Evaluate[{z2[t]} /. s]][[1]][[1]]]},
 {t, 0, Tmax}, PlotRange → All, Axes → False, Frame → True,
 FrameStyle → Directive[Bold, 12], PlotStyle → {Black, {Black, Dashed}, Thick}]
Graphz2 = ParametricPlot[{Re[Evaluate[{z2[t]} /. s]][[1]][[1]],
     Im[Evaluate[{z2[t]} /. s]][[1]][[1]]}, {t, 0, Tmax}, PlotRange → All,
    Axes → False, Frame → True, FrameStyle → Directive[Bold, 12],
    PlotStyle → {Black, Thick}, AspectRatio → 1 / 1.3];
Show[Graphz2, Graphics[
   {PointSize[Large], Text[Style[■, Medium, Blue], {Re[z20], Im[z20]}]
    }]]
PeriodGuess = N[2 Pi];
Plot[{{Re[Evaluate[{z2[t + PeriodGuess]} /. s]][[1]][[1]] -
     Re[Evaluate[{z2[t]} /. s]][[1]][[1]], PlotStyle → {Black, Thick}},
  Im[Evaluate[{z2[t + PeriodGuess]} /. s]][[1]][[1]] -
   Im[Evaluate[{z2[t]} /. s]][[1]][[1]]}, {t, 0, Tmax / 2}, PlotRange → All,
 Axes → True, AxesStyle → Directive[Bold, 12], PlotStyle → {Black, Thick}]

Plot[{{Re[Evaluate[{z3[t]} /. s]][[1]][[1]]], Im[Evaluate[{z3[t]} /. s]][[1]][[1]]]},
 {t, 0, Tmax}, PlotRange → All, Axes → False, Frame → True,
 FrameStyle → Directive[Bold, 12], PlotStyle → {Black, {Black, Dashed}, Thick}]
Graphz3 = ParametricPlot[{Re[Evaluate[{z3[t]} /. s]][[1]][[1]],
     Im[Evaluate[{z3[t]} /. s]][[1]][[1]]}, {t, 0, Tmax}, PlotRange → All,
    Axes → False, Frame → True, FrameStyle → Directive[Bold, 12],
    PlotStyle → {Black, Thick}, AspectRatio → 1 / 1.3];
Show[Graphz3, Graphics[
   {PointSize[Large], Text[Style[■, Medium, Blue], {Re[z30], Im[z30]}]
    }]]
PeriodGuess = N[2 Pi];
Plot[{{Re[Evaluate[{z3[t + PeriodGuess]} /. s]][[1]][[1]] -
     Re[Evaluate[{z3[t]} /. s]][[1]][[1]], PlotStyle → {Black, Thick}},
  Im[Evaluate[{z3[t + PeriodGuess]} /. s]][[1]][[1]] -
   Im[Evaluate[{z3[t]} /. s]][[1]][[1]]}, {t, 0, Tmax / 2}, PlotRange → All,
 Axes → True, AxesStyle → Directive[Bold, 12], PlotStyle → {Black, Thick}]
```

```
Plot[{{Re[Evaluate[{w1[t]} /. s]][[1]][[1]]}, Im[Evaluate[{w1[t]} /. s]][[1]][[1]]},
 {t, 0, Tmax / 2}, PlotRange → All, Axes → False, Frame → True,
 FrameStyle → Directive[Bold, 12], PlotStyle → {Black, {Black, Dashed}, Thick}]
Graphw1 = ParametricPlot[{Re[Evaluate[{w1[t]} /. s]][[1]][[1]],
    Im[Evaluate[{w1[t]} /. s]][[1]][[1]]}, {t, 0, Tmax / 2}, PlotRange → All,
   Axes → False, Frame → True, FrameStyle → Directive[Bold, 12],
   PlotStyle → {Black, Thick}, AspectRatio → 1 / 1.3];
Show[Graphw1, Graphics[
  {PointSize[Large], Text[Style[■, Medium, Blue], {Re[w10], Im[w10]}]
   }]]
PeriodGuess = N[2 Pi];
Plot[{{Re[Evaluate[{w1[t + PeriodGuess]} /. s]][[1]][[1]] -
     Re[Evaluate[{w1[t]} /. s]][[1]][[1]], PlotStyle → {Black, Thick}},
  Im[Evaluate[{w1[t + PeriodGuess]} /. s]][[1]][[1]] -
   Im[Evaluate[{w1[t]} /. s]][[1]][[1]]}, {t, 0, Tmax / 2}, PlotRange → All,
 Axes → True, AxesStyle → Directive[Bold, 12], PlotStyle → {Black, Thick}]

Plot[{{Re[Evaluate[{w2[t]} /. s]][[1]][[1]]}, Im[Evaluate[{w2[t]} /. s]][[1]][[1]]},
 {t, 0, Tmax / 2}, PlotRange → All, Axes → False, Frame → True,
 FrameStyle → Directive[Bold, 12], PlotStyle → {Black, {Black, Dashed}, Thick}]
Graphw2 = ParametricPlot[{Re[Evaluate[{w2[t]} /. s]][[1]][[1]],
    Im[Evaluate[{w2[t]} /. s]][[1]][[1]]}, {t, 0, Tmax / 2}, PlotRange → All,
   Axes → False, Frame → True, FrameStyle → Directive[Bold, 12],
   PlotStyle → {Black, Thick}, AspectRatio → 1 / 1.3];
Show[Graphw2, Graphics[
  {PointSize[Large], Text[Style[■, Medium, Blue], {Re[w20], Im[w20]}]
   }]]
PeriodGuess = N[2 Pi];
Plot[{{Re[Evaluate[{w2[t + PeriodGuess]} /. s]][[1]][[1]] -
     Re[Evaluate[{w2[t]} /. s]][[1]][[1]], PlotStyle → {Black, Thick}},
  Im[Evaluate[{w2[t + PeriodGuess]} /. s]][[1]][[1]] -
   Im[Evaluate[{w2[t]} /. s]][[1]][[1]]}, {t, 0, Tmax / 2}, PlotRange → All,
 Axes → True, AxesStyle → Directive[Bold, 12], PlotStyle → {Black, Thick}]

Plot[{{Re[Evaluate[{w3[t]} /. s]][[1]][[1]]}, Im[Evaluate[{w3[t]} /. s]][[1]][[1]]},
 {t, 0, Tmax / 2}, PlotRange → All, Axes → False, Frame → True,
 FrameStyle → Directive[Bold, 12], PlotStyle → {Black, {Black, Dashed}, Thick}]
Graphw3 = ParametricPlot[{Re[Evaluate[{w3[t]} /. s]][[1]][[1]],
    Im[Evaluate[{w3[t]} /. s]][[1]][[1]]}, {t, 0, Tmax}, PlotRange → All,
   Axes → False, Frame → True, FrameStyle → Directive[Bold, 12],
   PlotStyle → {Black, Thick}, AspectRatio → 1 / 1.3];
Show[Graphw3, Graphics[
  {PointSize[Large], Text[Style[■, Medium, Blue], {Re[w30], Im[w30]}]
   }]]
PeriodGuess = N[4 Pi];
Plot[{{Re[Evaluate[{w3[t + PeriodGuess]} /. s]][[1]][[1]] -
     Re[Evaluate[{w3[t]} /. s]][[1]][[1]], PlotStyle → {Black, Thick}},
  Im[Evaluate[{w3[t + PeriodGuess]} /. s]][[1]][[1]] -
   Im[Evaluate[{w3[t]} /. s]][[1]][[1]]}, {t, 0, Tmax / 3}, PlotRange → All,
 Axes → True, AxesStyle → Directive[Bold, 12], PlotStyle → {Black, Thick}]
```




\begin{thebibliography}{9}
\bibitem{C2015a} F. Calogero, "New solvable variants of the goldfish
many-body problem", Studies Appl. Math. (in press; published online
07.10.2015).

\bibitem{BC2015a} O. Bihun and F. Calogero, \textquotedblleft A new solvable
many-body problem of goldfish type\textquotedblright , J. Nonlinear Math.
Phys. (in press). arXiv:13749 [math-ph].

\bibitem{C2015b} F. Calogero, \textquotedblleft A solvable N-body problem of
goldfish type featuring $N^{2}$ arbitrary coupling
constants\textquotedblright , J. Phys. A: Math. Theor. (submitted to,
18.09.2015).

\bibitem{BC2015b} O. Bihun and F. Calogero, \textquotedblleft Generations of
monic polynomials such that the coefficients of the polynomials of the next
generation coincide with the zeros of a polynomial of the current generation,
and new solvable many-body problems\textquotedblright , Lett. Math. Phys.
(submitted to, 16.10.2015); arXiv: 1510.05017 [math-ph].

\bibitem{CG2008} F. Calogero and D. G\'{o}mez-Ullate, Asymptotically
isochronous systems, J. Nonlinear Math. Phys. \textbf{15}, 410-426 (2008).

\bibitem{C2008} F. Calogero, \textit{Isochronous systems}, Oxford University
Press, Oxford, 2008 (264 pages); marginally updated paperback edition (2012).

\bibitem{C2001} F. Calogero, \textit{Classical many-body problems amenable
to exact treatments}, Lectures Notes in Physics Monographs \textbf{m66},
Springer, Berlin Heidelberg, 2001 (749 pages).

\bibitem{GS2005} D. G\'{o}mez-Ullate and M. Sommacal, "Periods of the
goldfish many-body problem", J. Nonlinear Math. Phys. \textbf{12}, Suppl. 
\textbf{1}, 351--362 (2005).
\end{thebibliography}
\end{document}